\documentclass{elsarticle}
\usepackage[utf8]{inputenc}

\usepackage[font=normalsize, labelfont=bf]{caption}
\usepackage{subcaption}
\usepackage{overpic}
\usepackage{graphicx}
\usepackage{setspace}
\usepackage{lineno}
\usepackage{amssymb}
\usepackage{amsmath}
\usepackage{upgreek}
\usepackage{pgfplots}
\usepackage{bm}
\usepackage{soul,color}
\usepackage{epstopdf}
\usepackage{units}
\usepackage{lscape}
\usepackage{booktabs}
\usepackage{gensymb}
\usepackage{multirow}
\usepackage{floatrow}
\usepackage[flushleft]{threeparttable}
\usepackage{tikz}
\usepackage{placeins}
\usepackage{array}
\usepackage{tabularx}
\usepackage{algorithm}
\usepackage{algpseudocode}
\usepackage[a4paper]{geometry}
\usepackage{hyperref}

\modulolinenumbers[5]
\pgfplotsset{compat=1.8} % optional: slightly newer compat
\soulregister\cite7
\soulregister\ref7
\soulregister\pageref7

\hypersetup{
    colorlinks=true,
    linkcolor=blue,
    citecolor=blue,
    urlcolor=blue,
    pdftitle={Viscoelastic Droplet Impact on Surfaces with Sharp Wettability Contrast: Coupled Influence of Relaxation Time and Surface Tension},
}

\begin{document}
\begin{frontmatter}
\title{Viscoelastic Droplet Impact on Surfaces with Sharp Wettability Contrast: Coupled Influence of Relaxation Time and Surface Tension}

\author[1]{Mahmood Mousavi\corref{cor1}}
\author[1]{Parisa Tayerani}
\author[1]{Sebastian Stephens}
\author[1]{Cadence Ruskowski}
\author[2,3]{Bok Jik Lee\corref{cor2}}

\cortext[cor1]{Corresponding author: mousavs1@erau.edu}
\cortext[cor2]{Corresponding author: b.lee@snu.ac.kr}

\address[1]{School of Engineering, Embry-Riddle Aeronautical University, Daytona Beach, FL 32114, USA}
\address[2]{Department of Aerospace Engineering, Seoul National University, Seoul 08826, Republic of Korea}
\address[3]{Institute of Advanced Aerospace Technology, Seoul National University, Seoul 08826, Republic of Korea}

% Force the journal-required placement: email shown directly above the abstract
\vspace{0.5cm}

\begin{abstract}
The impact dynamics of viscoelastic droplets on solid surfaces play a critical role in numerous applications, including inkjet printing, spray coating, and microfluidics, where precise control of spreading, retraction, and rebound is essential. This numerical study investigates the coupled influence of fluid viscoelasticity, modeled via the Oldroyd-B constitutive equation, and gravitational-capillary balance on droplet behavior upon impact onto surfaces featuring sharp hybrid wettability. Employing a high-fidelity three-dimensional OpenFOAM-based solver that integrates the volume-of-fluid method, log-conformation formulation for improved numerical stability, and a velocity-dependent dynamic contact angle model, we simulated a 2~cm-diameter droplet impacting at 4~m/s across a range of relaxation times and surface tensions. Results demonstrate that increasing the relaxation time from 0.02~s to 0.12~s enhances elastic energy storage, leading to up to 12.9\% larger maximum spreading diameters (from $\sim$24.97~mm to 28.09--28.17~mm) and a 16.6\% reduction in minimum droplet height across uniform and hybrid surfaces. In contrast, increasing surface tension from 0.05~N/m to 0.15~N/m suppresses maximum spreading by about 1.1\% (from 27.21~mm to 26.90~mm) while increasing minimum height by 3.3\% (from 2.12~mm to 2.20~mm). On hybrid surfaces with static contact angles of $0^\circ$ and $160^\circ$, the sharp wettability contrast induces pronounced asymmetric spreading and directional fluid migration toward the hydrophilic region, ultimately producing distinctive dustpan- and shoe-like equilibrium morphologies. Variations in surface tension, which simultaneously modulate the Weber and E\"otv\"os numbers, reveal that stronger capillary forces suppress radial expansion while enhancing curvature-driven recoil and redistributing viscoelastic stresses.  These findings elucidate the intricate interplay among elastic stresses, capillary action, and surface wettability gradients, offering valuable insights for the design of advanced surfaces in droplet-based technologies.
\end{abstract}

\end{frontmatter}

\section{Introduction}

Droplet impact dynamics on solid surfaces is a fundamental phenomenon with wide-ranging applications in inkjet printing, spray cooling, pesticide application, and forensic science \cite{sotoudeh2021understanding}. The outcome of an impact---spreading, recoiling, splashing, or bouncing---is governed by a complex interplay of inertia, capillarity, and viscous dissipation. Early studies by Rein~\cite{rein1993phenomena} and Yarin and Weiss~\cite{yarin1995impact} established the primary impact regimes, while modern CFD and high-speed imaging have refined our understanding of gas-cushioning and energy budget partitioning~\cite{josserand2016drop}.More recently, attention has shifted toward complex fluids, as Newtonian models often fail to capture the behavior of polymer solutions, suspensions, and bioaerosols. For instance, recent studies on mucosalivary film fragmentation have shown that viscoelasticity can reduce droplet size during respiratory events by promoting the formation of thinner, more uniform liquid sheets prior to rupture~\cite{li2025viscoelasticity}. Similarly, the impact of fiber suspensions reveals that anisotropic particles significantly increase bulk viscosity and modify spreading dynamics, with the droplet size decreasing as the fiber volume fraction increases~\cite{rajesh2025impact}. These findings underscore the necessity of moving beyond Newtonian assumptions to accurately model real-world fluid interactions.

The inclusion of non-Newtonian properties, whether through shear-thinning or shear-thickening behaviors or viscoelastic additives, drastically alters the energy budget during impact. Mobaseri et al.~\cite{mobaseri2025maximum} established a general framework for predicting the maximum spreading of generalized Newtonian liquids by identifying a characteristic shear rate that governs viscous dissipation. When considering viscoelasticity, the stored elastic energy in polymer chains often promotes faster contact line advancement while suppressing splashing through high extensional viscosity~\cite{izbassarov2016effects}. However, the effects are highly dependent on the impact scenario; for example, in impacts onto deep viscoelastic pools, increasing the Weissenberg number can stabilize or destabilize droplet breakup depending on its magnitude~\cite{ss2025impact}. On inclined surfaces, fluid elasticity has been found to reduce maximum spreading while increasing the velocity and value of receding~\cite{norouzi2021experimental}. Despite these advances, a significant gap remains in understanding the non-linear coupling between internal viscoelastic stresses and extreme, heterogeneous wettability patterns.

Surface engineering through patterned superhydrophobic and superhydrophilic regions offers a promising route for passive droplet control. Hybrid surfaces can induce asymmetric spreading and directional transport, yet most existing numerical frameworks either simplify the fluid as Newtonian or isolate the wettability effect from gravitational forces \cite{mousavi2023impact,mousavi2022effect,mousavi2024potential,mousavi2023effects}. This limitation is particularly problematic for non-Newtonian liquids. For instance, adding Xanthan to water can reduce the contact time on superhydrophobic surfaces by up to 50\% by altering the separation morphology from vertical jetting to ``mushroom''-like structures~\cite{biroun2023impact}. In inkjet printing, spray coating, and 3D printing, viscoelastic inks frequently land on surfaces with sharp wettability contrasts, where precise control of asymmetric spreading and final morphology is essential for improving throughput and reducing waste \cite{pendar2022review,kaushik2018droplet,elkaseer2022effect,huynh2024efficient,bell2022anti}. Furthermore, recent axisymmetric simulations of viscoelastic drops on thin films show that fluid elasticity can enlarge crown dimensions, a process controlled by extensional forces in the crown wall at intermediate Weissenberg numbers~\cite{rezaie2023viscoelastic}.

Despite substantial progress in understanding droplet impact dynamics for both Newtonian and complex fluids, three key limitations persist in the literature. Most studies on viscoelastic droplets assume homogeneous wettability, the coupled influence of gravitational forces through systematic variation of the E\"otv\"os number has rarely been examined in the viscoelastic context, and the nonlinear coupling between elastic stress relaxation, contact-line pinning, and sharp wettability discontinuities remains largely uncharacterized. To address these limitations, the present study develops a high-fidelity three-dimensional numerical framework in OpenFOAM that simultaneously accounts for Oldroyd-B viscoelastic rheology, gravitational-capillary balance through systematic variation of the E\"otv\"os number, and dynamic contact angle hysteresis on hybrid hydrophilic--hydrophobic surfaces. This approach provides, for the first time, a fully coupled framework in which Oldroyd-B viscoelasticity, dynamic contact angle effects, and sharp wettability discontinuities are resolved together under gravitational loading---conditions representative of centimetre-scale droplets encountered in practical coating and printing operations.

\section{Physical Problem and Computational Domain}

\begin{figure}[H]
\centering
\includegraphics[width=0.8\linewidth]{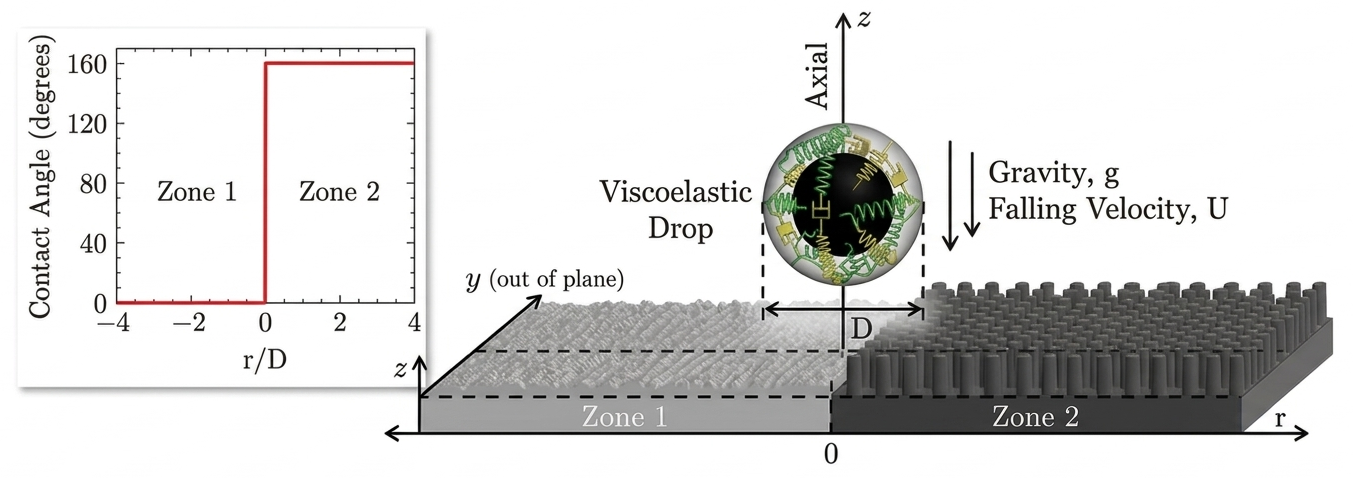}
\caption{A schematic representation of the physical domain. A three-dimensional viscoelastic droplet with a diameter of $D = 2~\mathrm{cm}$ impacts a hybrid wettability surface with an initial falling velocity of $U = 4~\mathrm{m/s}$ under the influence of gravity. Zone~1 is hydrophilic with $\mathrm{WCA} = 0^\circ$, while Zone~2 is superhydrophobic with $\mathrm{WCA} = 160^\circ$.}
\label{fig:physical_domain}
\end{figure}

This study investigates the three-dimensional time-dependent impact dynamics of a viscoelastic droplet moving through a surrounding Newtonian medium (air) and interacting with a solid surface featuring hybrid wettability, as illustrated in Fig.~\ref{fig:physical_domain}. The viscoelastic behavior of the droplet is described using the Oldroyd-B constitutive model~\cite{faroughi2020closure}, which accounts for polymeric stress evolution through relaxation and retardation mechanisms, while the liquid--gas interface is tracked using the volume-of-fluid (VOF) method~\cite{mousavi2023impact} coupled with a dynamic wall contact angle formulation. In the computational setup, a 3D droplet of diameter $D = 2~\mathrm{cm}$ is initially positioned at a height $H = 2D$ above the substrate and impacts with a velocity of $U = 4~\mathrm{m/s}$ under gravitational acceleration. The solid surface is divided into two distinct regions, Zone~1 and Zone~2, separated along the plane passing through the droplet impact center ($x=0$), thereby introducing a sharp wettability discontinuity in the lateral direction, with Zone~1 being hydrophilic ($\mathrm{WCA} = 0^\circ$) and Zone~2 being superhydrophobic ($\mathrm{WCA} = 160^\circ$). The computational domain is bounded by open pressure conditions at the top and lateral boundaries, while the bottom surface is treated as a no-slip wall subject to the imposed dynamic contact angle model. This three-dimensional configuration enables a detailed examination of the coupled effects of viscoelasticity and heterogeneous wettability on droplet deformation, asymmetric spreading, contact-line motion, and rebound dynamics by capturing complex flow features---such as lateral fluid migration and asymmetric stress distributions.

\subsection{Governing Equations and Dimensionless Analysis}

To simulate the three-dimensional two-phase flow system using the volume-of-fluid (VOF) method, the governing equations include the incompressibility condition $\nabla \cdot \mathbf{u} = 0$ and the momentum conservation equation
\begin{equation}
\rho \left( \frac{\partial \mathbf{u}}{\partial t} + \mathbf{u} \cdot \nabla \mathbf{u} \right) = -\nabla p + \nabla \cdot \boldsymbol{\tau} + \mathbf{f} + \rho \mathbf{g},
\end{equation}
where $\mathbf{u}$ is the velocity field, $p$ is the pressure, $\boldsymbol{\tau}$ is the total extra stress tensor, $\mathbf{f}$ is the surface tension force, and $\mathbf{g}$ is gravitational acceleration. The total extra stress tensor is decomposed as $\boldsymbol{\tau} = \boldsymbol{\tau}_s + \mathbf{T}$, where $\boldsymbol{\tau}_s = \eta_s (\nabla \mathbf{u} + (\nabla \mathbf{u})^T)$ denotes the Newtonian solvent contribution with solvent viscosity $\eta_s$, and $\mathbf{T}$ represents the polymeric viscoelastic stress governed by the Oldroyd-B constitutive equation
\begin{equation}
\mathbf{T} + \lambda \left( \frac{\partial \mathbf{T}}{\partial t} + \mathbf{u} \cdot \nabla \mathbf{T} - \mathbf{T} \cdot \nabla \mathbf{u} - (\nabla \mathbf{u})^T \cdot \mathbf{T} \right) = \eta_p (\nabla \mathbf{u} + (\nabla \mathbf{u})^T),
\end{equation}
with relaxation time $\lambda$, polymer viscosity $\eta_p$, and total liquid viscosity $\eta_l = \eta_s + \eta_p$. The retardation time is defined as $\lambda_r = \lambda \eta_s / \eta_l$. Fluid properties are interpolated using the liquid volume fraction $\alpha$ according to $\rho = \rho_l \alpha + \rho_g (1-\alpha)$ and $\eta = \eta_l \alpha + \eta_g (1-\alpha)$, where the subscripts $l$ and $g$ refer to the liquid and gas phases, respectively. The volume fraction evolves via the advection equation $\partial \alpha / \partial t + \mathbf{u} \cdot \nabla \alpha = 0$. Surface tension is incorporated through the continuum surface force model $\mathbf{f} = \sigma \kappa \frac{\bar{\rho} \nabla \alpha}{\frac{1}{2}(\rho_l + \rho_g)}$, where $\sigma$ is the surface tension, $\kappa = -\nabla \cdot \mathbf{n}$ is the interface curvature, and $\bar{\rho}$ is the averaged density. 

The dynamic contact angle model captures the well-known hydrodynamic dependence of the contact angle on contact-line speed: the advancing angle $\theta_d$ increases above the static value as the contact line advances ($\mathrm{Ca} > 0$), while the receding angle $\theta_r$ decreases, both effects reflecting the viscous dissipation in the wedge region near the moving contact line. Dynamic contact line behavior is modeled using a velocity-dependent formulation based on Kistler's implementation of the Hoffman function~\cite{hoffman1975study, kistler1993hydrodynamics}, with advancing and receding conditions determined by the sign of the contact line velocity $V_{cl} = \mathbf{n}_w \times \mathbf{u}_{cell} \times \mathbf{n}_w$. The dynamic contact angles are given by
\begin{equation}
\theta_d = f_H\left(Ca + f_H^{-1}(\theta_s)\right), \quad
\theta_r = (\theta_s - 72 Ca)^{1/3},
\end{equation}
where $Ca = \eta_l U / \sigma$ is the capillary number based on the local contact line speed $U$, $\theta_s$ is the static contact angle, and the Hoffman function is expressed as
\begin{equation}
f_H(x) = \arccos\left(1 - 2 \tanh\left(5.16 \left( \frac{x}{1 - 1.31 x^{0.99}} \right)^{0.706} \right) \right).
\end{equation}

To characterize the balance of forces, the E\"otv\"os number $Eo = (\rho_l - \rho_g) g D^2 / \sigma$ quantifies the relative importance of gravitational to capillary forces, the Weber number $We = \rho_l U^2 D / \sigma$ represents inertial to capillary forces, and the Morton number $Mo = g \mu_g^4 (\rho_l - \rho_g) / (\sigma^3 \rho_g^2)$ is introduced with fluid properties $\rho_l = 1000~\mathrm{kg/m^3}$, $\rho_g = 1~\mathrm{kg/m^3}$, $g = 9.81~\mathrm{m/s^2}$, $D = 0.02~\mathrm{m}$, and $\mu_g = 1 \times 10^{-5}~\mathrm{Pa \cdot s}$. In this study, variations in surface tension $\sigma$ simultaneously modify both $We$ and $Eo$, enabling systematic investigation of capillary dominance across inertia-dominated to capillarity-dominated regimes while the Morton number remains small, indicating negligible gas viscosity effects. This dimensionless framework provides a consistent basis for interpreting the interactions among inertia, gravity, capillarity, and elasticity during viscoelastic droplet impact on heterogeneous surfaces.

\subsection{Numerical implementation}
The numerical framework employed in this study is based on the rheoInterFoam solver from the RheoTool package implemented in OpenFOAM version 9 \cite{pimenta2021conjugate}. Designed specifically for complex rheological flows, the solver has been adapted for three-dimensional viscoelastic droplet impact under laminar conditions. To ensure numerical stability at elevated relaxation times, a log-conformation formulation of the stress tensor is adopted following Afonso et al.~\cite{afonso2009log}, which improves robustness by transforming the constitutive equations into logarithmic space. Furthermore, the dynamic contact angle model described earlier is directly coupled with the solver, allowing the contact angle to evolve according to local hydrodynamic conditions and thereby ensuring accurate resolution of wetting, spreading, and recoiling at the contact line. The model updates the contact angle at each time step based on the local velocity and capillary effects, maintaining full consistency with the momentum equation. Temporal discretization is performed using a first-order implicit Euler scheme \cite{jasak1996error}, while spatial derivatives are evaluated using Gauss linear discretization. Convective terms are treated with specialized high-resolution schemes: the \texttt{limitedLinearV} scheme for momentum \cite{sweby1984high}, the van Leer scheme for phase fraction transport \cite{vanleer1974towards}, and the CUBISTA scheme for viscoelastic stress transport \cite{alves2003convergent}. Pressure--velocity coupling is handled using the SIMPLEC algorithm \cite{vandoormaal1984enhancements} with one outer and one inner corrector per time step. In addition, the volume fraction equation is stabilized through two sub-cycles and MULES correction to preserve a sharp interface \cite{zalesak1979fully}. Linear systems are solved using the Preconditioned Conjugate Gradient (PCG) method with Diagonal Incomplete Cholesky (DIC) preconditioning for pressure, and the Preconditioned Bi-Conjugate Gradient (PBiCG) method with Diagonal Incomplete LU (DILU) preconditioning for velocity and stress fields \cite{jasak2007openfoam}. A strict convergence tolerance of $10^{-10}$ is imposed for all equations to ensure high numerical accuracy. To further enhance stability, the maximum Courant--Friedrichs--Lewy (CFL) number is limited to 0.01, a constraint that is essential for resolving viscoelastic flows where strong normal stresses and sharp gradients near the interface and contact line can induce numerical instability. Consequently, the time step is automatically adjusted during the simulation, ranging from $10^{-8}$~s to $10^{-10}$~s, with the smallest values occurring during the initial impact and rapid deformation stages where velocity gradients and elastic stresses are highest.

\section{Results and Discussion}\label{sec:results}
\subsection{Grid Study and Model Verification}
\begin{figure}[H]
\centering
\caption{Grid sensitivity study of the droplet interface profile for viscoelastic droplet impact on a uniformly hydrophilic surface (\( \mathrm{WCA} = 0^\circ \)) with relaxation time \( \lambda = 0.02~\mathrm{s} \). The interface height is plotted as a function of wetted radius for four mesh resolutions: \( \Delta x / D = 2 \times 10^{-3} \), \( 1.5 \times 10^{-3} \), \( 1 \times 10^{-3} \), and \( 0.8 \times 10^{-3} \). Results are shown at (a) \( t = 0.01~\mathrm{s} \), corresponding to maximum spreading, and (b) \( t = 0.02~\mathrm{s} \), where the droplet approaches its equilibrium configuration.}
\label{fig:grid_study_plots}

    % Row 1: $\lambda$ = 0.01
    \begin{subfigure}[b]{0.475\linewidth}
        \centering
        \includegraphics[width=\linewidth]{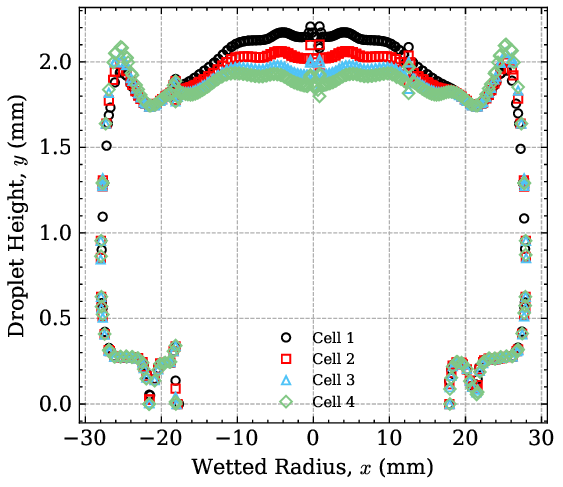}
            \subcaption{}
    \end{subfigure}
    \hfill
    \begin{subfigure}[b]{0.475\linewidth}
        \centering
        \includegraphics[width=\linewidth]{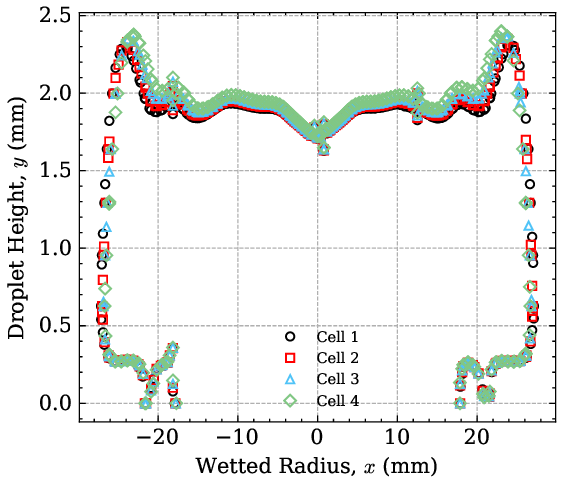}
            \subcaption{}
    \end{subfigure}
    % Row 2: $\lambda$ = 0.02

\end{figure}

Figure~\ref{fig:grid_study_plots} demonstrates the influence of mesh resolution on the predicted droplet interface profile at two representative stages of the impact process. At \( t = 0.01~\mathrm{s} \), corresponding to maximum spreading (Fig.~\ref{fig:grid_study_plots}a), the coarse mesh (\( \Delta x / D = 2 \times 10^{-3} \)) exhibits noticeable deviations in the interface shape, particularly near the contact line region where curvature and gradients are highest. In contrast, the finer meshes show improved agreement, capturing the detailed interface structure with reduced numerical error. At \( t = 0.02~\mathrm{s} \), when the droplet approaches its equilibrium state (Fig.~\ref{fig:grid_study_plots}b), all meshes predict a similar overall profile; however, small discrepancies persist in regions of high curvature and near the contact line for coarser resolutions. The solutions obtained using \( \Delta x / D = 1 \times 10^{-3} \) and \( 0.8 \times 10^{-3} \) demonstrate mesh convergence due to the consistency between results. Based on this observation, the mesh with \( \Delta x / D = 1 \times 10^{-3} \) is selected for subsequent simulations, as it provides a balance between computational efficiency and sufficient accuracy in resolving interface dynamics and contact line behavior.

\begin{figure}[H]
\centering
\caption{Cell quality distribution of the computational mesh during droplet impact on a uniformly hydrophilic surface (\( \mathrm{WCA} = 0^\circ \)) with \( \lambda = 0.02~\mathrm{s} \). The results are shown at (a) \( t = 0.005~\mathrm{s} \) (early spreading stage) and (b) \( t = 0.01~\mathrm{s} \) (maximum receding stage). The mesh resolution corresponds to \( \Delta x / D = 1.5 \times 10^{-3} \). The majority of cells maintain high quality (values below 3), while slightly elevated values appear near refinement transitions and high-deformation regions close to the interface.}
\label{fig:mesh_quality}

\begin{subfigure}[b]{0.48\linewidth}
    \centering
    \includegraphics[width=\linewidth]{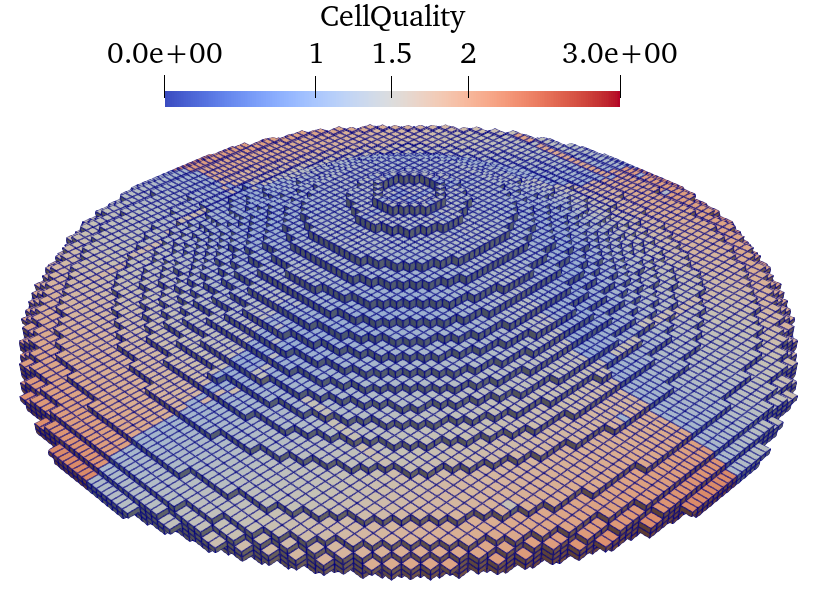}
    \subcaption{$t = 0.005~\mathrm{s}$}
\end{subfigure}
\hspace{0.02\linewidth}
\begin{subfigure}[b]{0.48\linewidth}
    \centering
    \includegraphics[width=\linewidth]{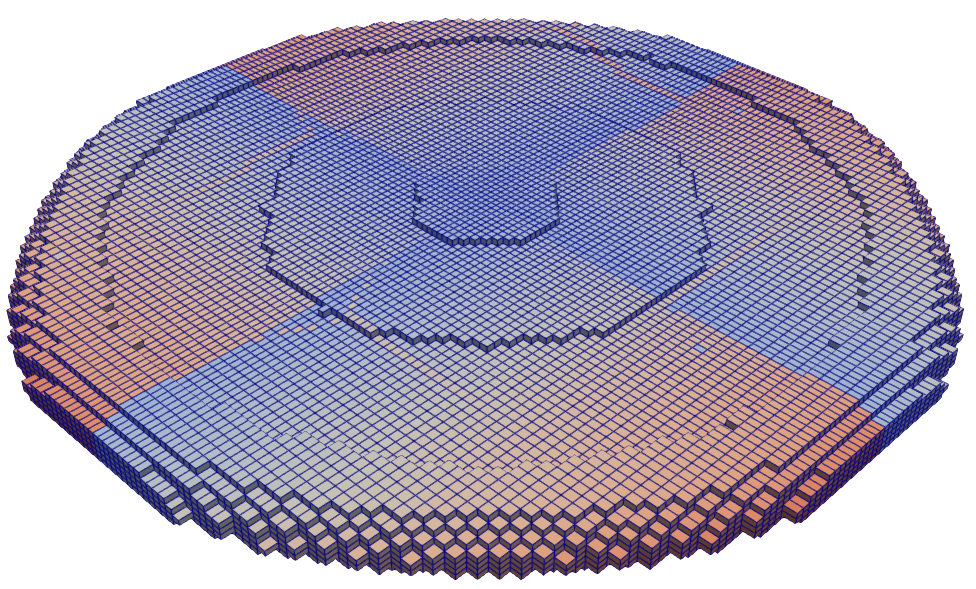}
    \subcaption{$t = 0.01~\mathrm{s}$}
\end{subfigure}

\end{figure}

Figure~\ref{fig:mesh_quality} illustrates the evolution of mesh quality during different stages of droplet dynamics. At early time (\( t = 0.005~\mathrm{s} \)), corresponding to the initial spreading phase, the mesh remains highly uniform with cell quality values predominantly below 2. This indicates the elements are well conditioned across the domain. As the droplet evolves toward maximum receding (\( t = 0.01~\mathrm{s} \)), slight variations in cell quality appear near the interface and refinement regions due to strong deformation and localized gradients. Nevertheless, the cell quality remains within an acceptable range (below approximately 3), confirming that the mesh maintains good numerical properties throughout the simulation. The absence of highly distorted cells ensures stable computation of viscoelastic stresses and accurate resolution of interface dynamics, particularly in regions of large curvature and contact line motion.

%*************************************************************************************

\section{Numerical Validation and Verification}

\begin{figure}[H]
\centering

\begin{subfigure}{0.48\textwidth}
\centering
\includegraphics[width=\linewidth]{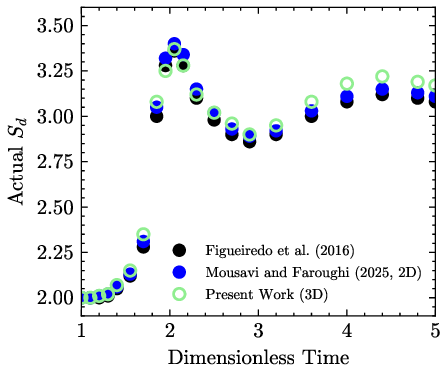}
\caption{}
\label{fig:comparison_plot}
\end{subfigure}
\hfill
\begin{subfigure}{0.48\textwidth}
\centering
\includegraphics[width=\linewidth]{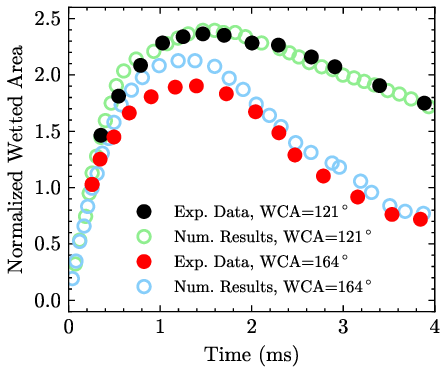}
\caption{}
\label{fig:wetted_area_graph}
\end{subfigure}

\vspace{1.5em}

\begin{subfigure}{\textwidth}
\centering
\includegraphics[width=0.6\linewidth]{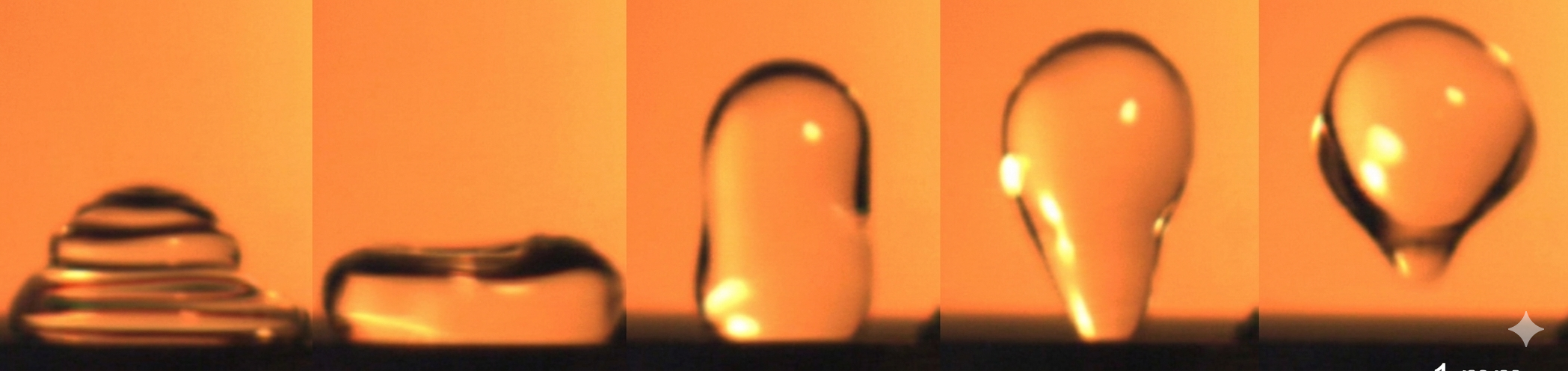}\\

\includegraphics[width=0.6\linewidth]{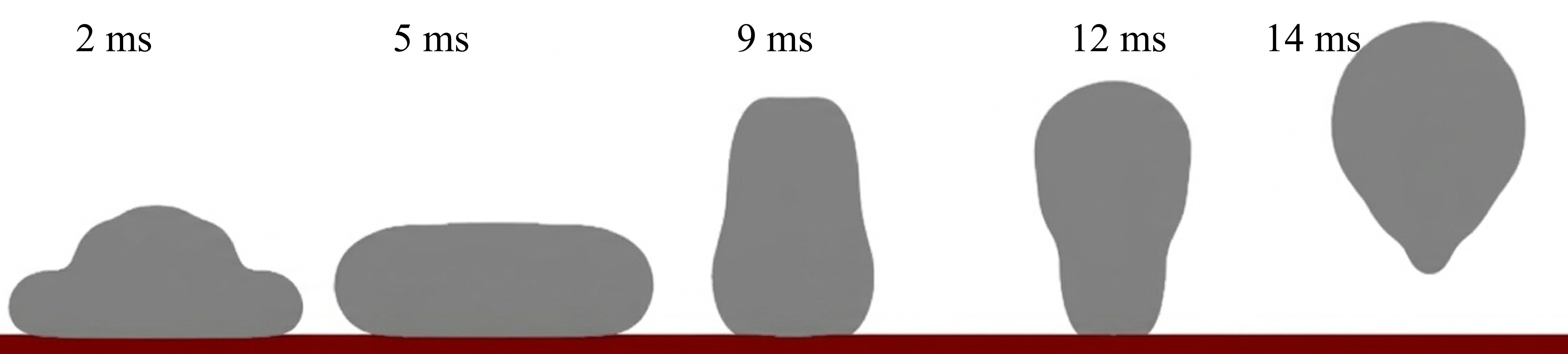}
\caption{}
\label{fig:snapshot_comparison}
\end{subfigure}

\caption{Validation of the numerical approach: (a) spreading diameter evolution benchmarked against Figueiredo et al. \cite{figueiredo2016two}, Mousavi and Faroughi \cite{mousavi2025spreading} (2D), and present 3D results; (b) normalized wetted area versus time for wall contact angles of $121^\circ$ and $164^\circ$ (Kim et al. \cite{kim2012drop}); (c) qualitative snapshot comparison between experimental results (top) and present numerical simulations (bottom) for Newtonian droplet impact on a superhydrophobic surface.}
\label{fig:validation_combined}
\end{figure}

To ensure the precision and reliability of the developed numerical framework, a multi-stage validation process was conducted that included benchmarking against established numerical studies for viscoelastic droplets and direct comparisons with experimental data from the literature for Newtonian impacts by setting the relaxation time $\lambda = 0$ in the solver. As illustrated in Fig.~\ref{fig:validation_combined}, the present numerical predictions exhibit excellent agreement with the benchmark data of Figueiredo et al. \cite{figueiredo2016two} as well as the 2D results of Mousavi and Faroughi \cite{mousavi2025spreading} and the present 3D simulations for the spreading diameter evolution of viscoelastic droplets (Fig.~\ref{fig:validation_combined}a). Quantitative experimental validation is further demonstrated in Fig.~\ref{fig:validation_combined}b through comparison of the normalized wetted area against the measurements of Kim et al. \cite{kim2012drop} for wall contact angles of $121^\circ$ and $164^\circ$. Finally, Fig.~\ref{fig:validation_combined}c provides qualitative validation by direct comparison of temporal snapshots between the experimental observations of Wang et al. \cite{wang2007impact} (top row) and the current numerical simulations (bottom row), accurately reproducing the morphological evolution during spreading, maximum extension, and rebound phases. This consistent agreement across numerical benchmarks, quantitative plots, and visual comparisons confirms the robustness and accuracy of the numerical scheme for simulating droplet impact physics.

%*************************************************************

%%%%%%%%%%%%%%%%%%%%%%%%%%%%%%%%%%%%%%%%%%%%
\subsection{Droplet Dynamics under Coupled Wettability and Viscoelastic Effects}

\begin{figure}[H]
\centering
    \begin{subfigure}[b]{0.32\linewidth}
        \centering
        \includegraphics[width=\linewidth]{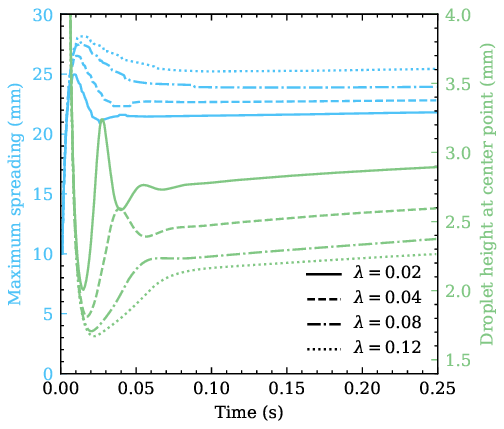}
        \subcaption{\( WCA = 0^\circ \)}
    \end{subfigure}
    \hfill
    \begin{subfigure}[b]{0.32\linewidth}
        \centering
        \includegraphics[width=\linewidth]{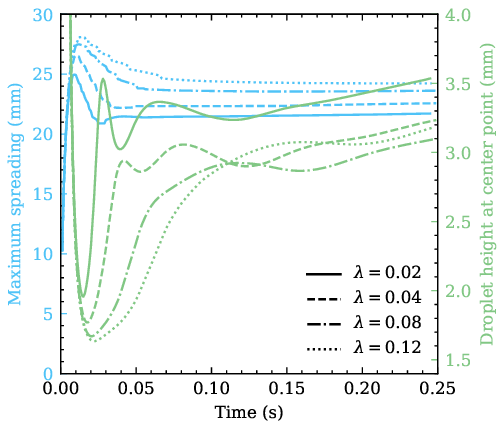}
        \subcaption{\( WCA = 0^\circ\!-\!160^\circ \)}
    \end{subfigure}
    \hfill
    \begin{subfigure}[b]{0.33\linewidth}
        \centering
        \includegraphics[width=\linewidth]{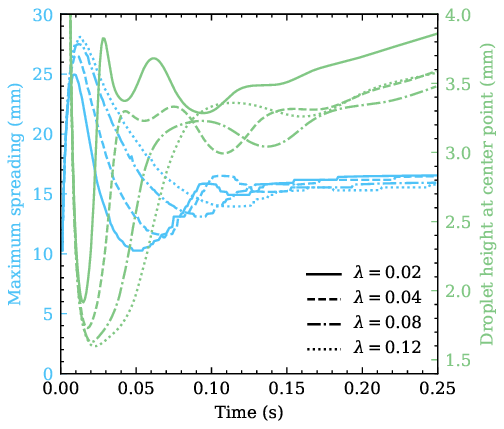} % Placeholder
        \subcaption{\( WCA = 160^\circ \)}
    \end{subfigure}
    \caption{Effect of relaxation time \( \lambda \) on the time evolution of the actual spreading diameter \( S_d \) for the three surface configurations.}
    \label{fig:lambda_effect_lowEo}
\end{figure}

This section examines the coupled effects of surface wettability and fluid viscoelasticity on droplet impact dynamics across uniformly hydrophilic ($\mathrm{WCA} = 0^\circ$), hybrid ($\mathrm{WCA} = 0^\circ$--$160^\circ$), and uniformly hydrophobic ($\mathrm{WCA} = 160^\circ$) surfaces. The results in Fig.~\ref{fig:lambda_effect_lowEo} show that increasing the relaxation time from $\lambda = 0.02~\mathrm{s}$ to $\lambda = 0.12~\mathrm{s}$ enlarges the maximum spreading diameter by 12.7--12.9\% (from approximately 24.94~mm to 28.09--28.17~mm) across all configurations. This increase occurs because longer relaxation times allow greater elastic energy storage, which delays viscous dissipation and extends the inertial spreading phase \cite{bartolo2007dynamics,bergeron2000controlling,wang2017impact}. On the uniformly hydrophilic surface, strong wetting enhances contact line mobility and largely suppresses recoil, so the effect of increasing $\lambda$ mainly appears as greater spreading, consistent with wetting-dominated dynamics where capillary adhesion governs the final state \cite{rafai2004spreading,josserand2016drop}. In contrast, the uniformly hydrophobic surface exhibits reduced spreading and more pronounced recoil, where stored elastic energy is efficiently converted into kinetic energy due to weaker solid--liquid adhesion, thereby amplifying retraction and internal oscillations \cite{crooks2000influence,bertola2009experimental}.

On the hybrid surface, the sharp wettability discontinuity creates an asymmetric distribution of capillary forces and contact line resistance. Zone~1 ($0^\circ$) promotes spreading through strong capillary attraction, while Zone~2 ($160^\circ$) resists wetting and enhances recoil, generating a lateral capillary pressure gradient that drives fluid migration toward the hydrophilic region \cite{darhuber2005principles,quere2008wetting}. The viscoelastic nature of the fluid amplifies this asymmetry because polymer stretching near the wettability interface produces localized normal stresses that delay stress relaxation and modify contact line motion in regions of high shear and extension \cite{wang2017impact,yue2012phase}. As a result, the hybrid configuration sustains spreading and delays recoil at higher $\lambda$, with the wettability jump acting as a capillary--elastic coupling zone. This enables a time-dependent redistribution of mass and energy rather than simple partitioning of the droplet, thereby extending classical Newtonian impact behavior where spreading is primarily governed by inertia, viscosity, and surface tension \cite{josserand2016drop,yue2010sharp}.

\FloatBarrier

%&&&&&&&&&&&&&&&&&&&&&&&&&&&&&&&&&&&&&&&&&&&&&&&&&&&&&&&&&&&&&&&&&&&

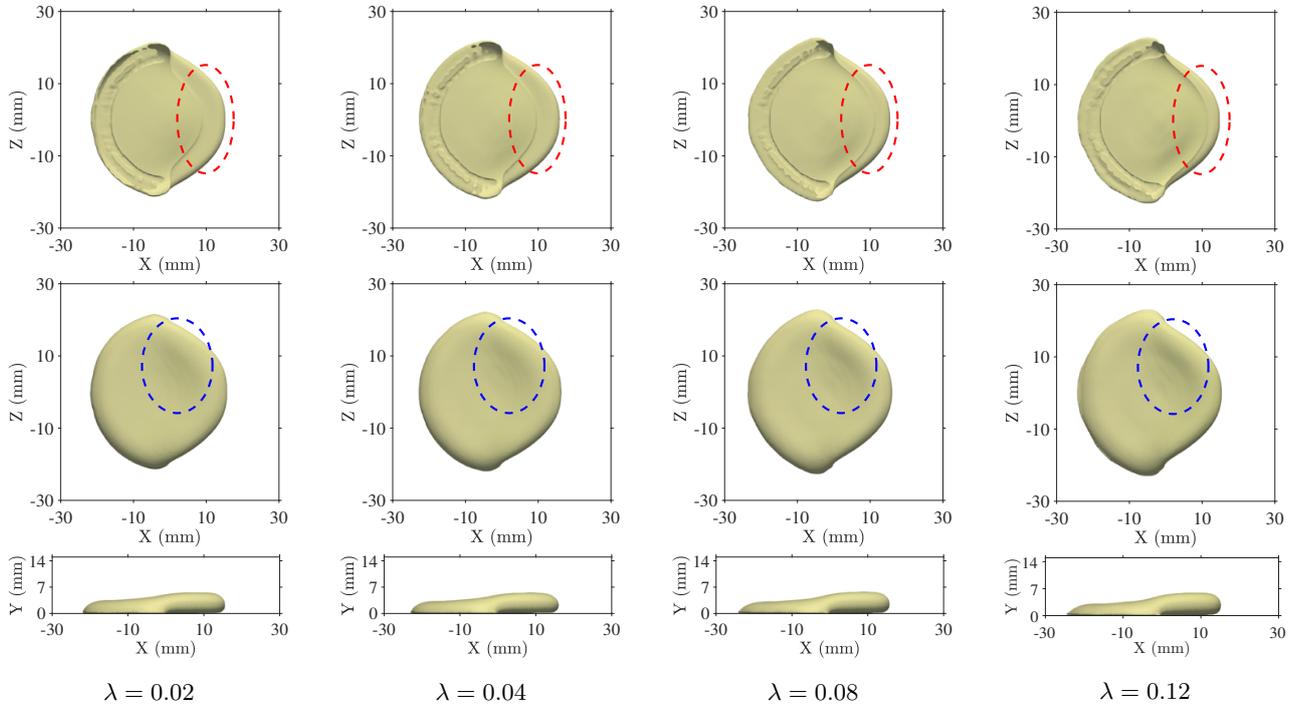
\begin{figure}[htbp]
\centering

% -------- Figure 1 --------
\begin{minipage}{0.22\linewidth}
\centering
\begin{tikzpicture}
\node[anchor=south west, inner sep=0] (img) at (0,0)
    {\includegraphics[width=\linewidth]{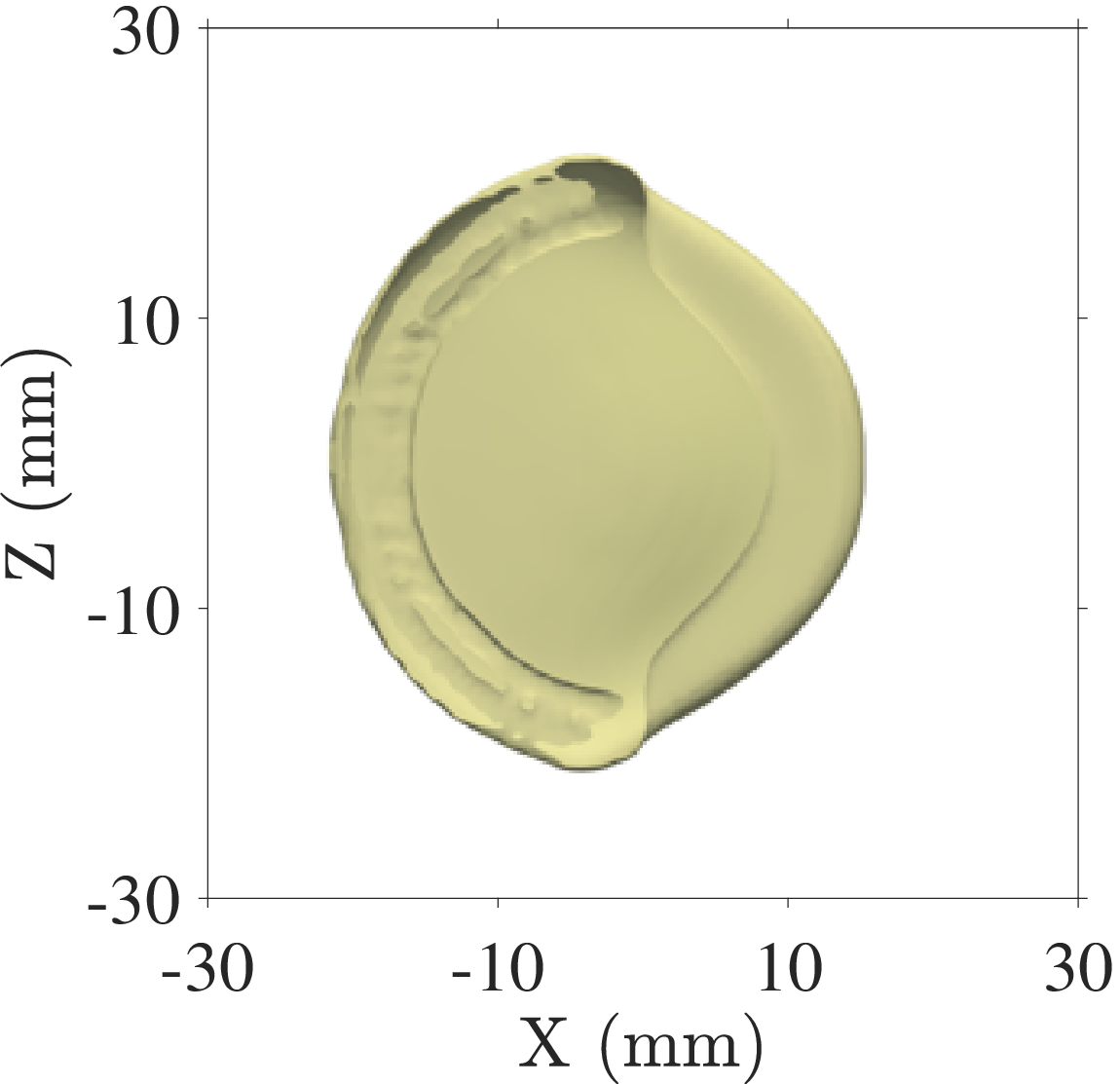}};
\begin{scope}[x={(img.south east)},y={(img.north west)}]
\draw[red, dashed, thick] (0.7, 0.575) ellipse (0.1 and 0.2);
\end{scope}
\end{tikzpicture}
\vspace{2pt}
\begin{tikzpicture}
\node[anchor=south west, inner sep=0] (img) at (0,0)
    {\includegraphics[width=\linewidth]{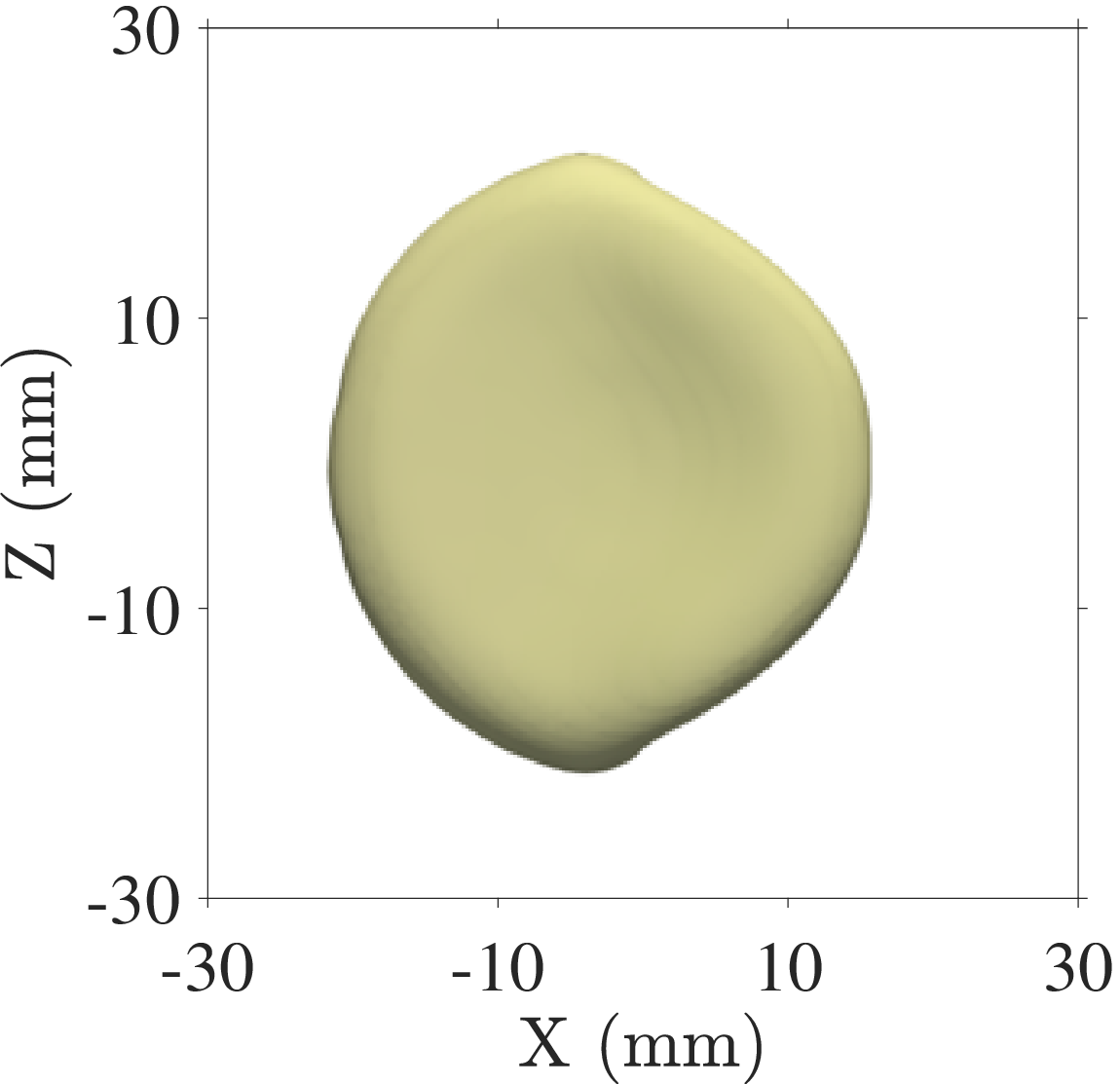}};
\begin{scope}[x={(img.south east)},y={(img.north west)}]
\draw[blue, dashed, thick] (0.6, 0.67) ellipse (0.125 and 0.175);

\end{scope}
\end{tikzpicture}
\vspace{2pt}
\includegraphics[width=\linewidth]{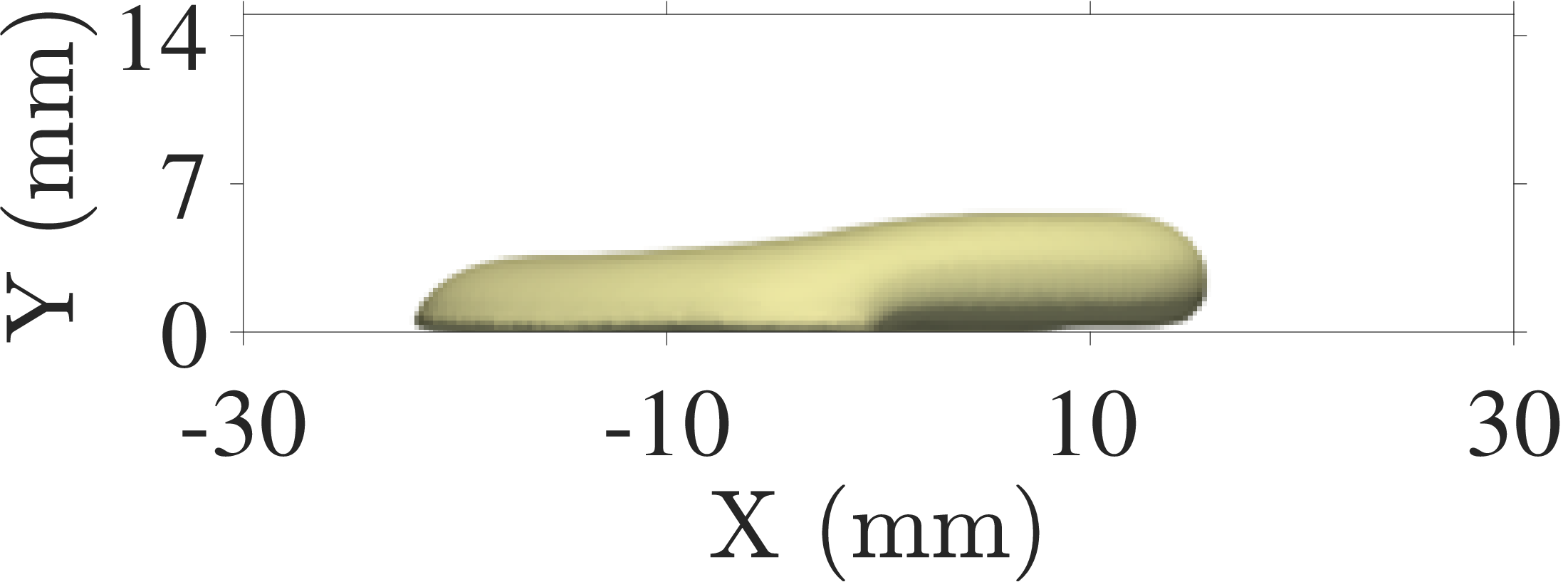}
\subcaption*{$\lambda = 0.02$}

\end{minipage}
\hfill
% -------- Figure 2 --------
\begin{minipage}{0.22\linewidth}
\centering
\begin{tikzpicture}
    \node[anchor=south west, inner sep=0] (img) at (0,0)
        {\includegraphics[width=\linewidth]{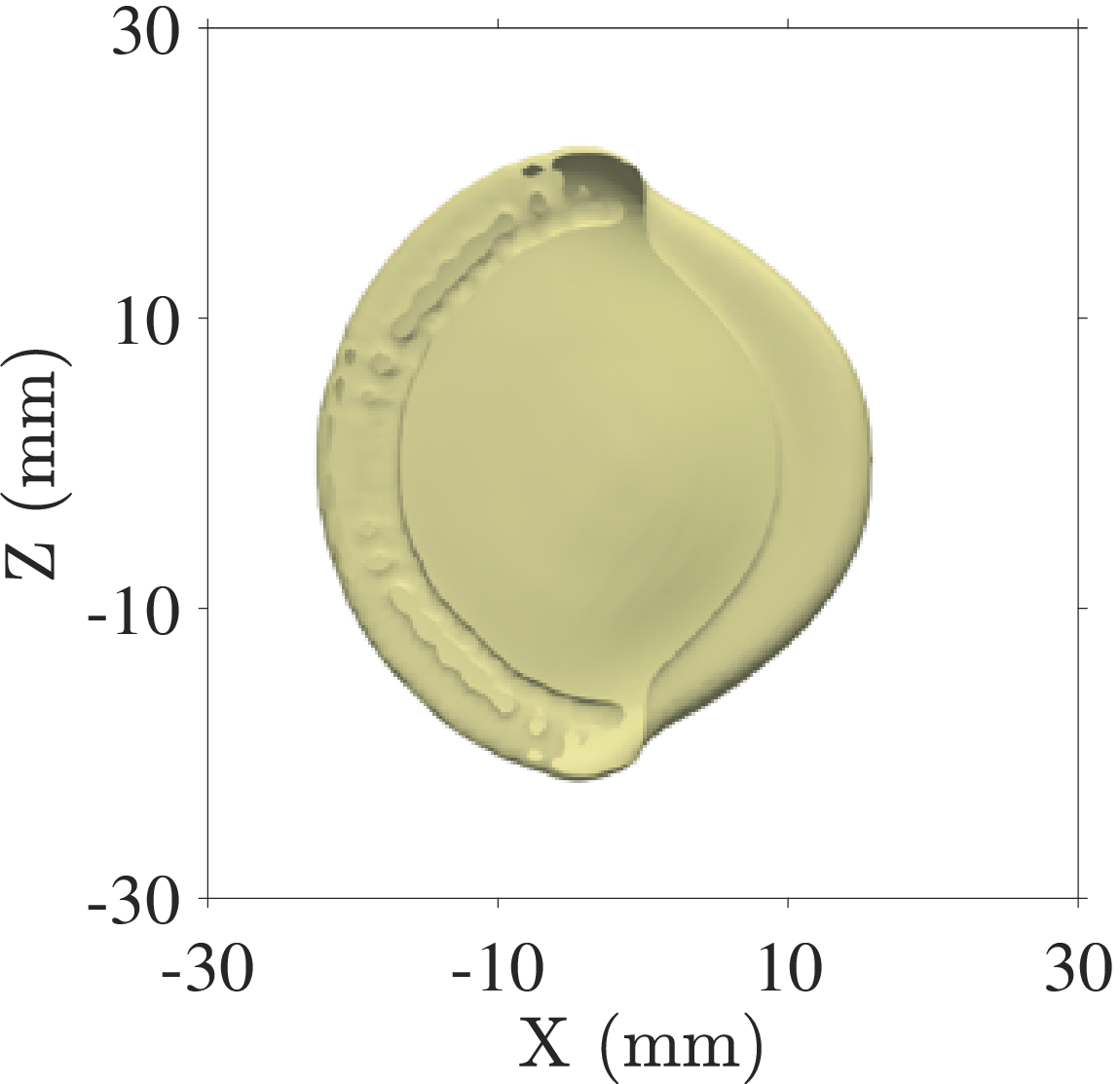}};
    \begin{scope}[x={(img.south east)},y={(img.north west)}]
        \draw[red, dashed, thick] (0.7, 0.575) ellipse (0.1 and 0.2);
    \end{scope}
\end{tikzpicture}
\vspace{2pt}
\begin{tikzpicture}
\node[anchor=south west, inner sep=0] (img) at (0,0)
{\includegraphics[width=\linewidth]{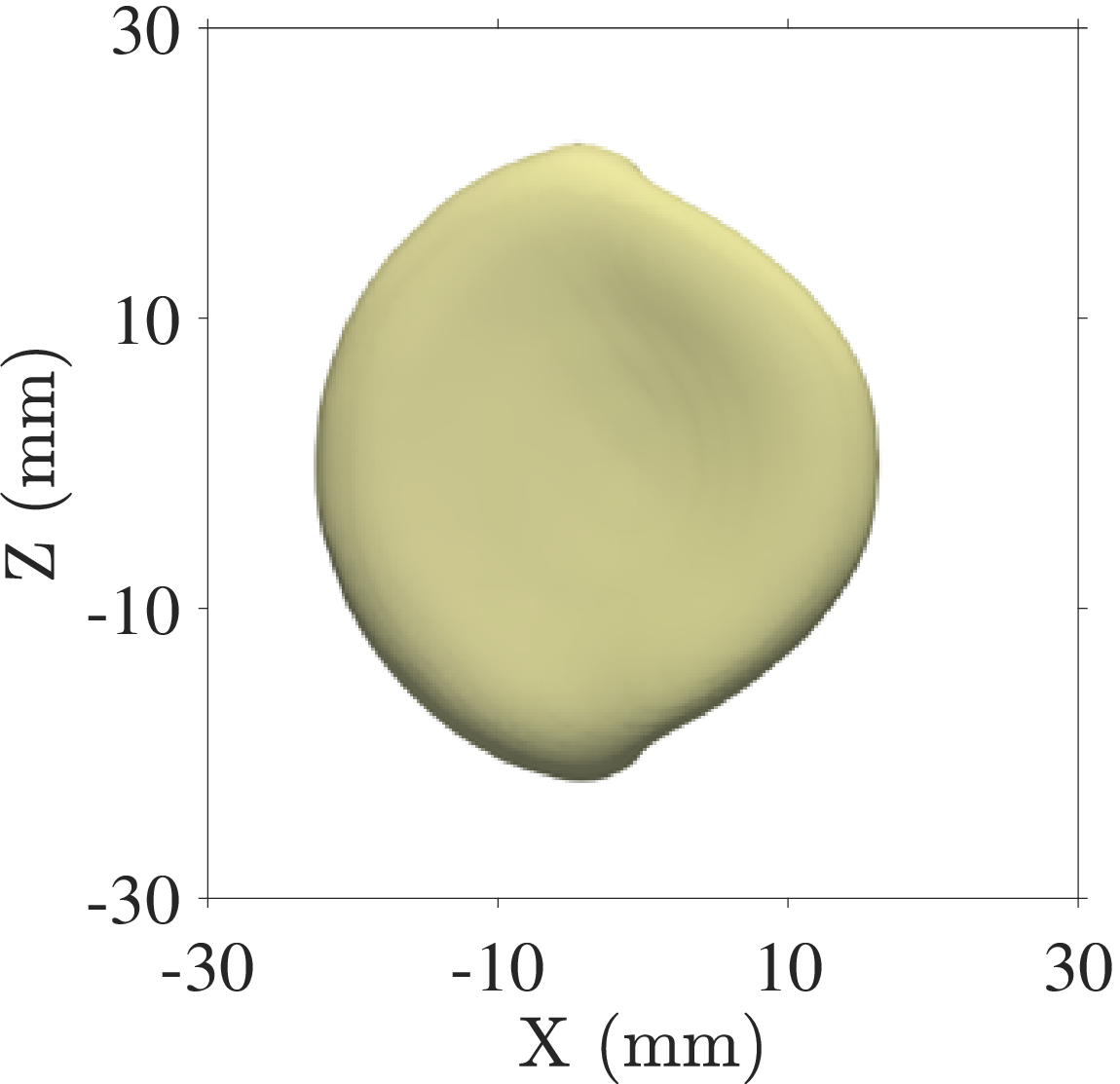}};
\begin{scope}[x={(img.south east)},y={(img.north west)}]
\draw[blue, dashed, thick] (0.6, 0.67) ellipse (0.125 and 0.175);
\end{scope}
\end{tikzpicture}
\vspace{2pt}
\includegraphics[width=\linewidth]{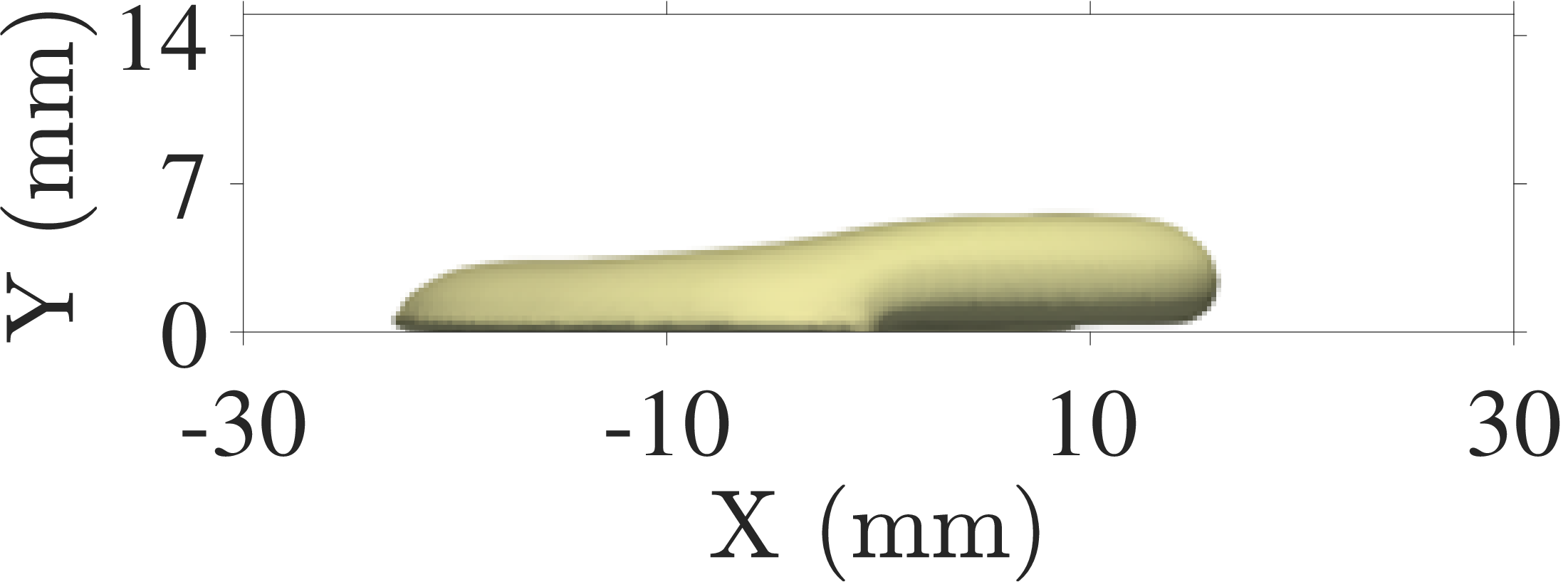}
\subcaption*{$\lambda = 0.04$}
\end{minipage}
\hfill
% -------- Figure 3 --------
\begin{minipage}{0.22\linewidth}
\centering
\begin{tikzpicture}
    \node[anchor=south west, inner sep=0] (img) at (0,0)
        {\includegraphics[width=\linewidth]{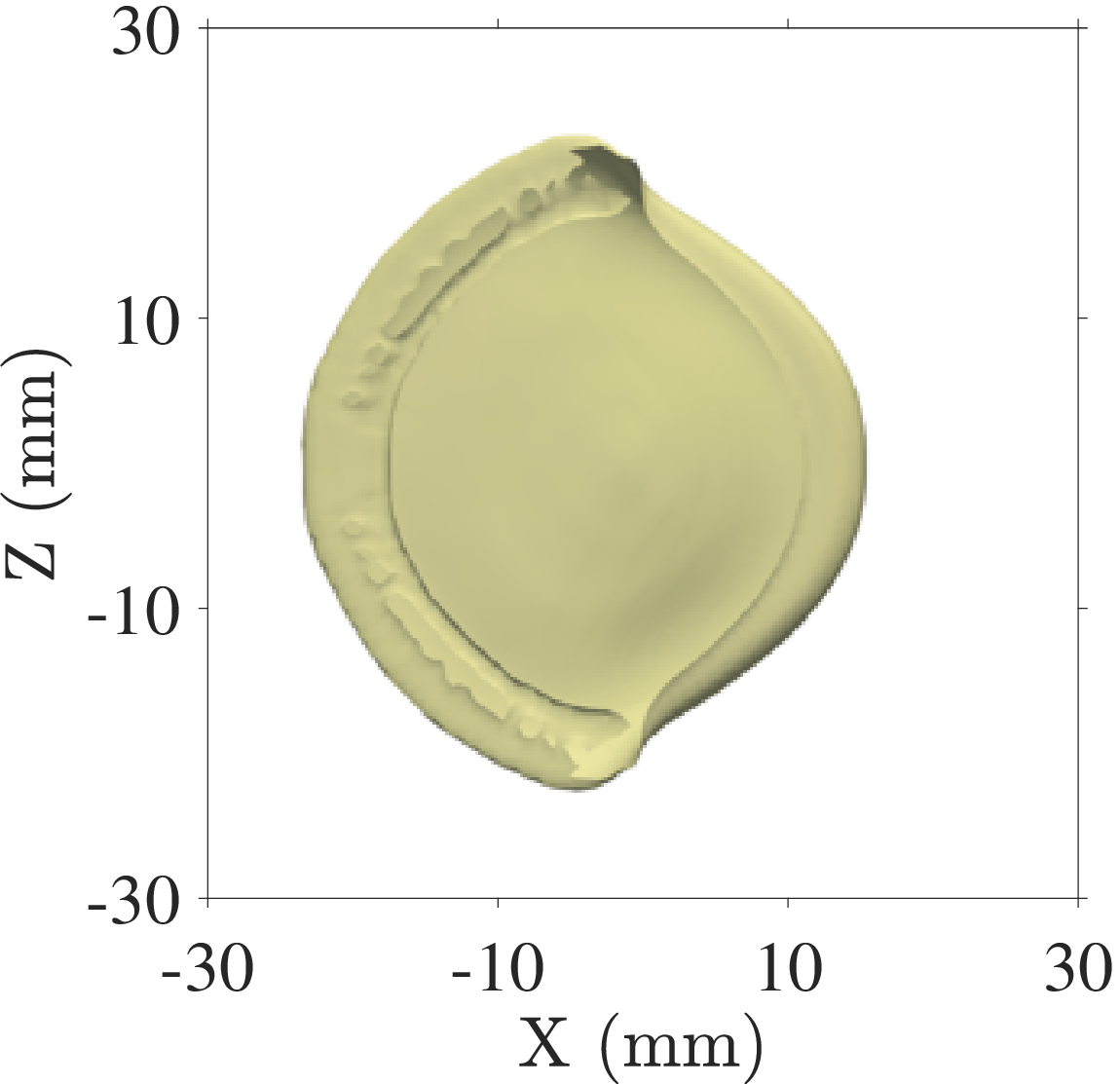}};
    \begin{scope}[x={(img.south east)},y={(img.north west)}]
        \draw[red, dashed, thick] (0.7, 0.575) ellipse (0.1 and 0.2);
    \end{scope}
\end{tikzpicture}
\vspace{2pt}
\begin{tikzpicture}
    \node[anchor=south west, inner sep=0] (img) at (0,0)
        {\includegraphics[width=\linewidth]{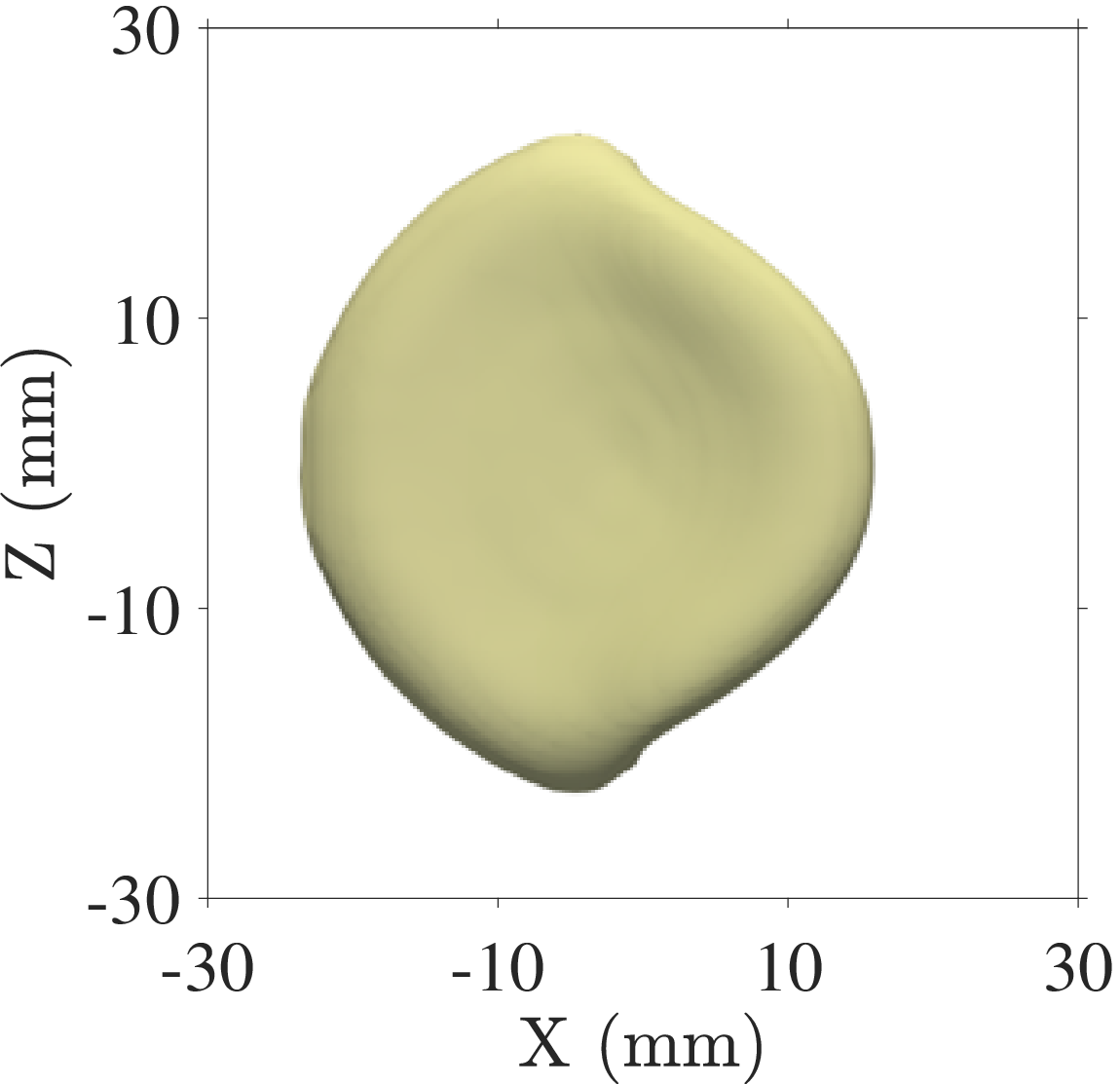}};
\begin{scope}[x={(img.south east)},y={(img.north west)}]
\draw[blue, dashed, thick] (0.6, 0.67) ellipse (0.125 and 0.175);
\end{scope}
\end{tikzpicture}
\vspace{2pt}
\includegraphics[width=\linewidth]{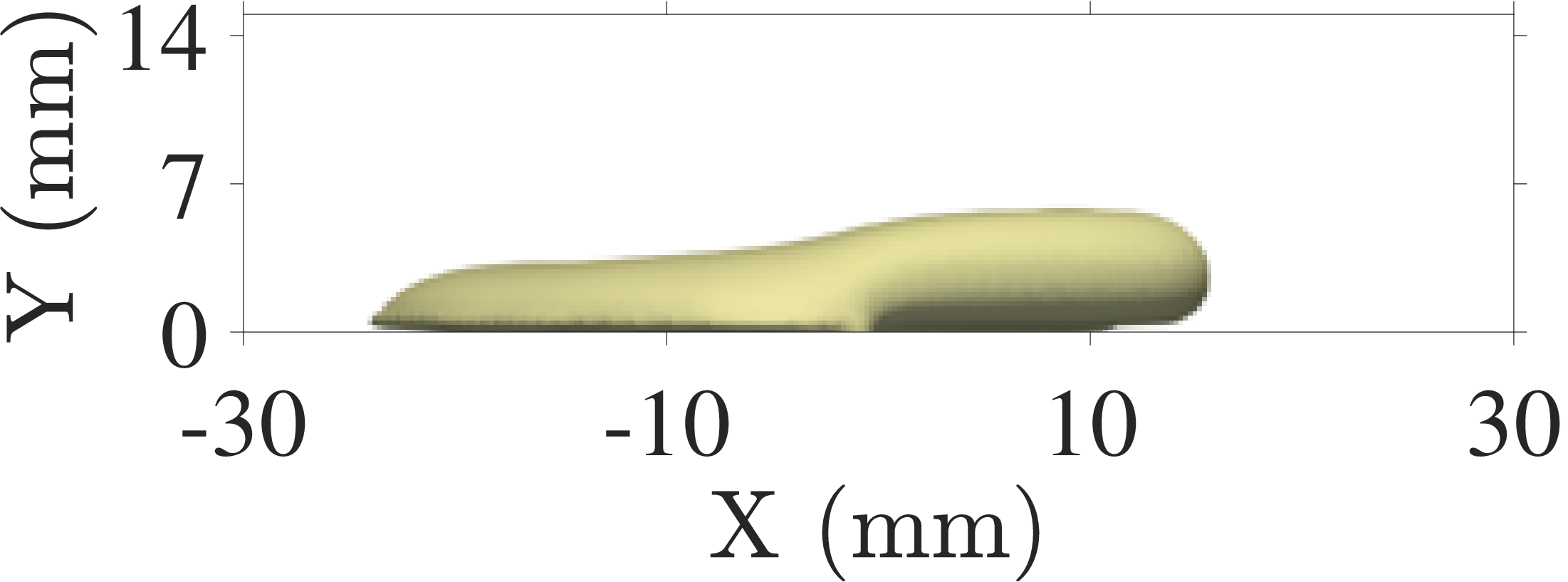}
\subcaption*{$\lambda = 0.08$}
\end{minipage}
\hfill
% -------- Figure 4 --------
\begin{minipage}{0.22\linewidth}
\centering
\begin{tikzpicture}
    \node[anchor=south west, inner sep=0] (img) at (0,0)
        {\includegraphics[width=\linewidth]{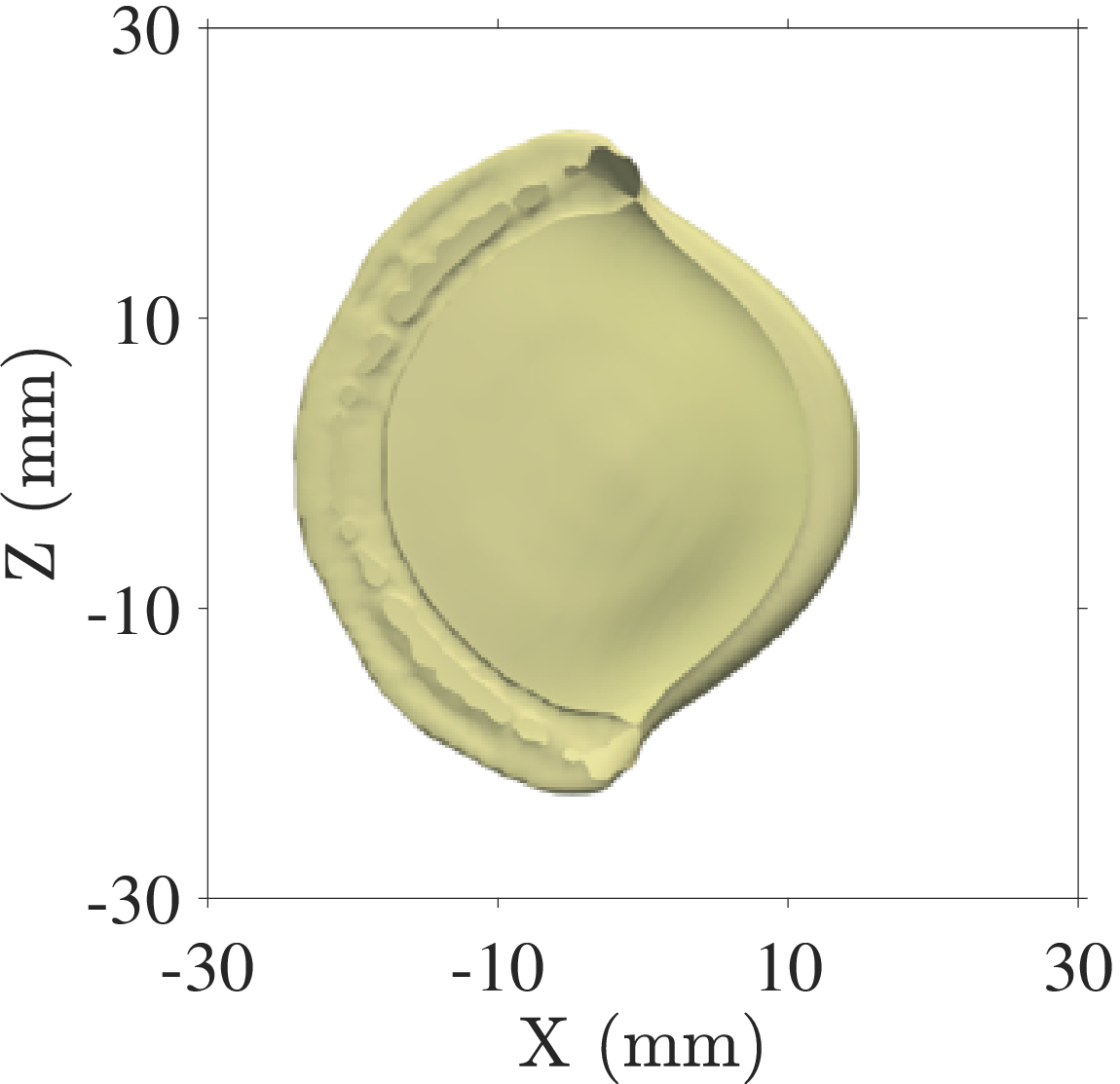}};
    \begin{scope}[x={(img.south east)},y={(img.north west)}]
        \draw[red, dashed, thick] (0.7, 0.575) ellipse (0.1 and 0.2);
    \end{scope}
\end{tikzpicture}
\vspace{2pt}
\begin{tikzpicture}
    \node[anchor=south west, inner sep=0] (img) at (0,0)
        {\includegraphics[width=\linewidth]{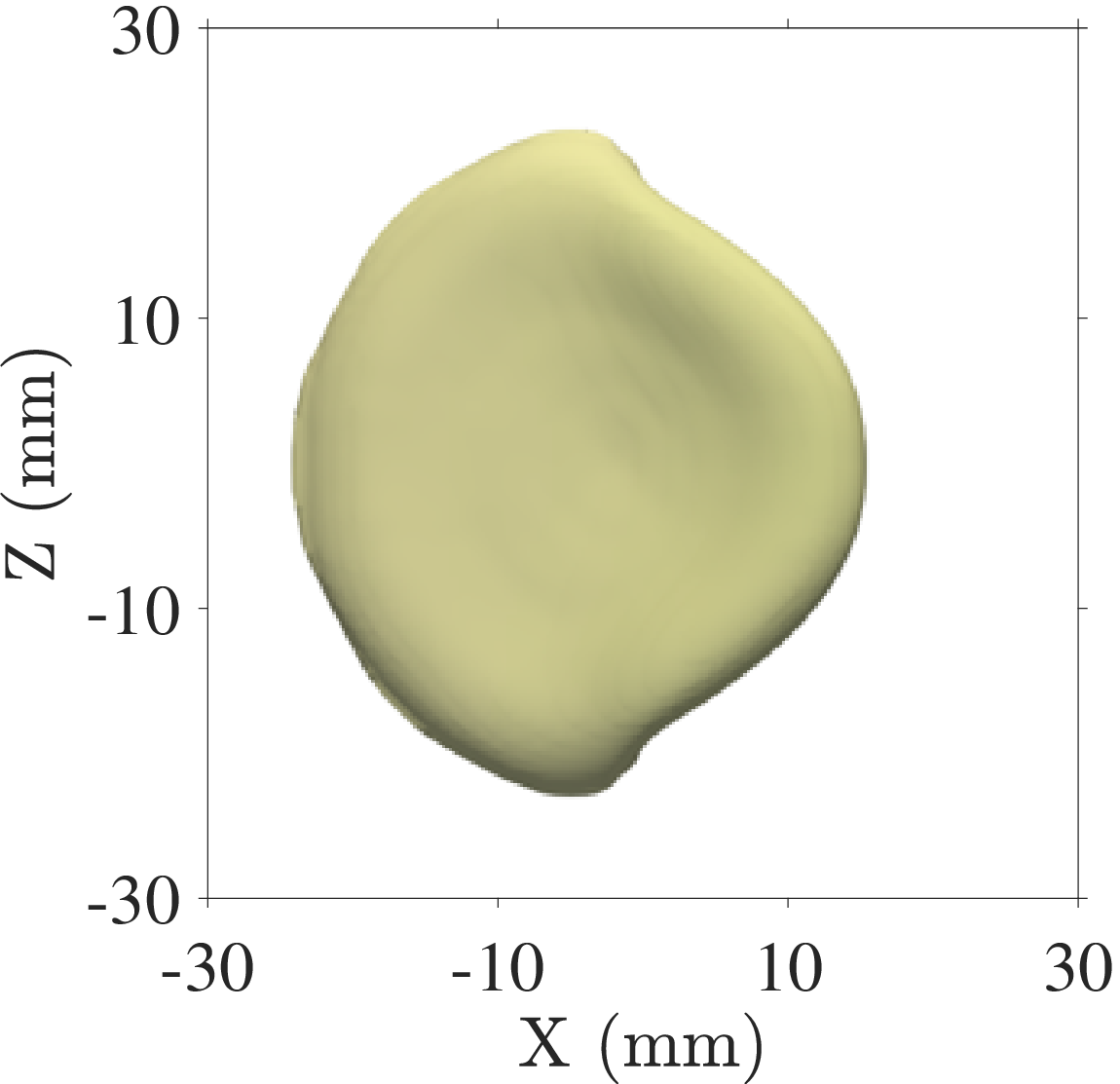}};
\begin{scope}[x={(img.south east)},y={(img.north west)}]
\draw[blue, dashed, thick] (0.6, 0.67) ellipse (0.125 and 0.175);
\end{scope}
\end{tikzpicture}
\vspace{2pt}
\includegraphics[width=\linewidth]{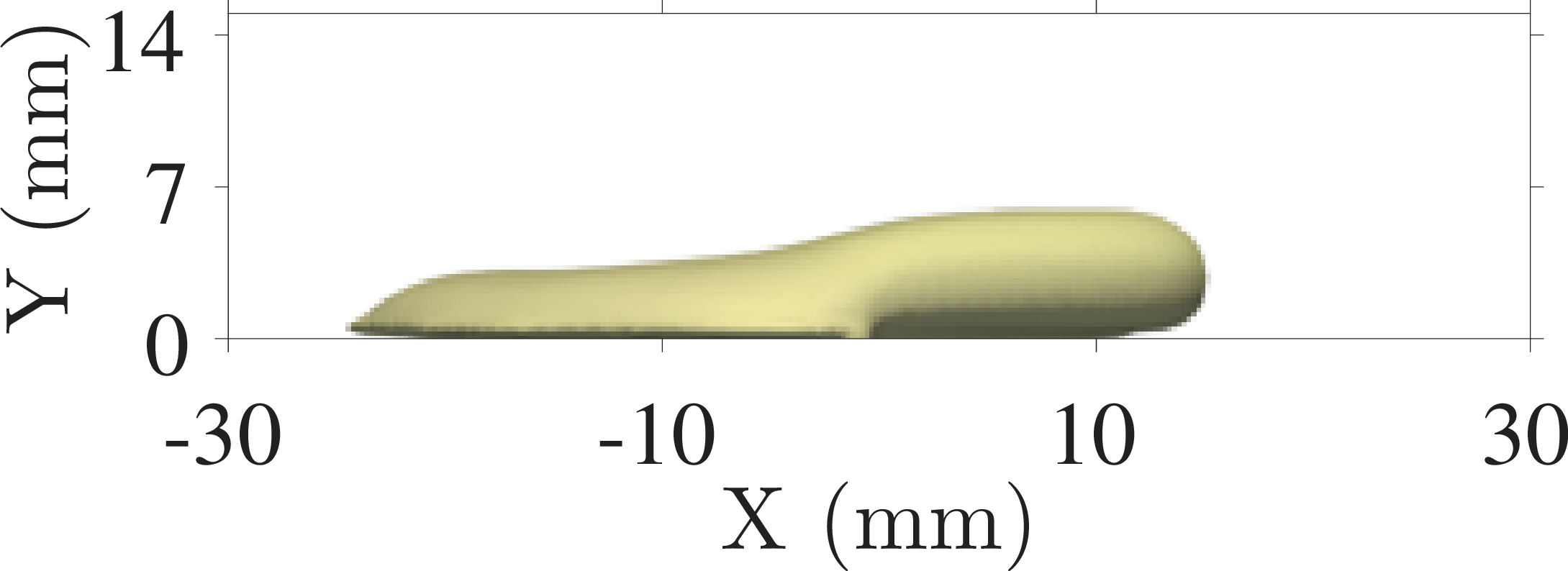}
\subcaption*{$\lambda = 0.12$}
\end{minipage}
\caption{Final equilibrium configuration of the viscoelastic droplet at $t = 0.25$ s, where the droplet is fully at rest ($\mathbf{u}=0$). Each column corresponds to a different relaxation time $\lambda$ and shows the effect of viscoelasticity on the final droplet morphology. For each case, the first row shows the bottom view (contact region with the solid surface), the second row presents the top view, and the third row provides the side view. These views highlight the wetted footprint, overall surface shape, and final droplet thickness, respectively.}
\label{fig:rest_shape_lambda}
\end{figure}

Figure~\ref{fig:rest_shape_lambda} provides further insight into the final state corresponding to the spreading and retraction dynamics discussed in Fig.~\ref{fig:lambda_effect_lowEo}. While the earlier results showed that increasing $\lambda$ enhances the maximum spreading and delays recoil, the present figure reveals how these effects are ultimately encoded in the equilibrium morphology. In the bottom view (first row), the dashed regions indicate that the contact boundary attached to the wall becomes progressively narrower with increasing $\lambda$, suggesting a reduction in the effective wetted perimeter despite the larger transient spreading. This behavior reflects the delayed but more directed retraction induced by elastic stress relaxation. In the top view (second row), the droplet edge on the advancing side exhibits a noticeable upward displacement and a more pronounced curvature for larger $\lambda$, leading to a bowl-like shape that indicates stronger accumulation of elastic stresses near the interface. This observation is consistent with the asymmetric redistribution of fluid seen in the hybrid configuration, where elastic memory sustains deformation even after motion ceases. Finally, the side view confirms that the droplet retains a thicker and less uniformly distributed profile as $\lambda$ increases, indicating that the relaxation process does not fully restore a symmetric cap-like shape. Together, these features demonstrate that viscoelasticity not only modifies transient dynamics but also leaves a persistent imprint on the final equilibrium configuration, particularly through localized curvature changes and reduced contact area on the solid surface.

%&&&&&&&&&&&&&&&&&&&&&&&&&&&&&&&&&&&&&&&&&&&&&&&&&&&&&&&&&&&&&&&&&&&

%%%%%%% with frame

%%%%%%%%%%%%%%%%%%%%%%%%%%%%end of with frame
%NEW FIGURS FOR 3D CASE

\begin{figure}[H]
\centering

    \includegraphics[width=0.45\linewidth]{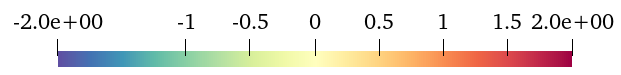}\\[3pt]
% ================= Row 1: WCA = 10° =================
\begin{minipage}{0.22\linewidth}
    \centering
    {\scriptsize $t=0.01\,\mathrm{s}$}\\[2pt]
    \includegraphics[width=\linewidth]{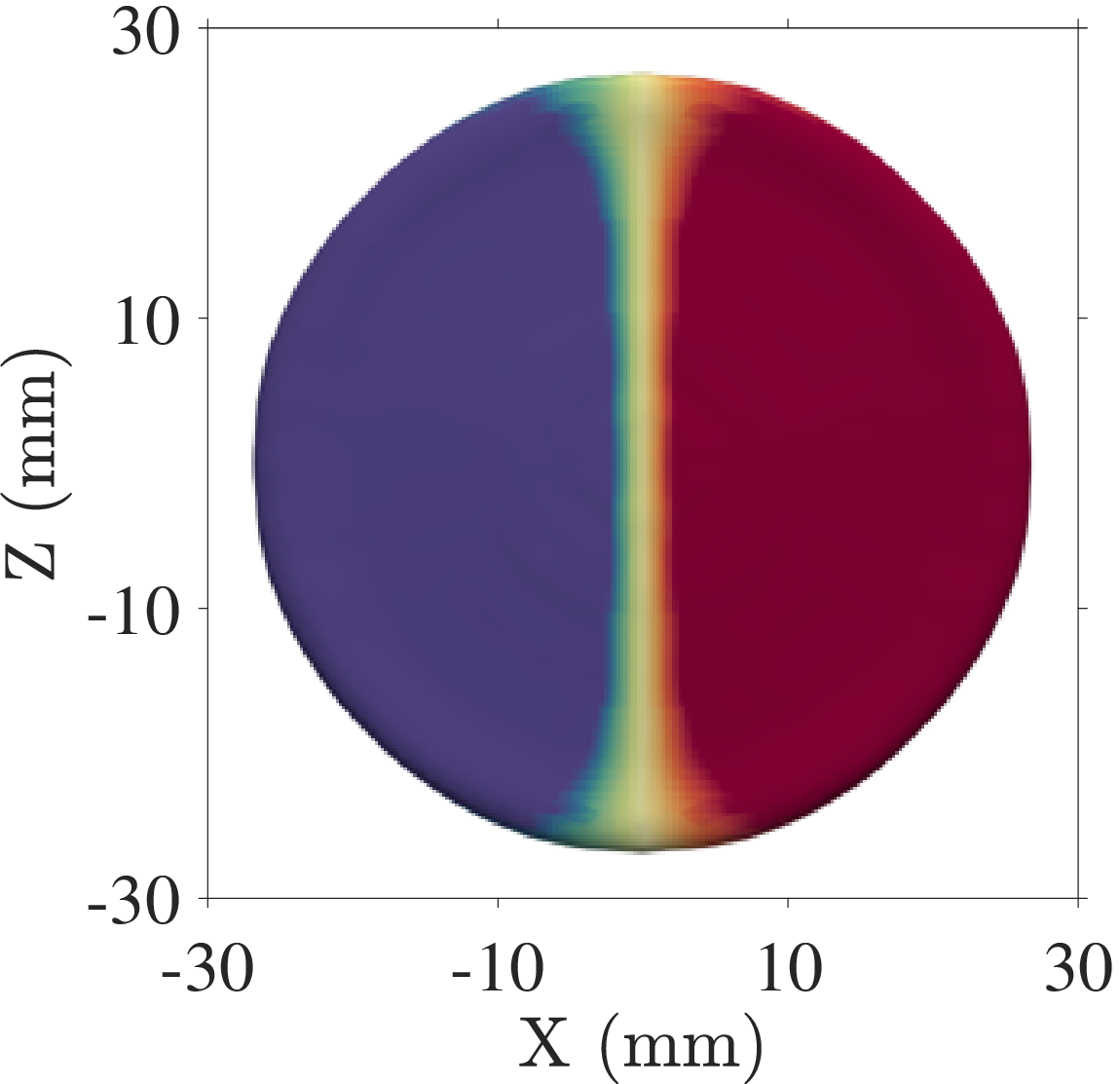}\\[3pt]
    \includegraphics[width=\linewidth]{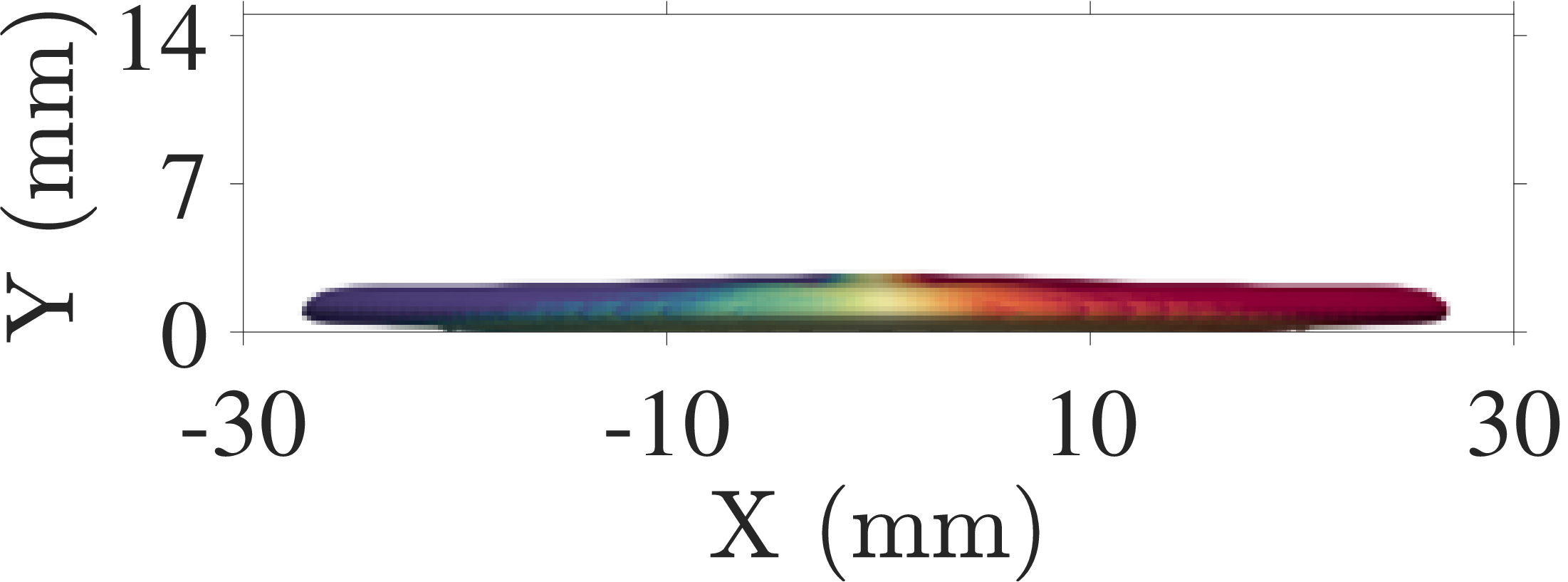}
\end{minipage}\hfill
\begin{minipage}{0.22\linewidth}
    \centering
    {\scriptsize $t=0.02\,\mathrm{s}$}\\[2pt]
    \includegraphics[width=\linewidth]{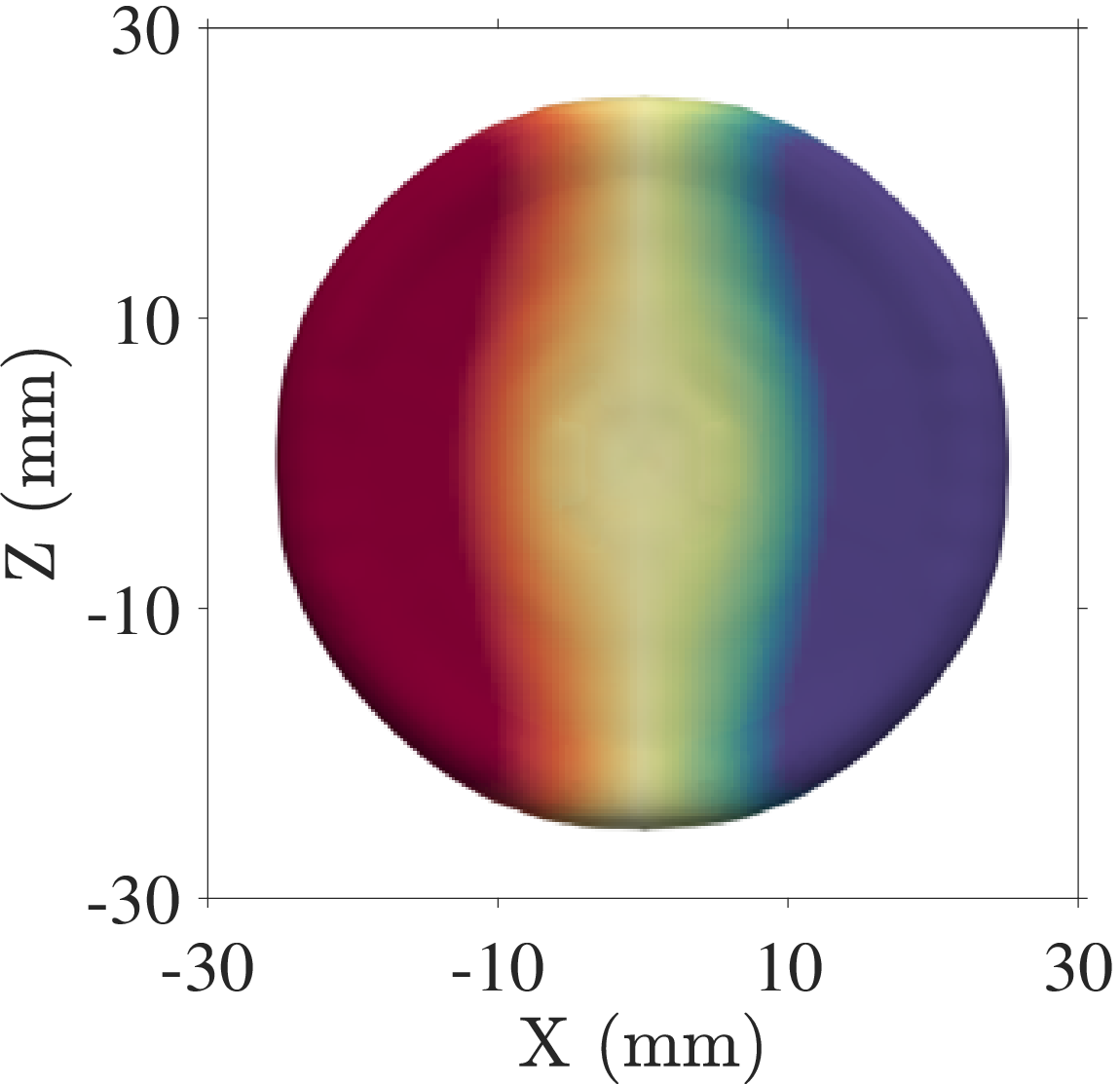}\\[3pt]
    \includegraphics[width=\linewidth]{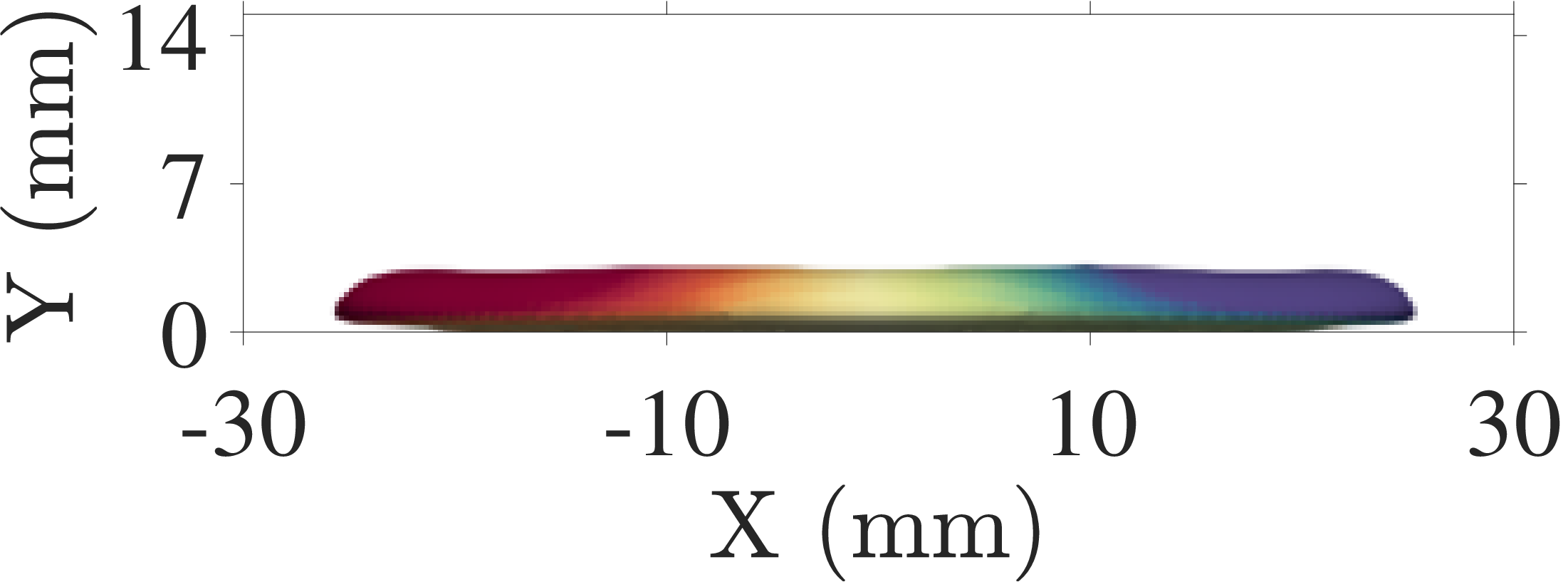}
\end{minipage}\hfill
\begin{minipage}{0.22\linewidth}
    \centering
    {\scriptsize $t=0.03\,\mathrm{s}$}\\[2pt]
    \includegraphics[width=\linewidth]{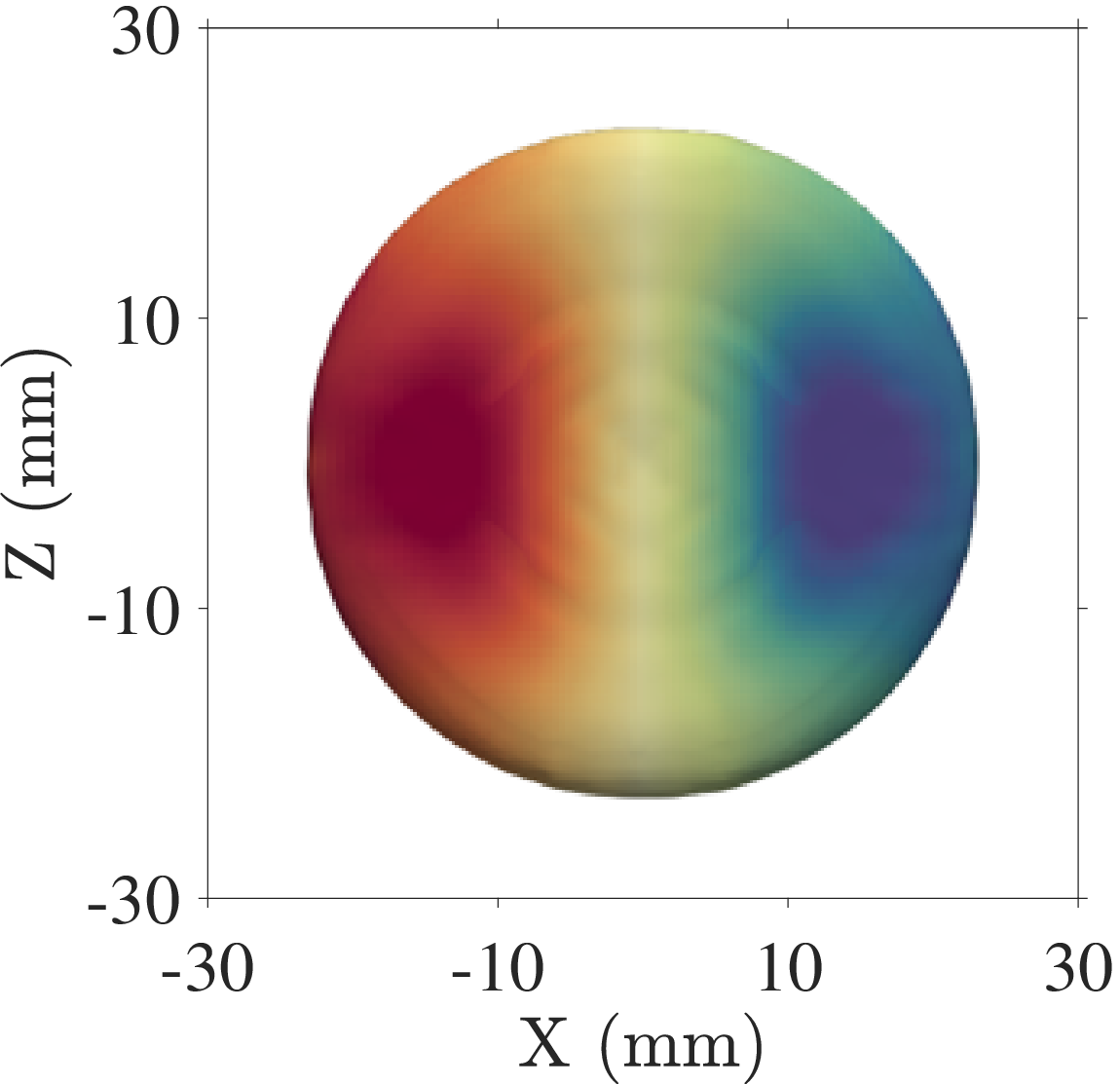}\\[3pt]
    \includegraphics[width=\linewidth]{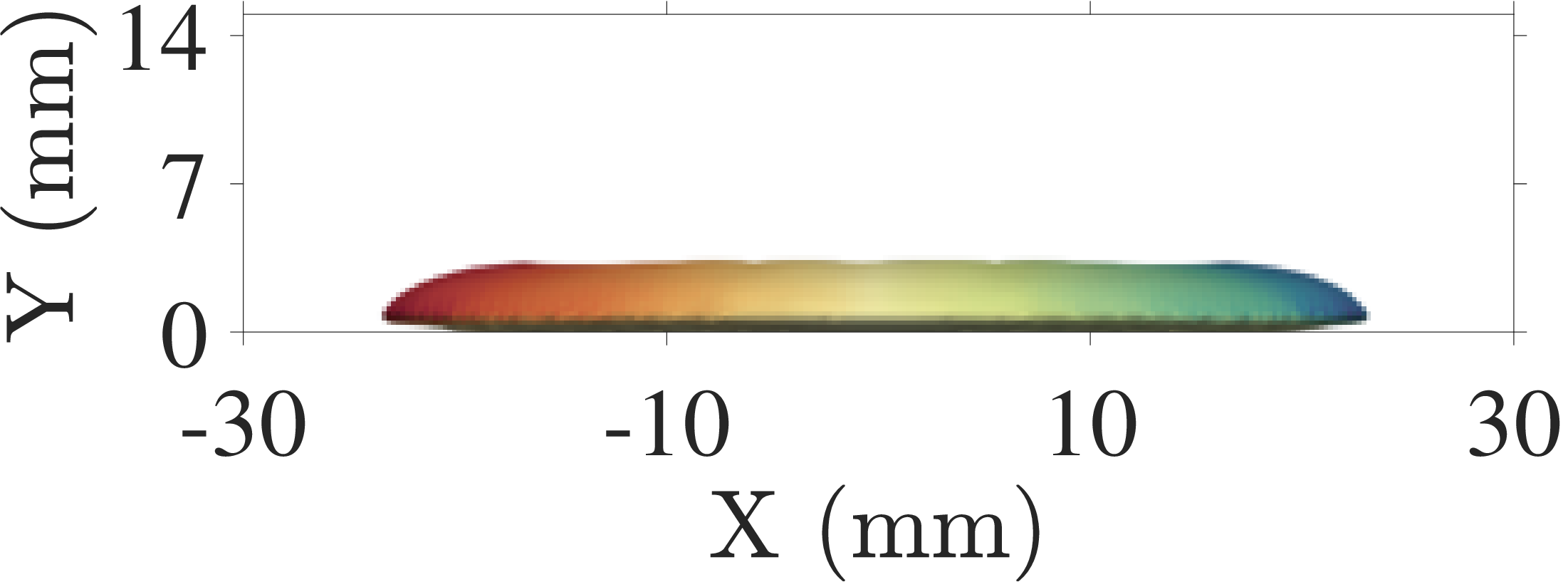}
\end{minipage}\hfill
\begin{minipage}{0.22\linewidth}
    \centering
    {\scriptsize $t=0.04\,\mathrm{s}$}\\[2pt]
    \includegraphics[width=\linewidth]{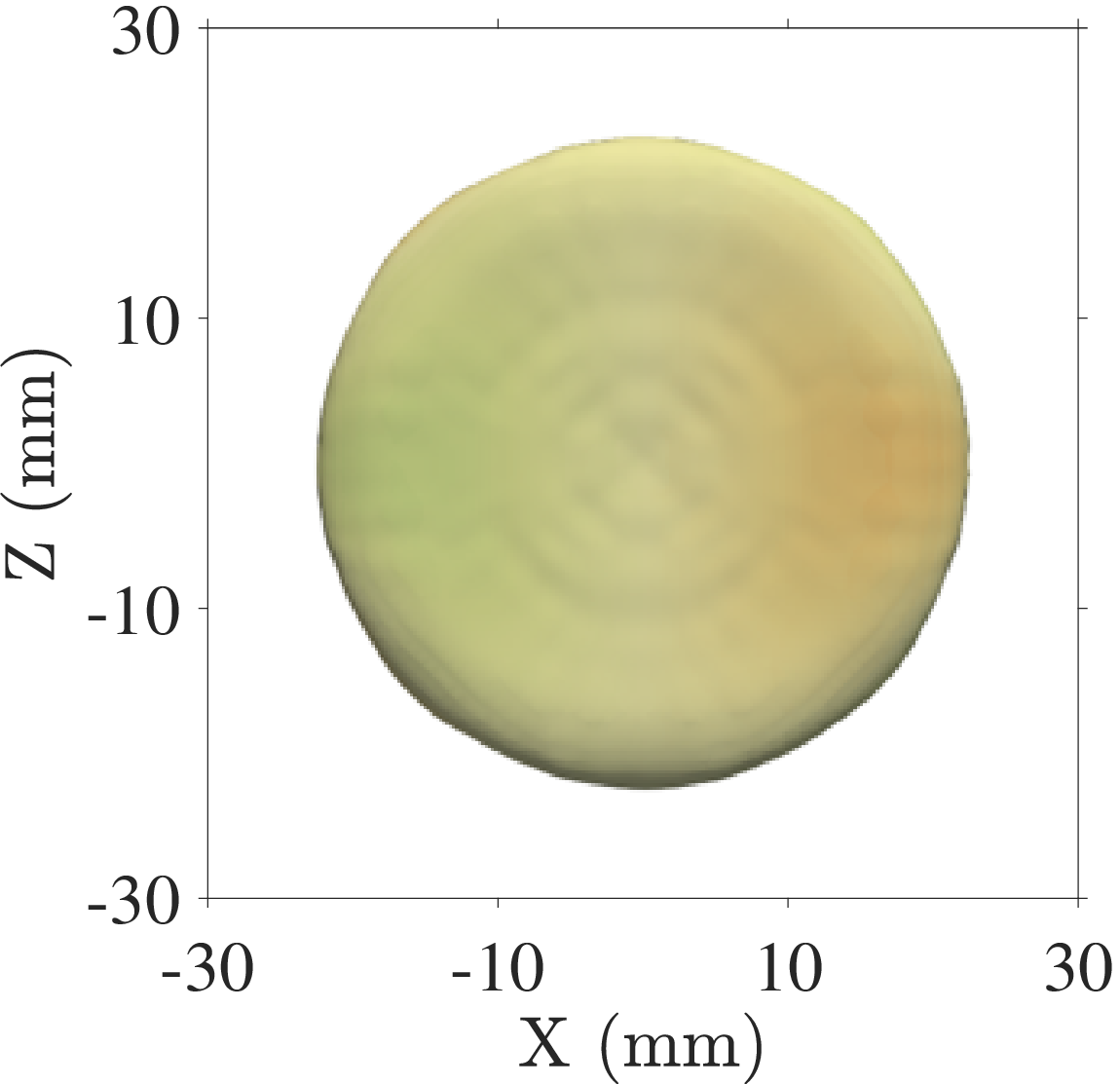}\\[3pt]
    \includegraphics[width=\linewidth]{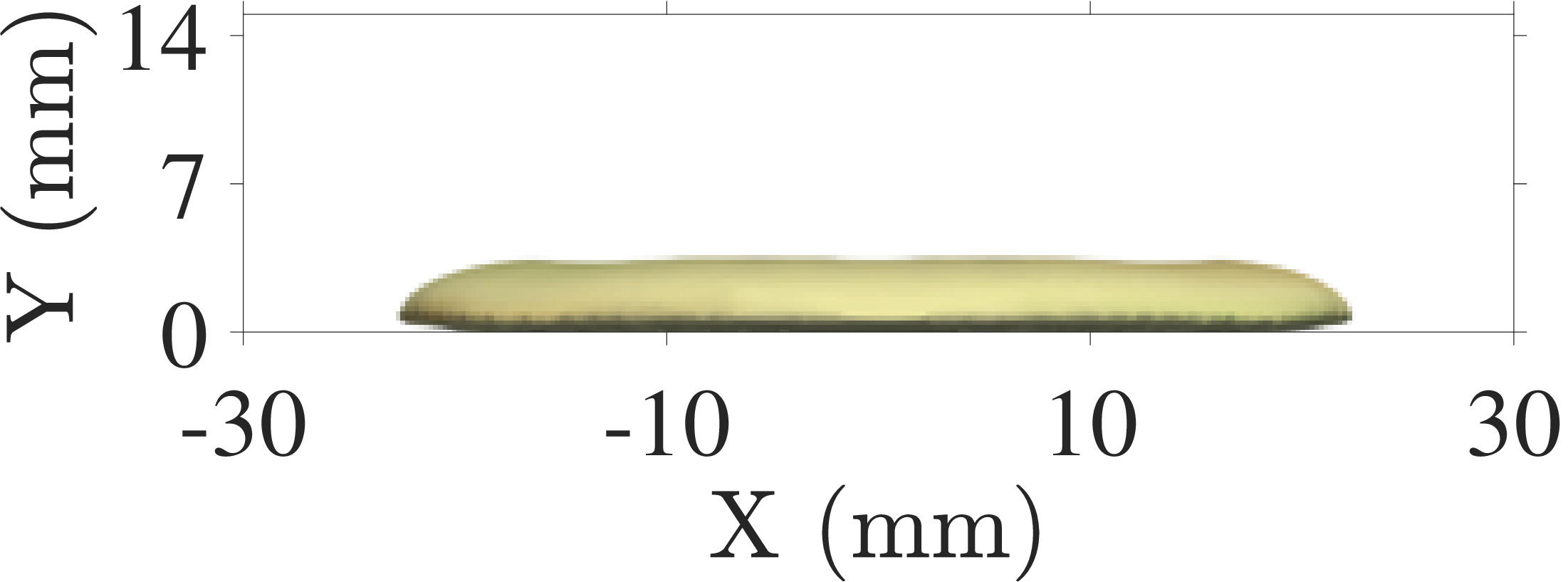}
\end{minipage}

\vspace{2pt}
\begin{center}\text{WCA = $0^\circ$}\end{center}

\vspace{2pt}

% ================= Row 2: Hybrid =================
\begin{minipage}{0.22\linewidth}
    \centering
    {\scriptsize $t=0.01\,\mathrm{s}$}\\[2pt]
    \includegraphics[width=\linewidth]{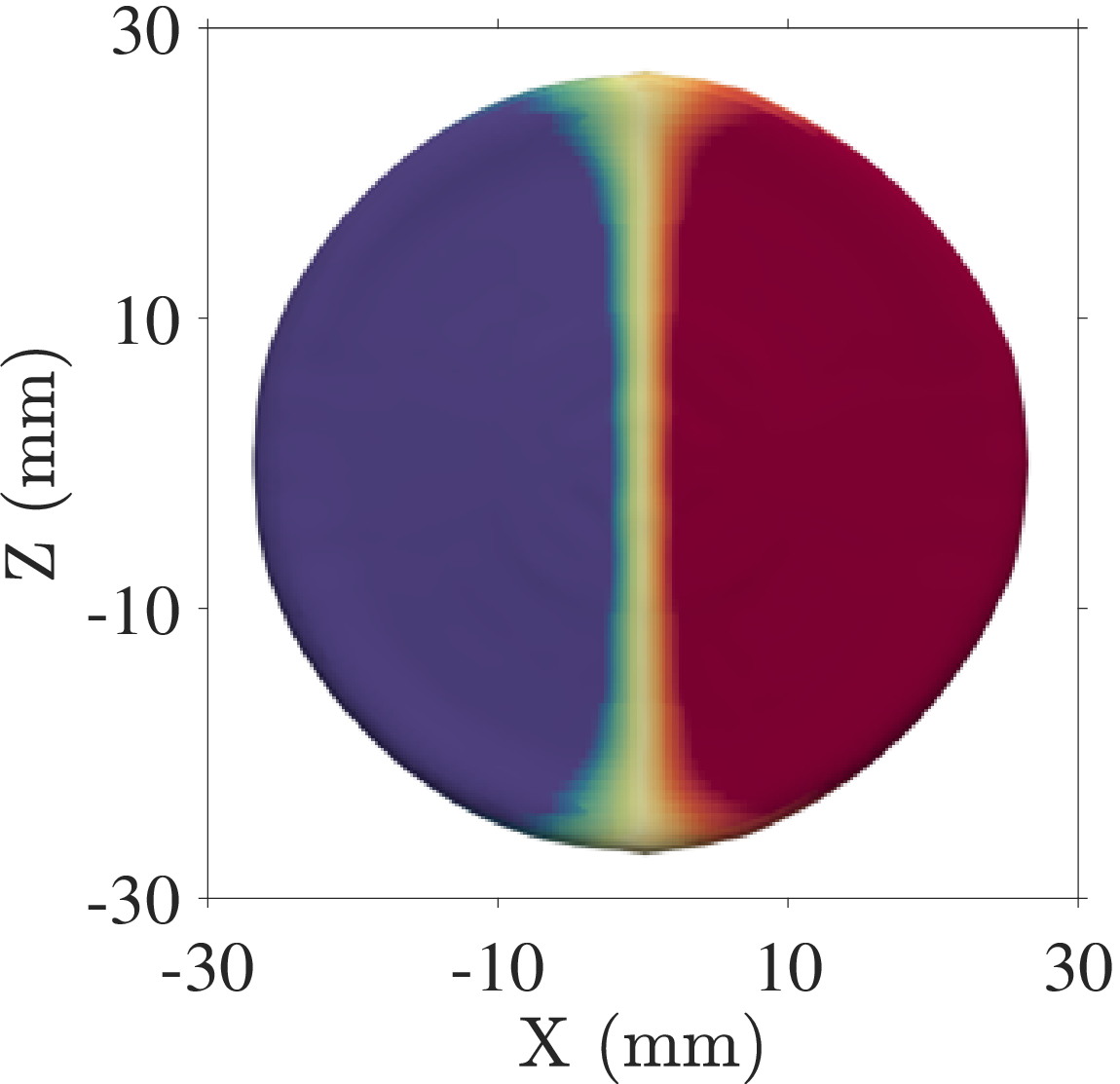}\\[3pt]
    \includegraphics[width=\linewidth]{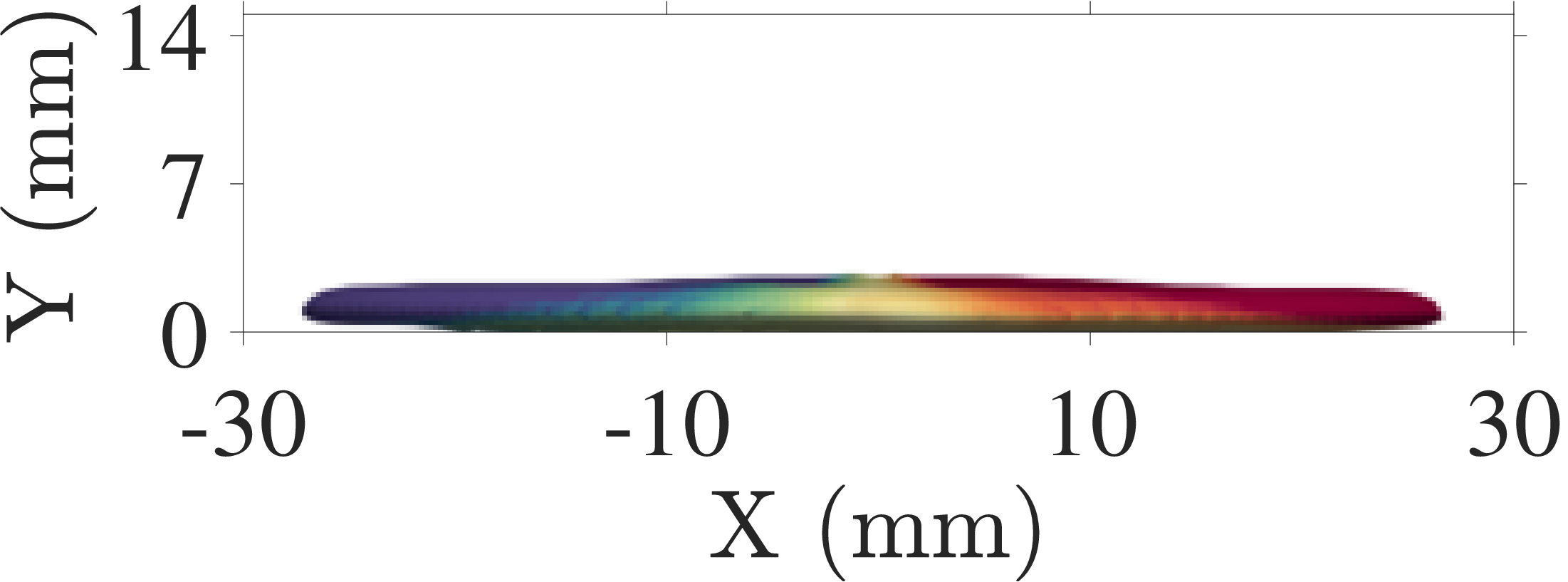}
\end{minipage}\hfill
\begin{minipage}{0.22\linewidth}
    \centering
    {\scriptsize $t=0.02\,\mathrm{s}$}\\[2pt]
    \includegraphics[width=\linewidth]{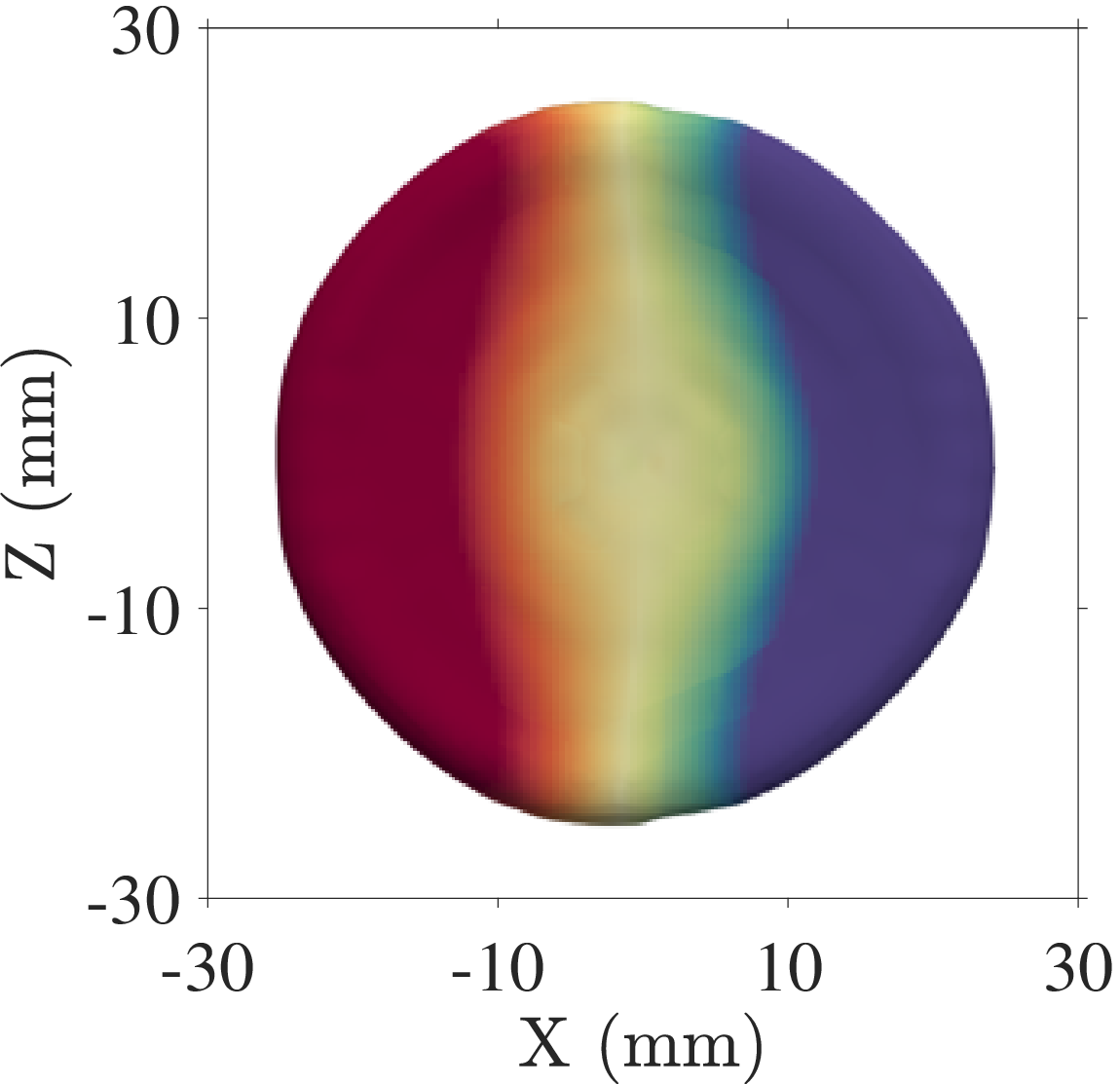}\\[3pt]
    \includegraphics[width=\linewidth]{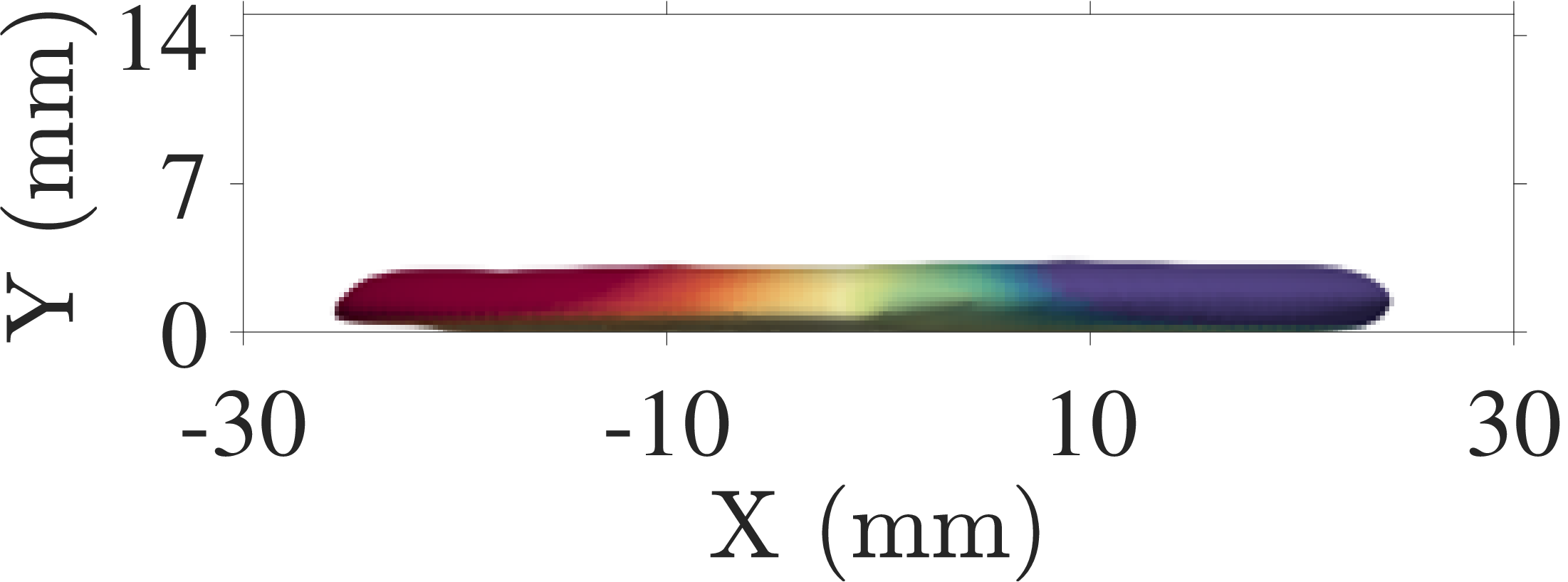}
\end{minipage}\hfill
\begin{minipage}{0.22\linewidth}
    \centering
    {\scriptsize $t=0.03\,\mathrm{s}$}\\[2pt]
    \includegraphics[width=\linewidth]{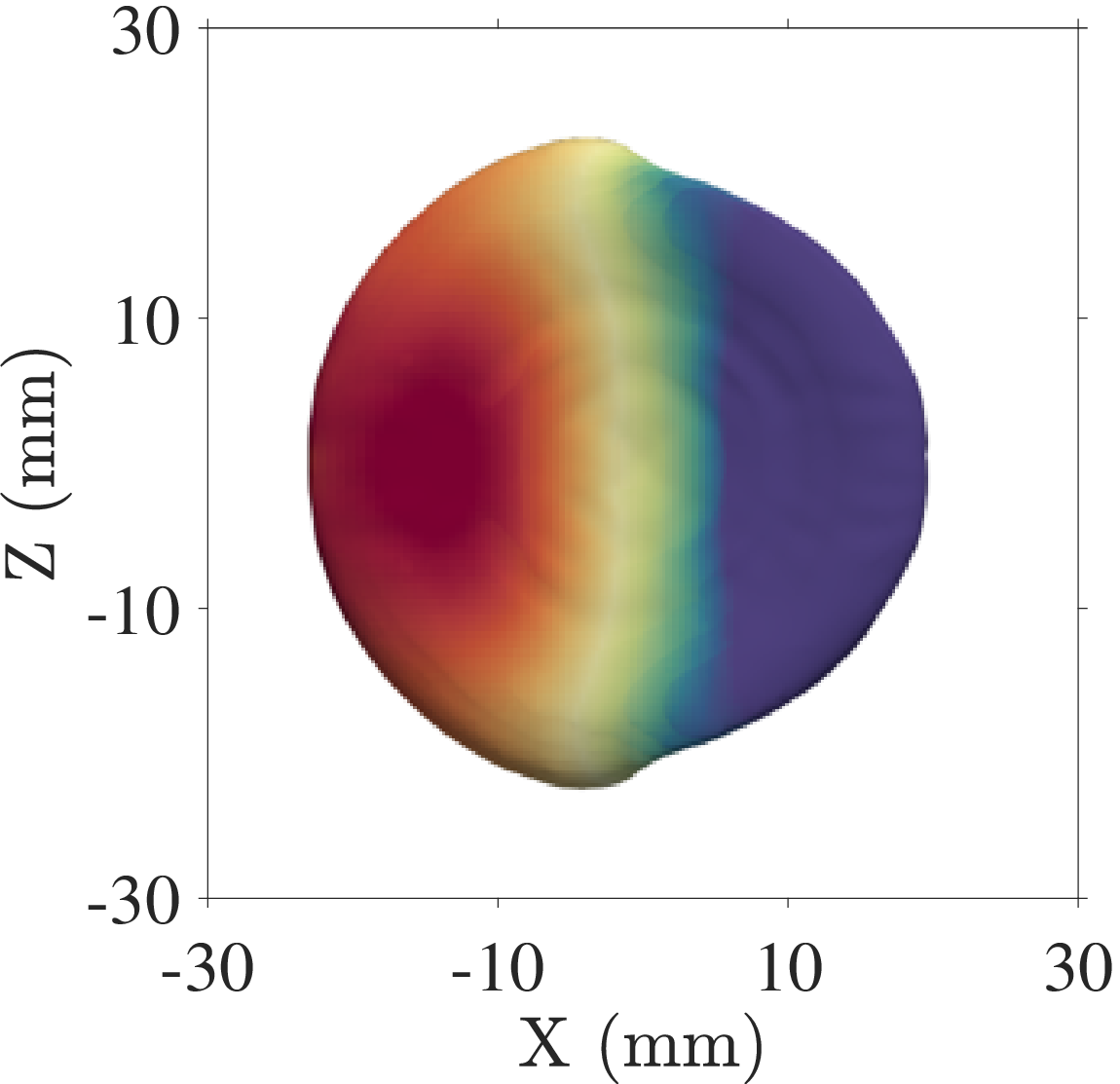}\\[3pt]
    \includegraphics[width=\linewidth]{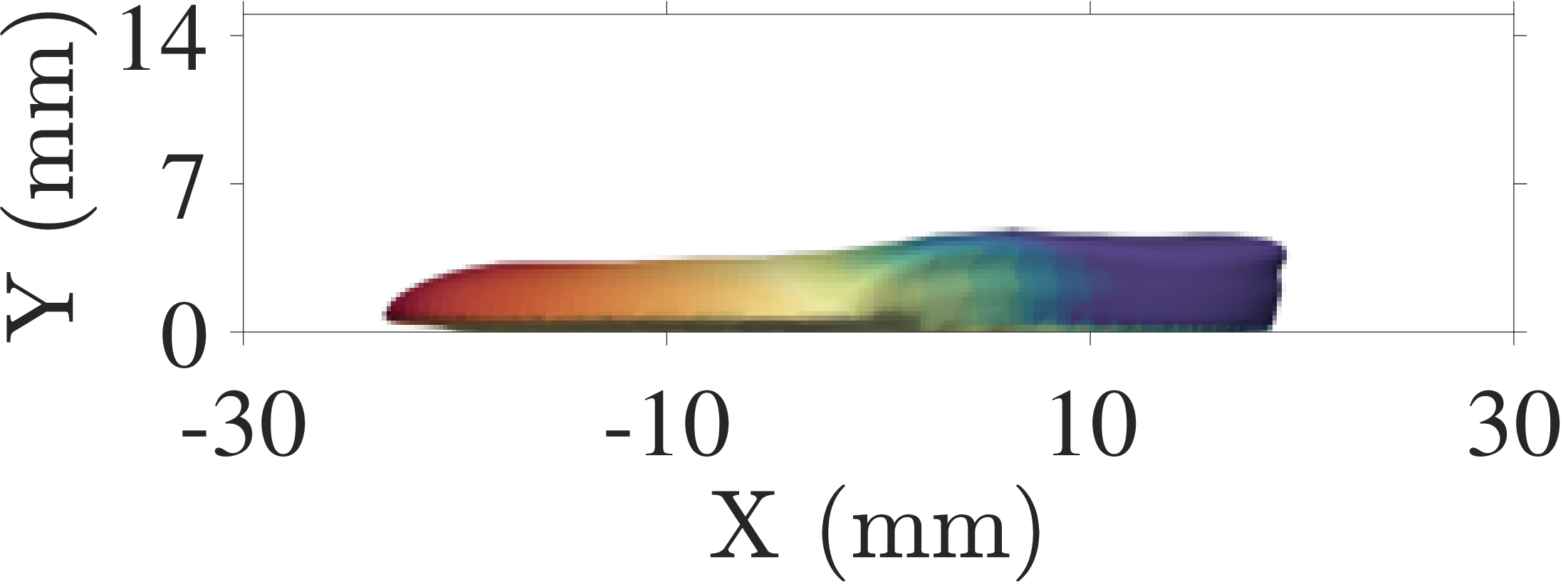}
\end{minipage}\hfill
\begin{minipage}{0.22\linewidth}
    \centering
    {\scriptsize $t=0.04\,\mathrm{s}$}\\[2pt]
    \includegraphics[width=\linewidth]{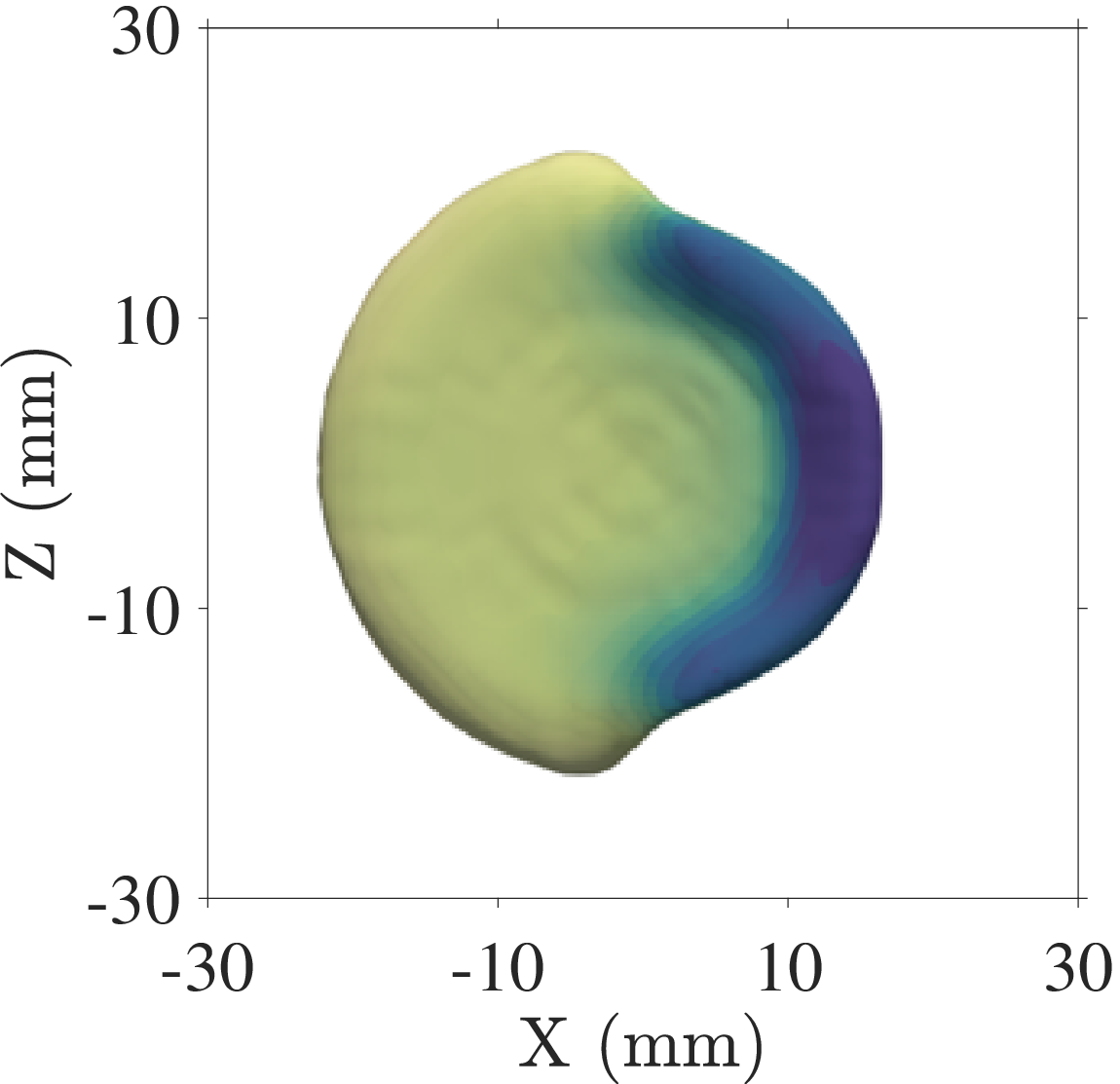}\\[3pt]
    \includegraphics[width=\linewidth]{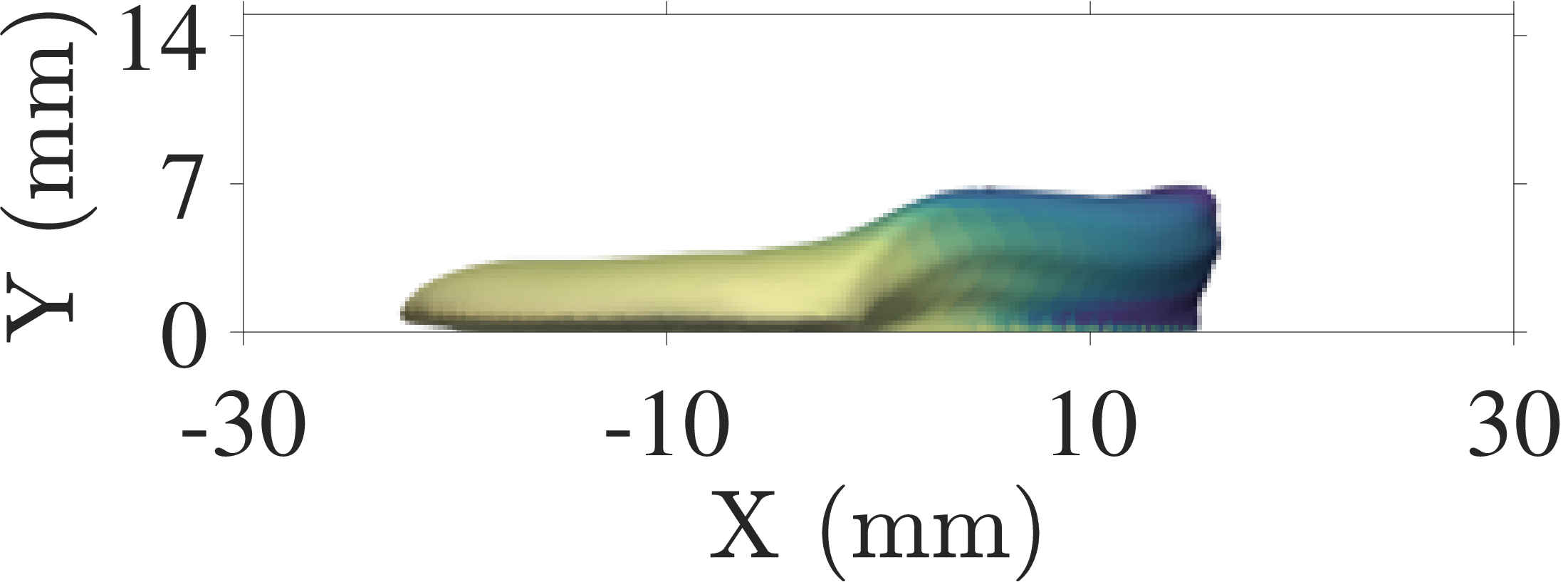}
\end{minipage}

\vspace{2pt}
\begin{center}\text{WCA = Hybrid $0^\circ$--$160^\circ$}\end{center}

\vspace{2pt}

% ================= Row 3: WCA = 160° =================
\begin{minipage}{0.22\linewidth}
    \centering
    {\scriptsize $t=0.01\,\mathrm{s}$}\\[2pt]
    \includegraphics[width=\linewidth]{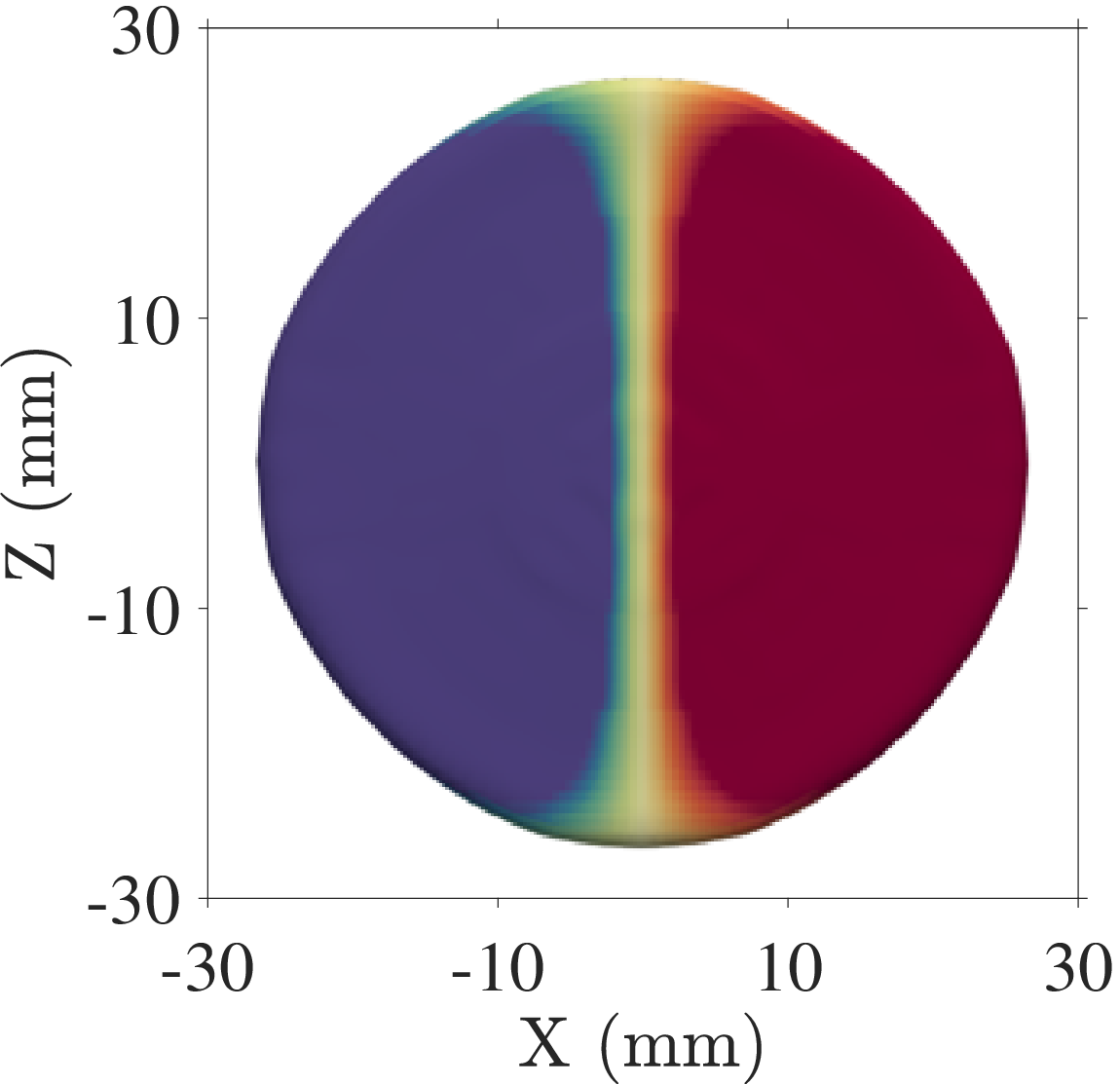}\\[3pt]
    \includegraphics[width=\linewidth]{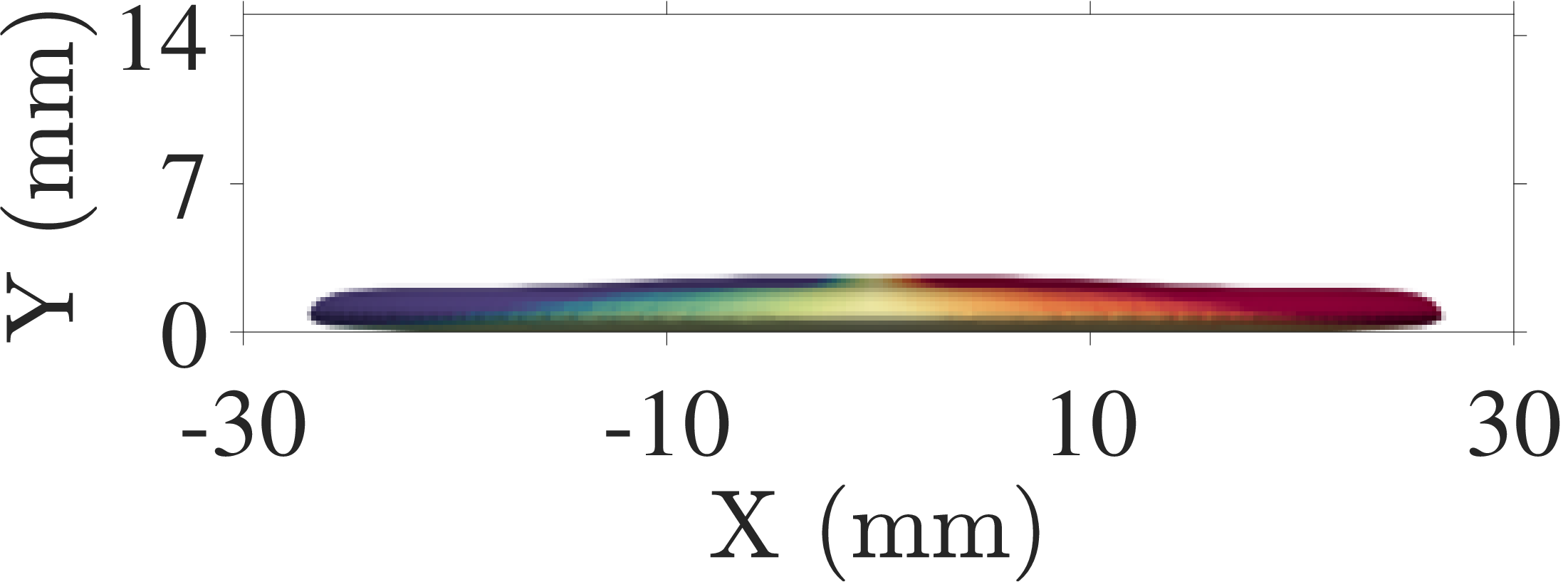}
\end{minipage}\hfill
\begin{minipage}{0.22\linewidth}
    \centering
    {\scriptsize $t=0.02\,\mathrm{s}$}\\[2pt]
    \includegraphics[width=\linewidth]{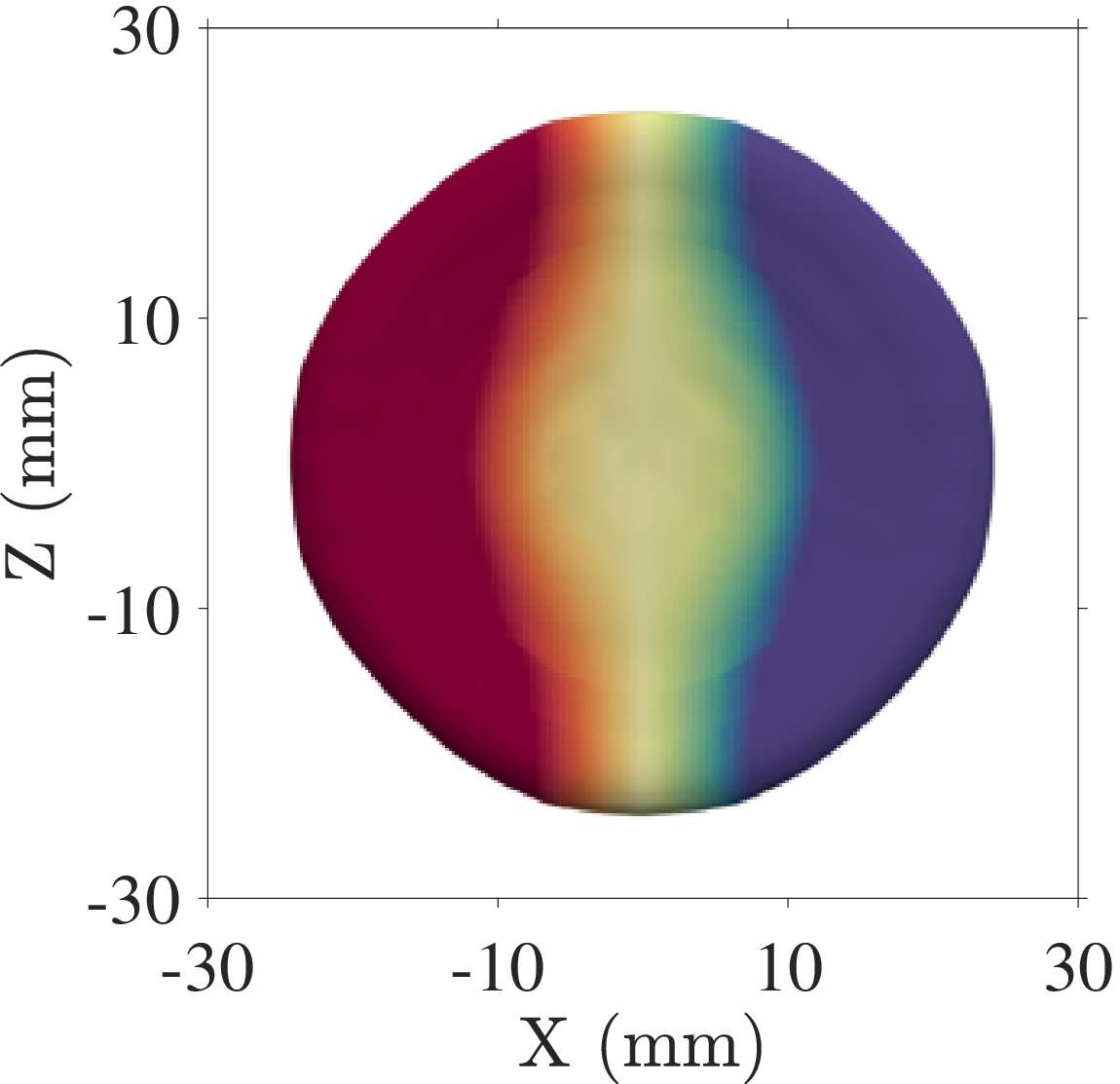}\\[3pt]
    \includegraphics[width=\linewidth]{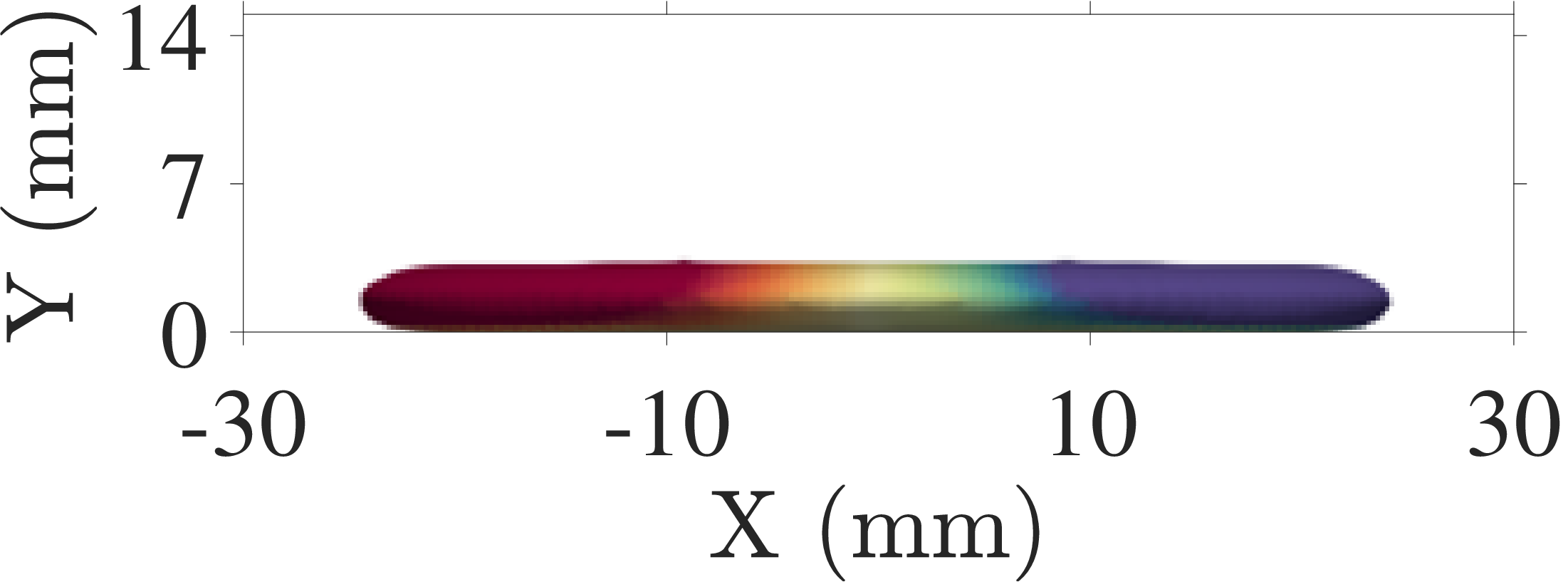}
\end{minipage}\hfill
\begin{minipage}{0.22\linewidth}
    \centering
    {\scriptsize $t=0.03\,\mathrm{s}$}\\[2pt]
    \includegraphics[width=\linewidth]{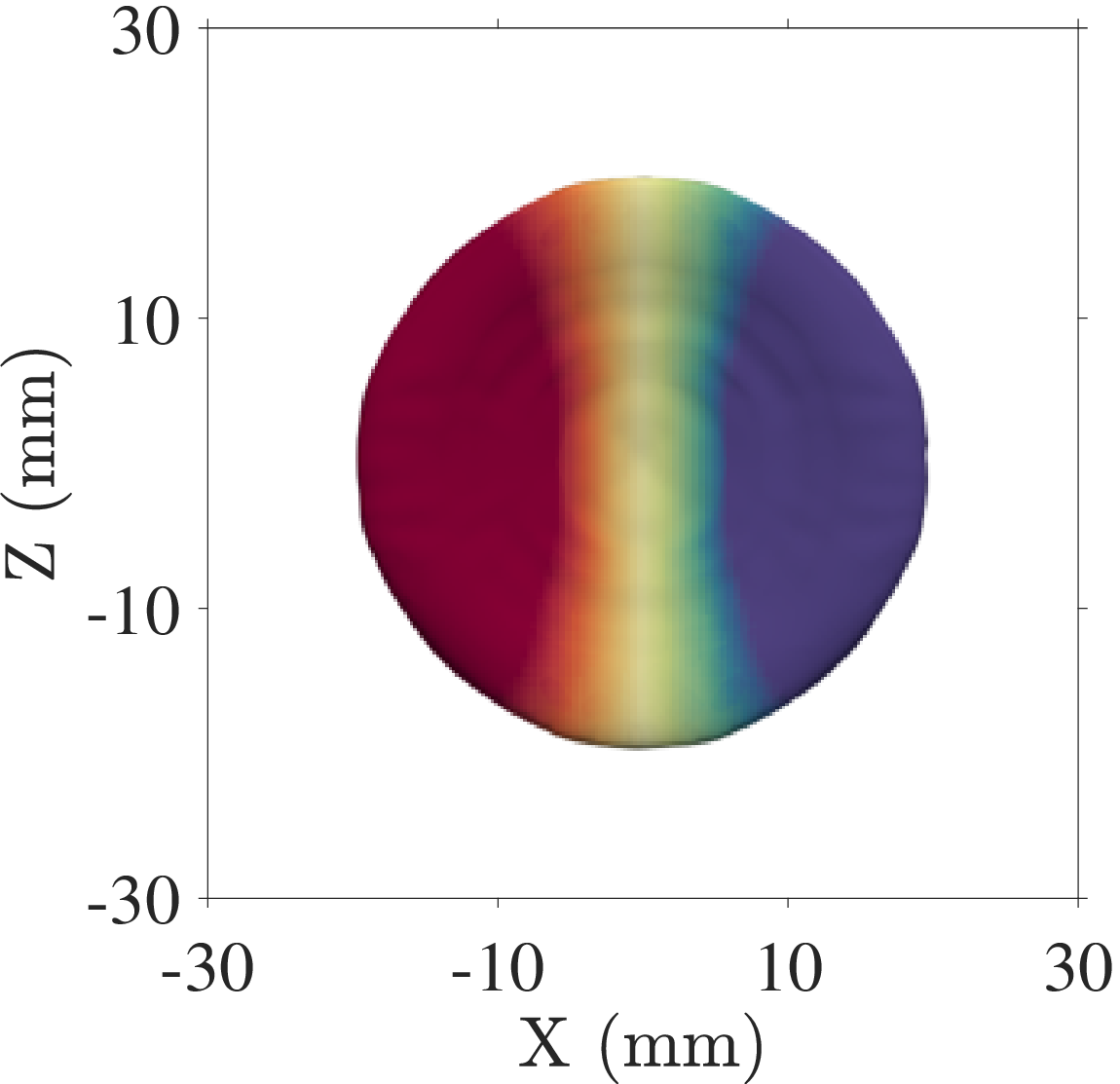}\\[3pt]
    \includegraphics[width=\linewidth]{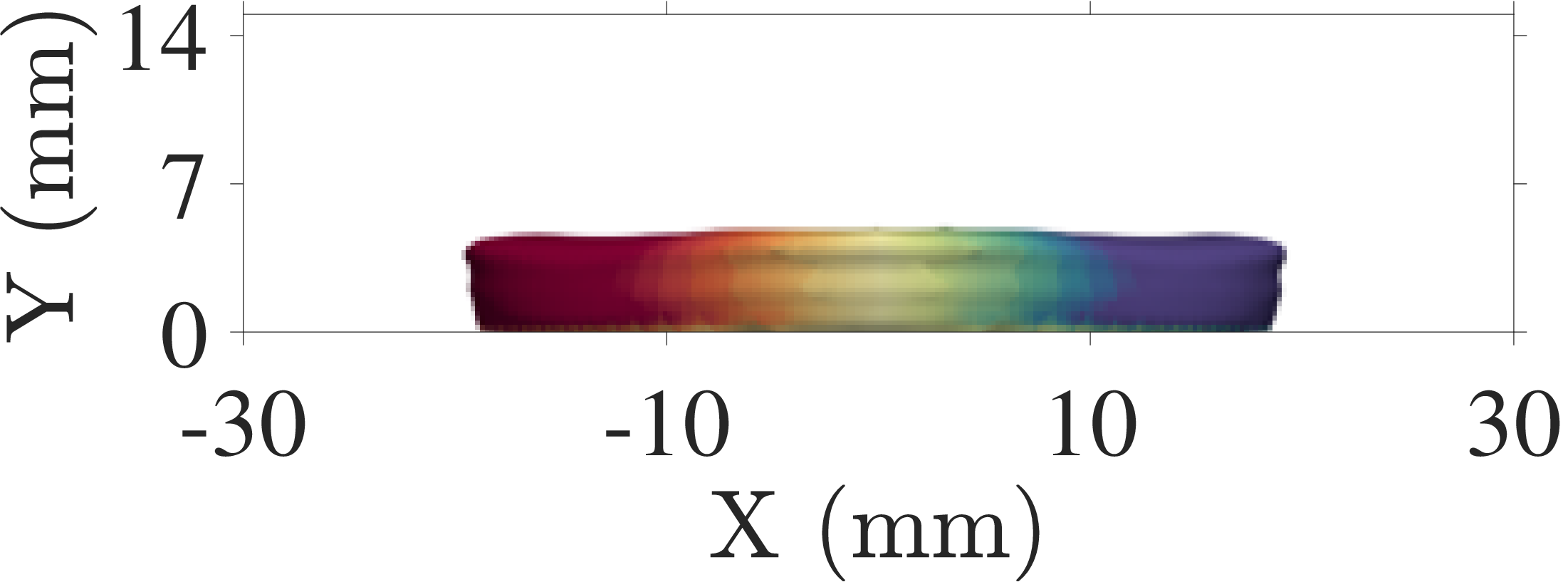}
\end{minipage}\hfill
\begin{minipage}{0.22\linewidth}
    \centering
    {\scriptsize $t=0.04\,\mathrm{s}$}\\[2pt]
    \includegraphics[width=\linewidth]{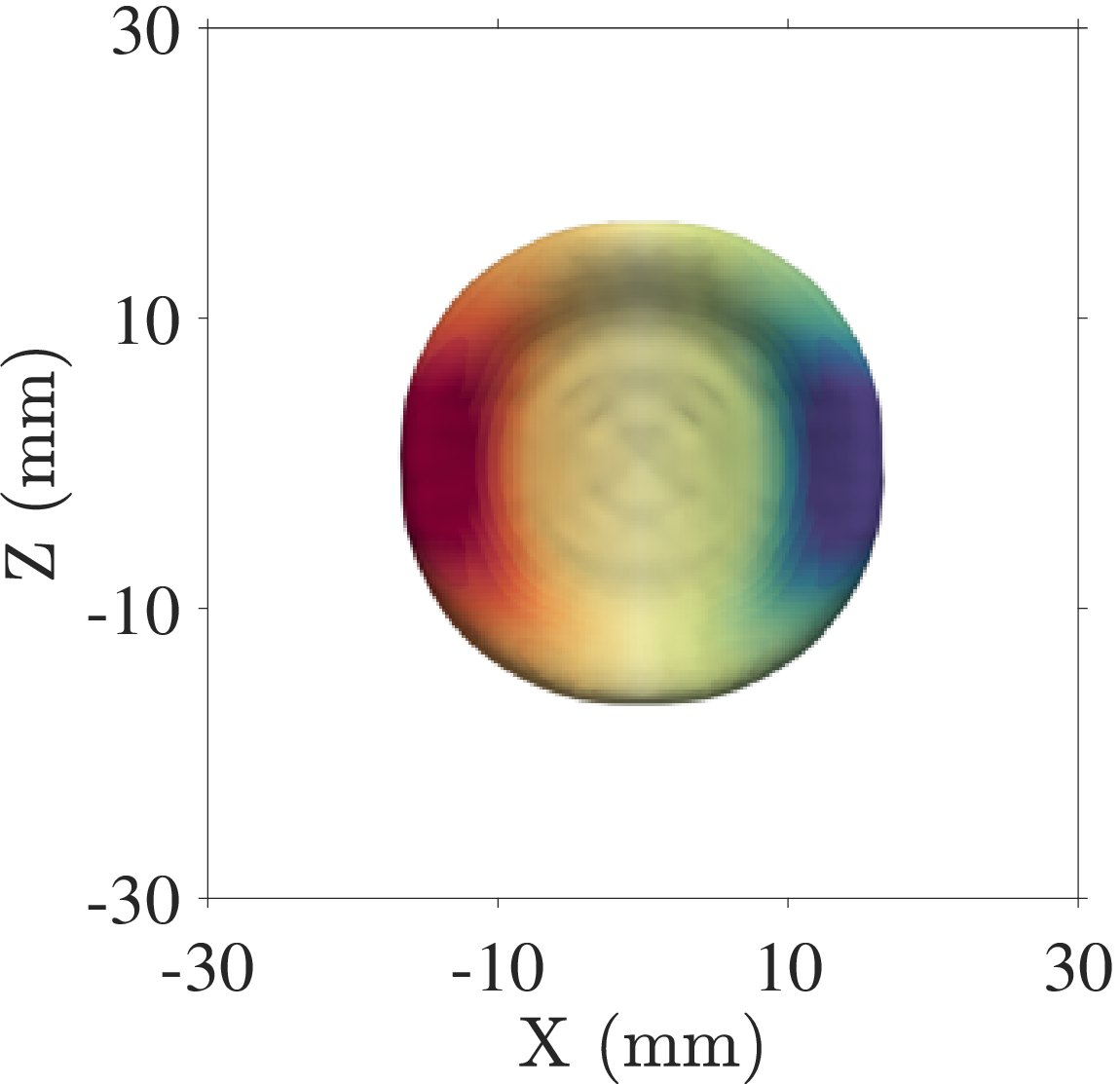}\\[3pt]
    \includegraphics[width=\linewidth]{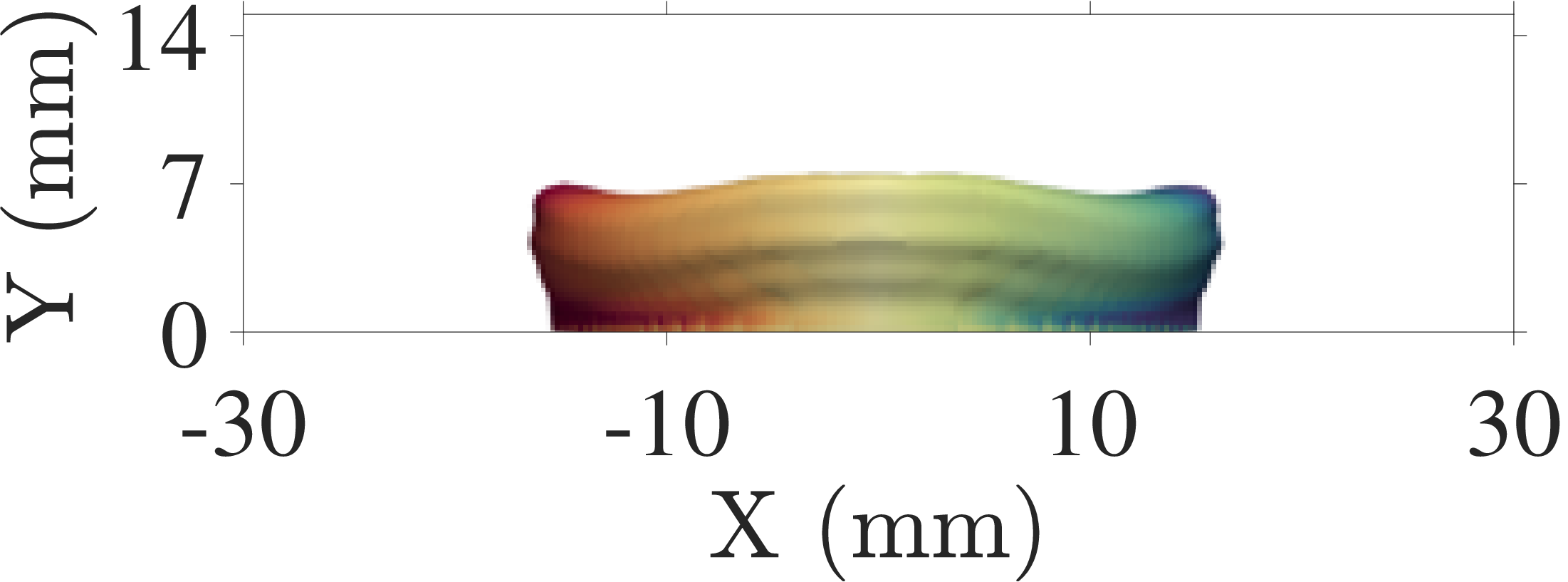}
\end{minipage}
\begin{center}\text{WCA = $160^\circ$}\end{center}

% -------- Figure for lambda = 0.04 --------
\caption{Temporal evolution of the axial velocity ($u_x$) distribution on the iso-surface of $\alpha = 0.5$ for $\lambda = 0.04$, from the instant of maximum spreading toward the final equilibrium state. The columns correspond to increasing time ($t = 0.01$ s to $t = 0.04$ s), while the rows represent different surface wettability conditions: hydrophilic (WCA = $0^\circ$), hybrid (WCA = $0^\circ$--$160^\circ$), and hydrophobic (WCA = $160^\circ$). The velocity field illustrates the transition from strong internal flow during recoil to a quiescent state as the droplet approaches equilibrium.}
\label{fig:ux_lambda_004}

\end{figure}

\begin{figure}[H]
\centering
    \includegraphics[width=0.45\linewidth]{ulegend.png}\\[3pt]
% ================= Row 1: WCA = 10° =================
\begin{minipage}{0.22\linewidth}
    \centering
    {\scriptsize $t=0.01\,\mathrm{s}$}\\[2pt]
    \includegraphics[width=\linewidth]{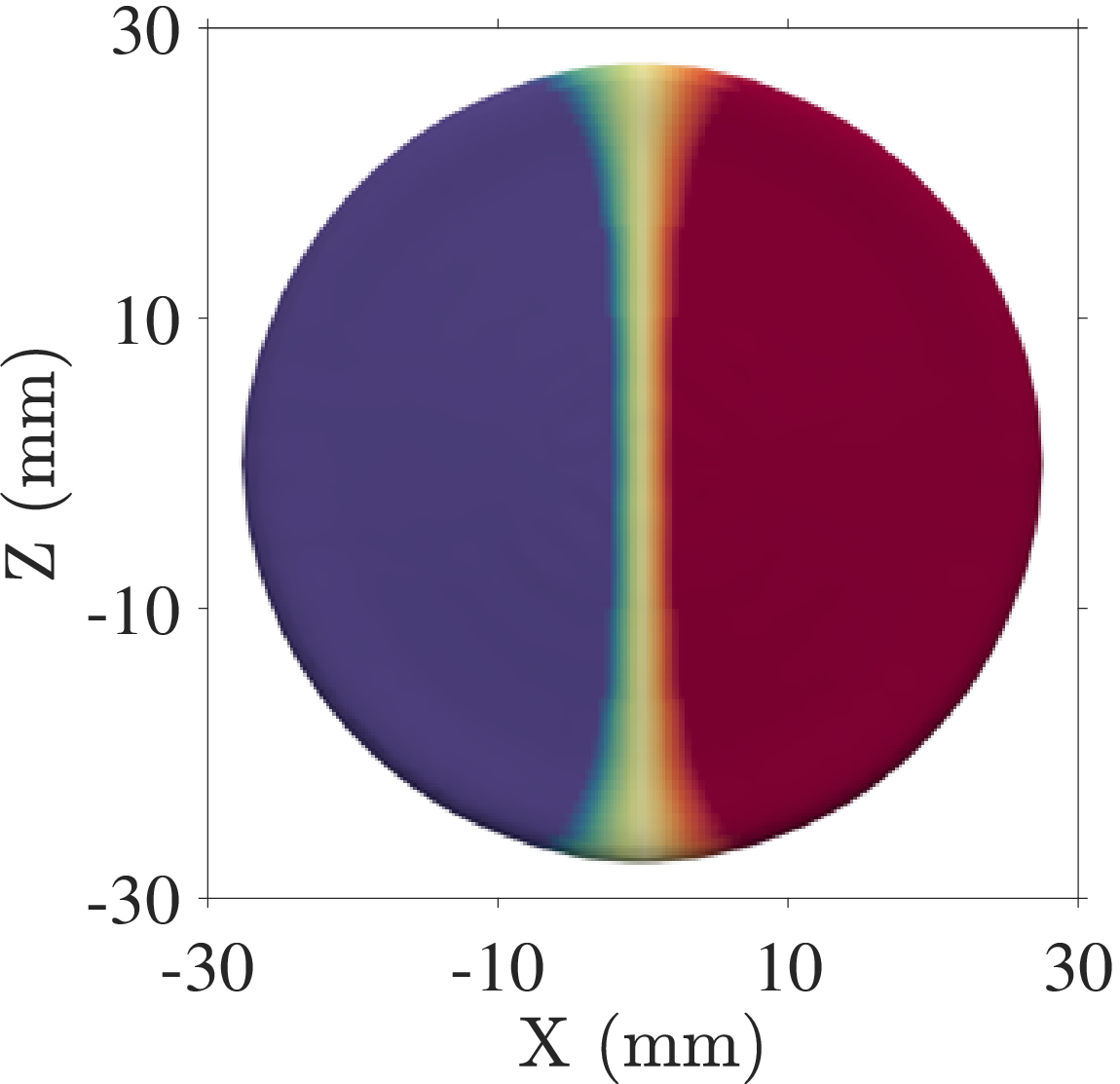}\\[3pt]
    \includegraphics[width=\linewidth]{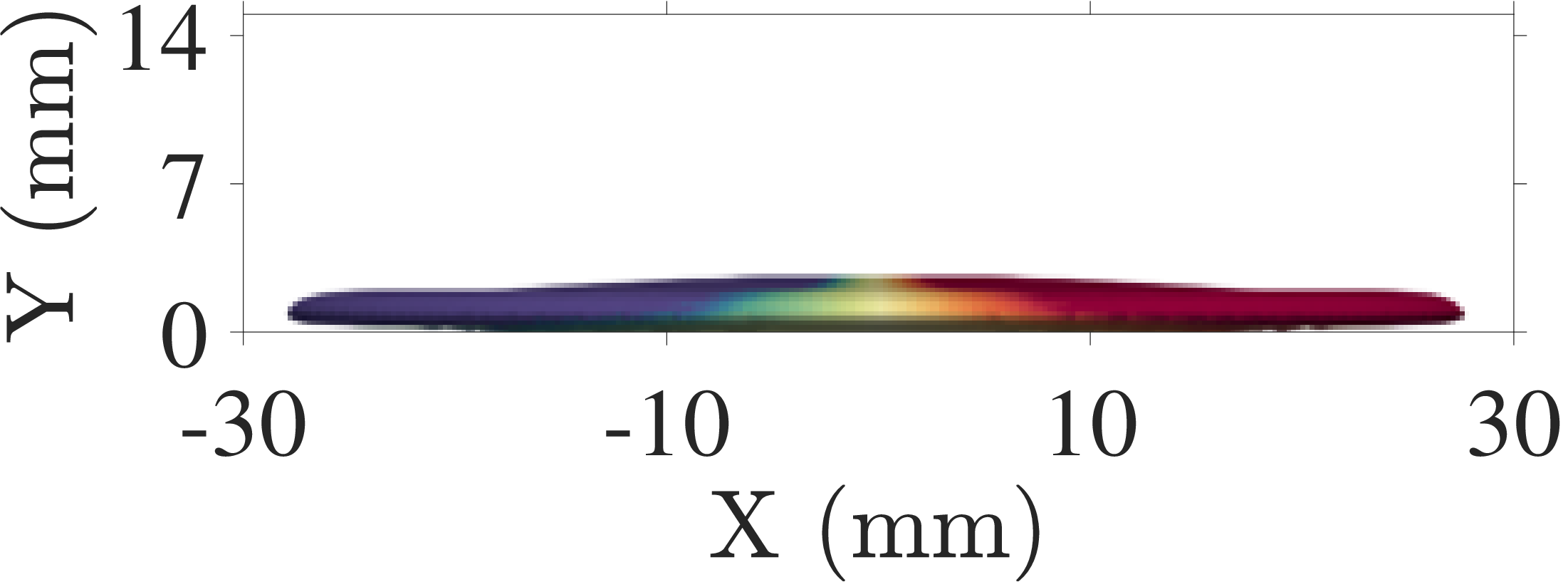}
\end{minipage}\hfill
\begin{minipage}{0.22\linewidth}
    \centering
    {\scriptsize $t=0.02\,\mathrm{s}$}\\[2pt]
    \includegraphics[width=\linewidth]{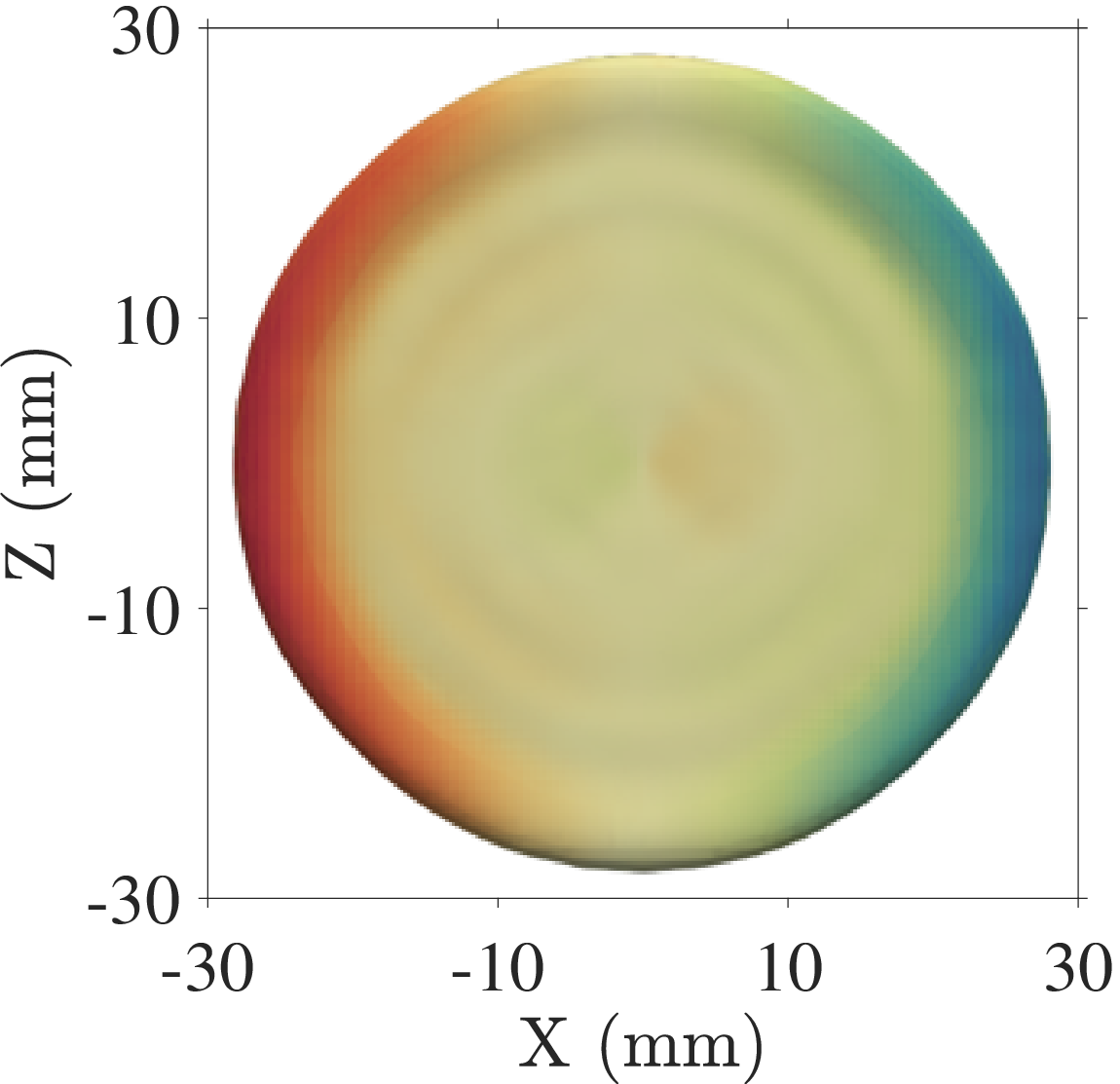}\\[3pt]
    \includegraphics[width=\linewidth]{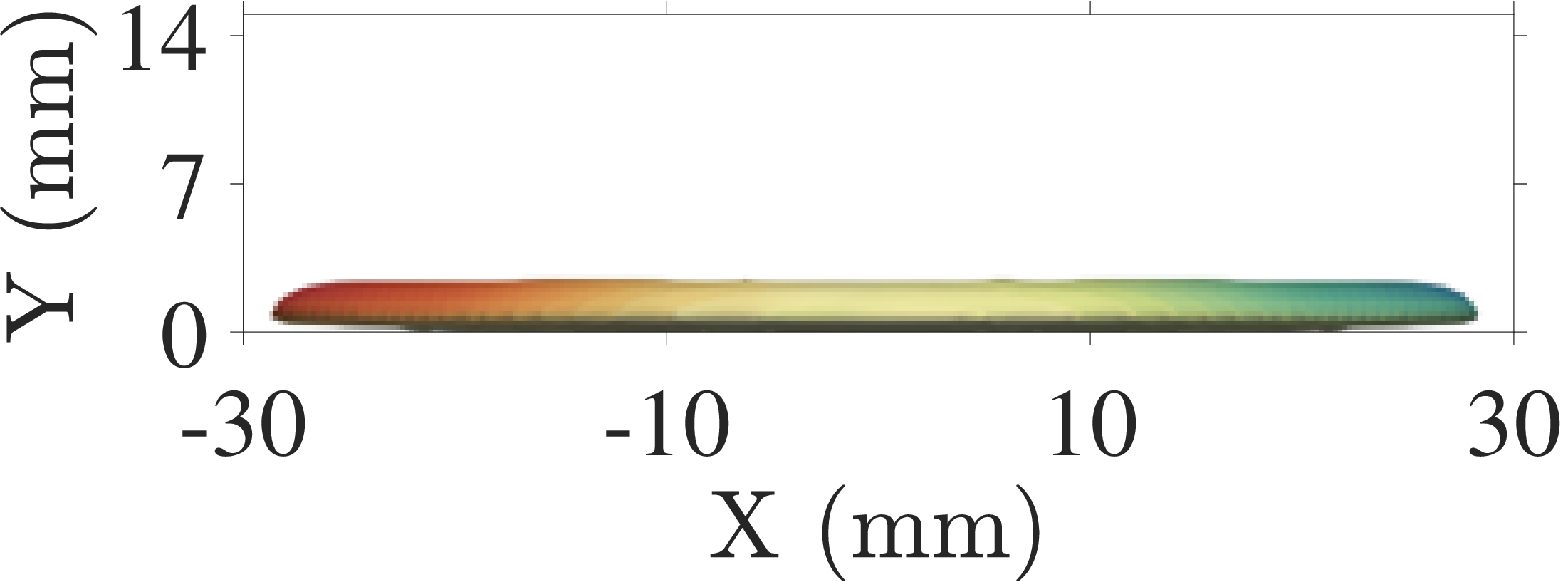}
\end{minipage}\hfill
\begin{minipage}{0.22\linewidth}
    \centering
    {\scriptsize $t=0.03\,\mathrm{s}$}\\[2pt]
    \includegraphics[width=\linewidth]{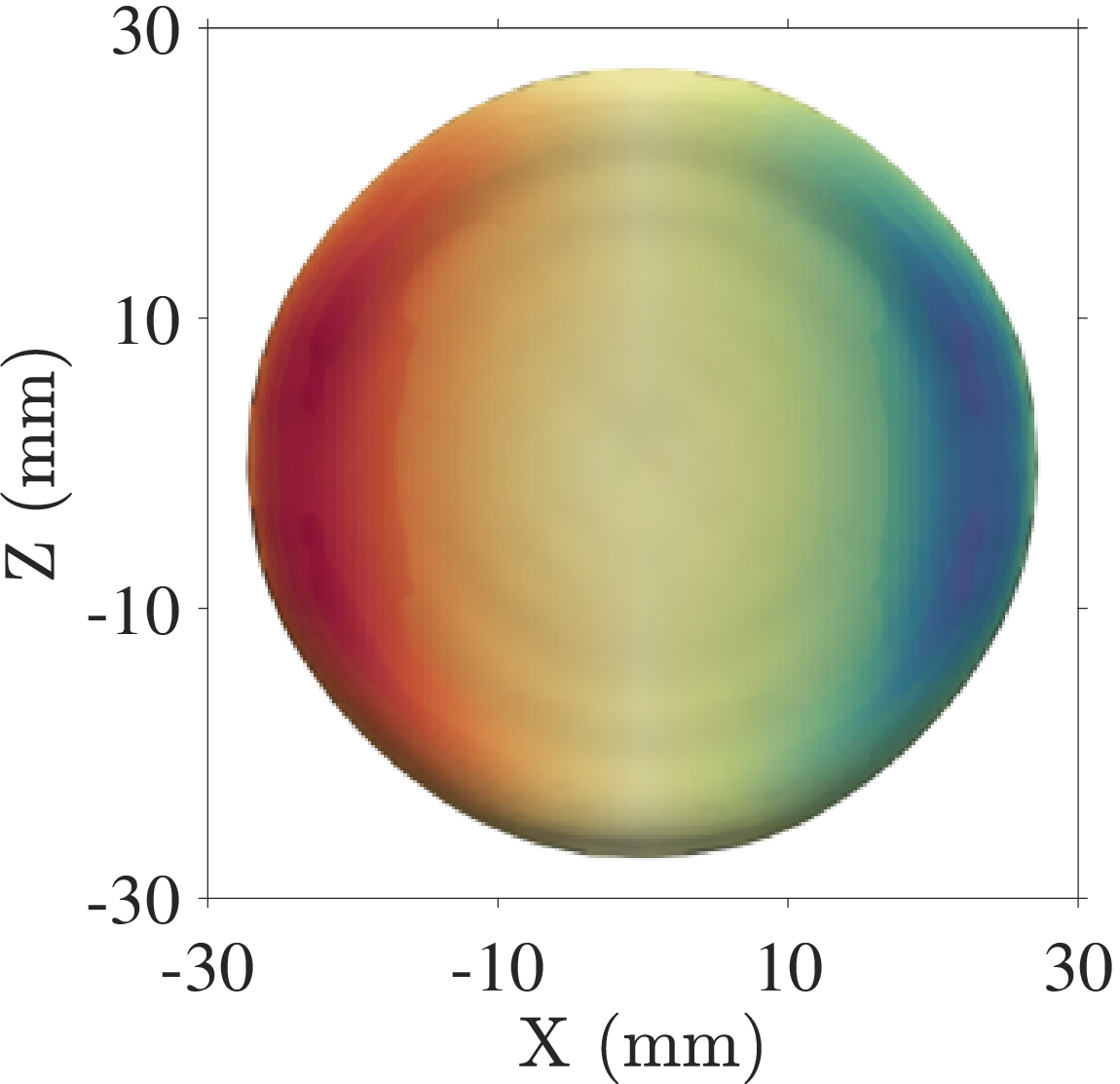}\\[3pt]
    \includegraphics[width=\linewidth]{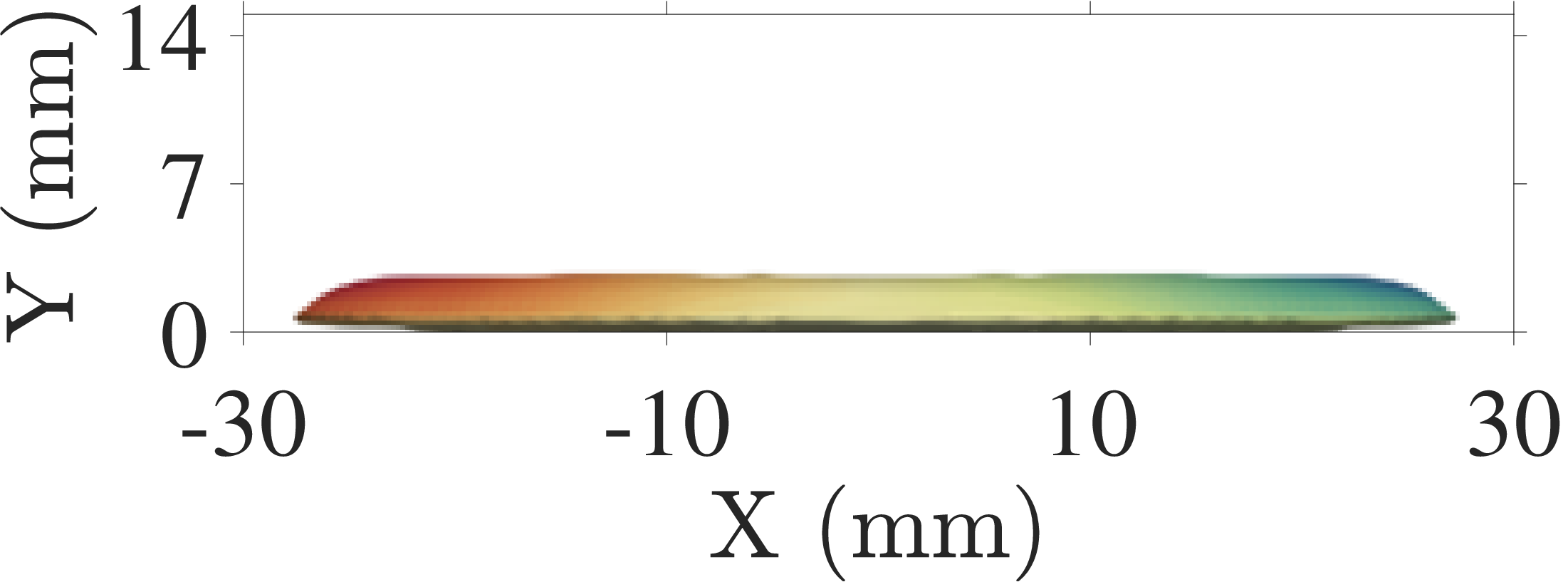}
\end{minipage}\hfill
\begin{minipage}{0.22\linewidth}
    \centering
    {\scriptsize $t=0.04\,\mathrm{s}$}\\[2pt]
    \includegraphics[width=\linewidth]{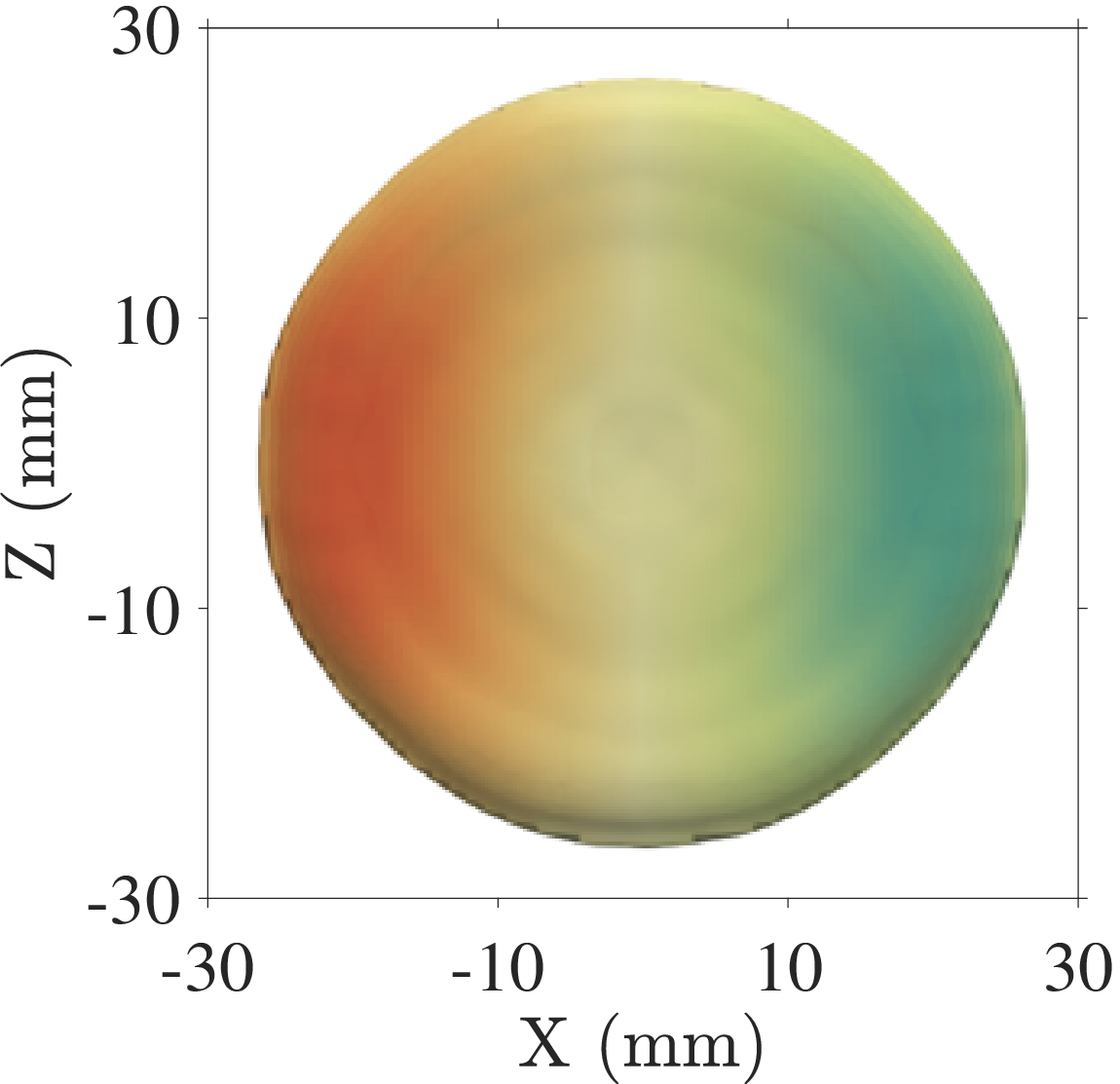}\\[3pt]
    \includegraphics[width=\linewidth]{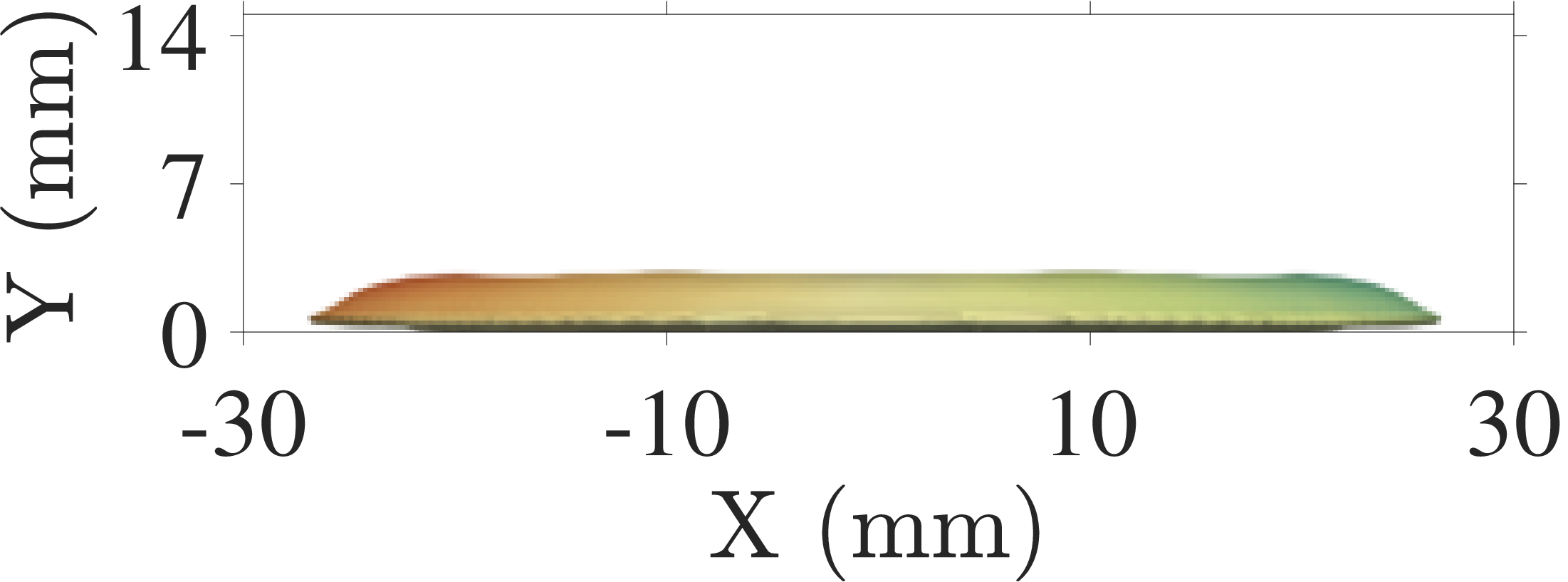}
\end{minipage}

\vspace{2pt}
\begin{center}\text{WCA = $0^\circ$}\end{center}

\vspace{2pt}

% ================= Row 2: Hybrid =================
\begin{minipage}{0.22\linewidth}
    \centering
    {\scriptsize $t=0.01\,\mathrm{s}$}\\[2pt]
    \includegraphics[width=\linewidth]{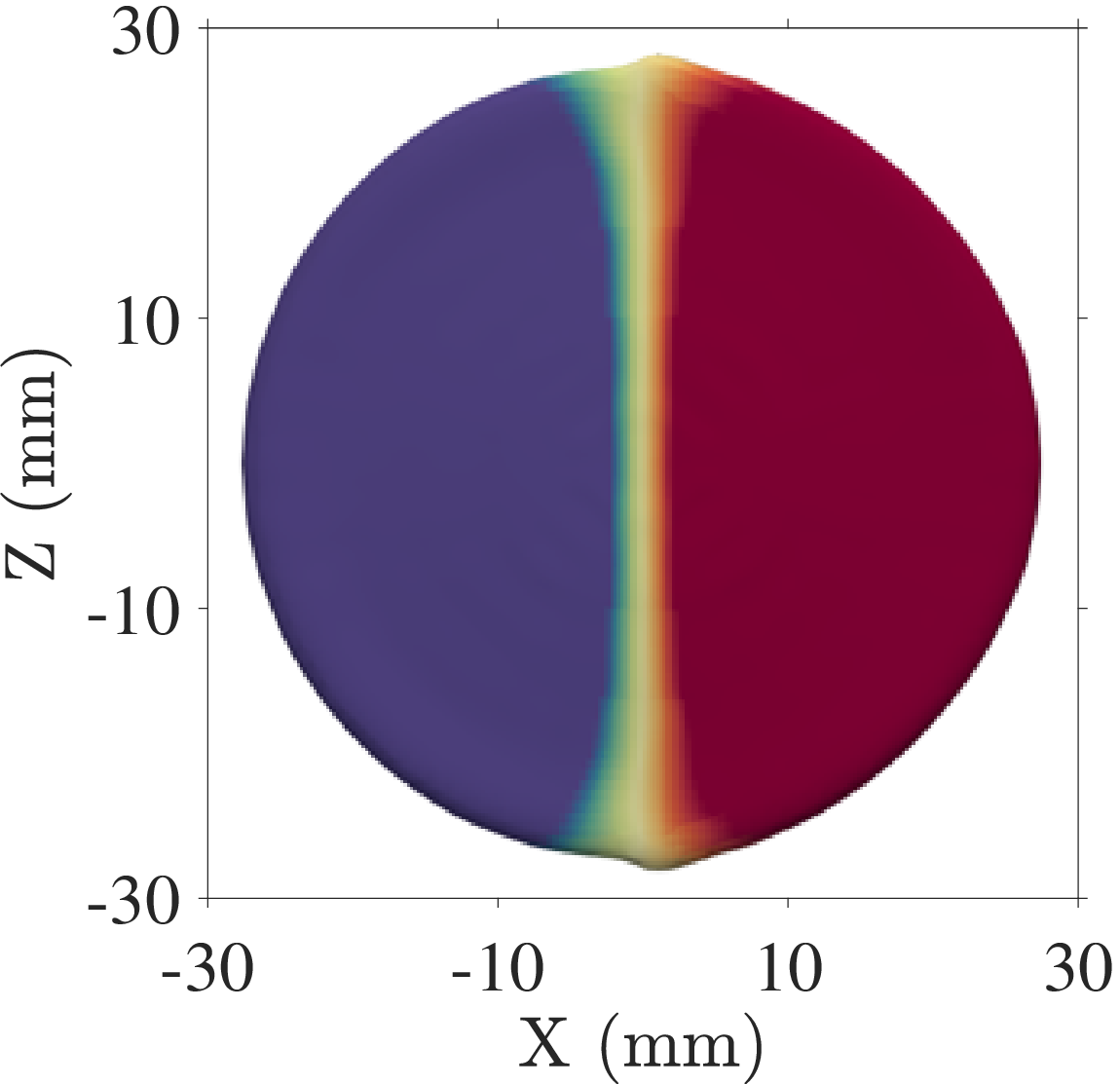}\\[3pt]
    \includegraphics[width=\linewidth]{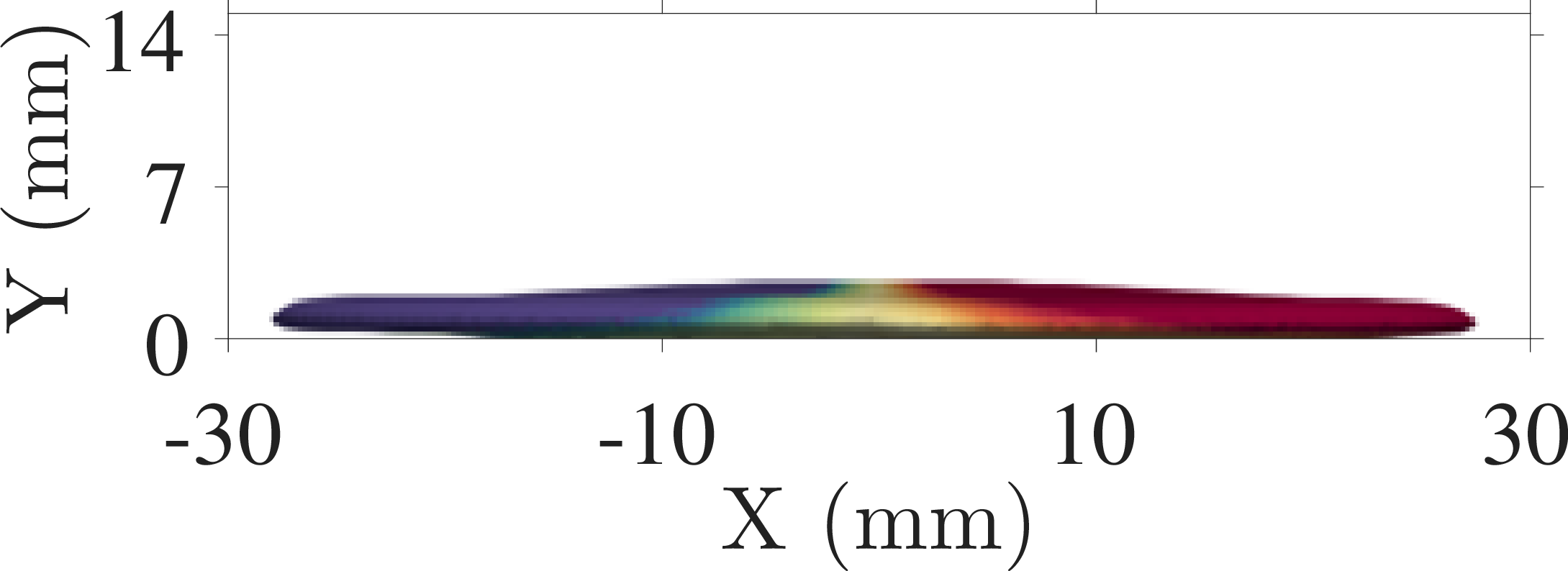}
\end{minipage}\hfill
\begin{minipage}{0.22\linewidth}
    \centering
    {\scriptsize $t=0.02\,\mathrm{s}$}\\[2pt]
    \includegraphics[width=\linewidth]{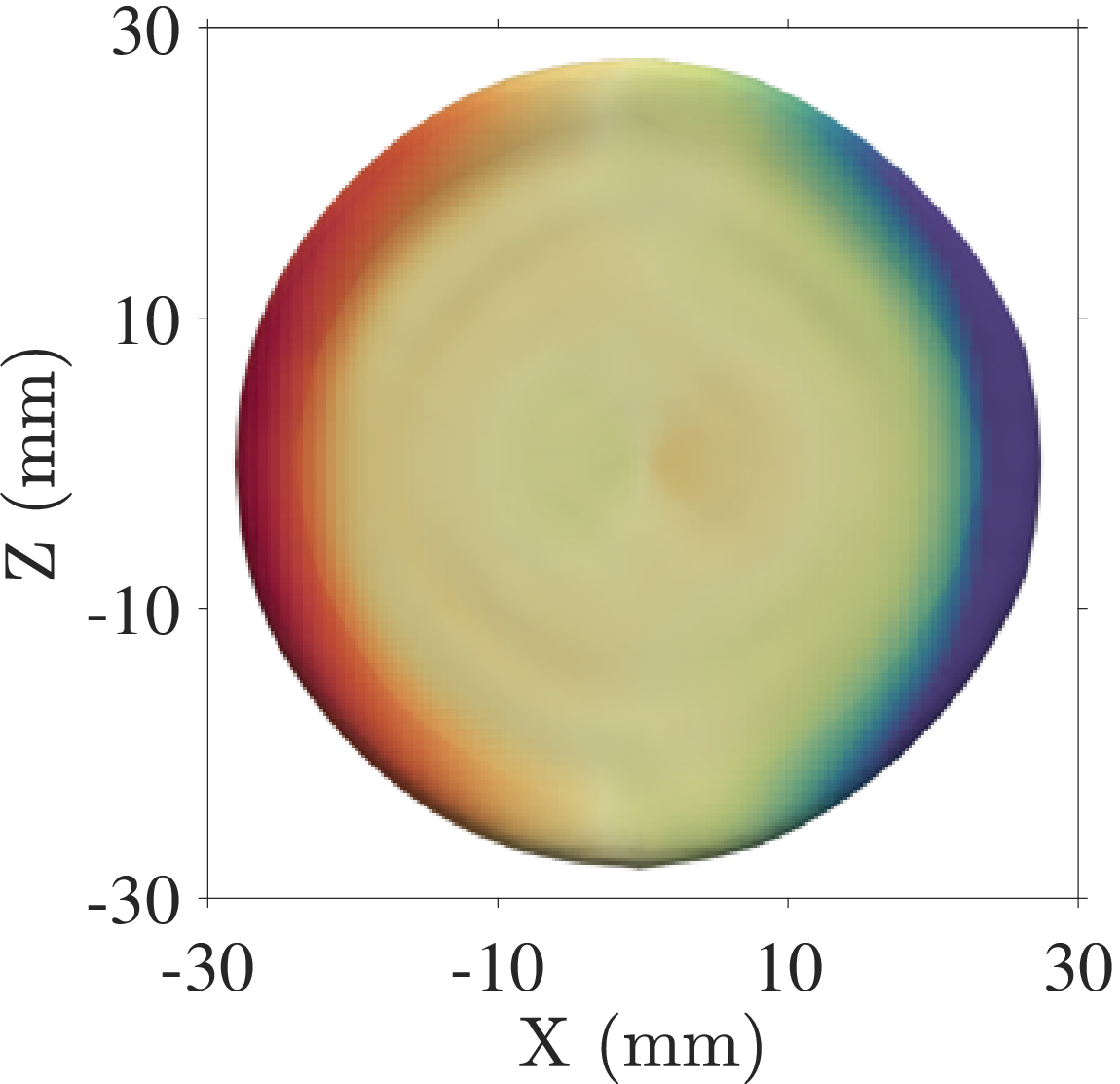}\\[3pt]
    \includegraphics[width=\linewidth]{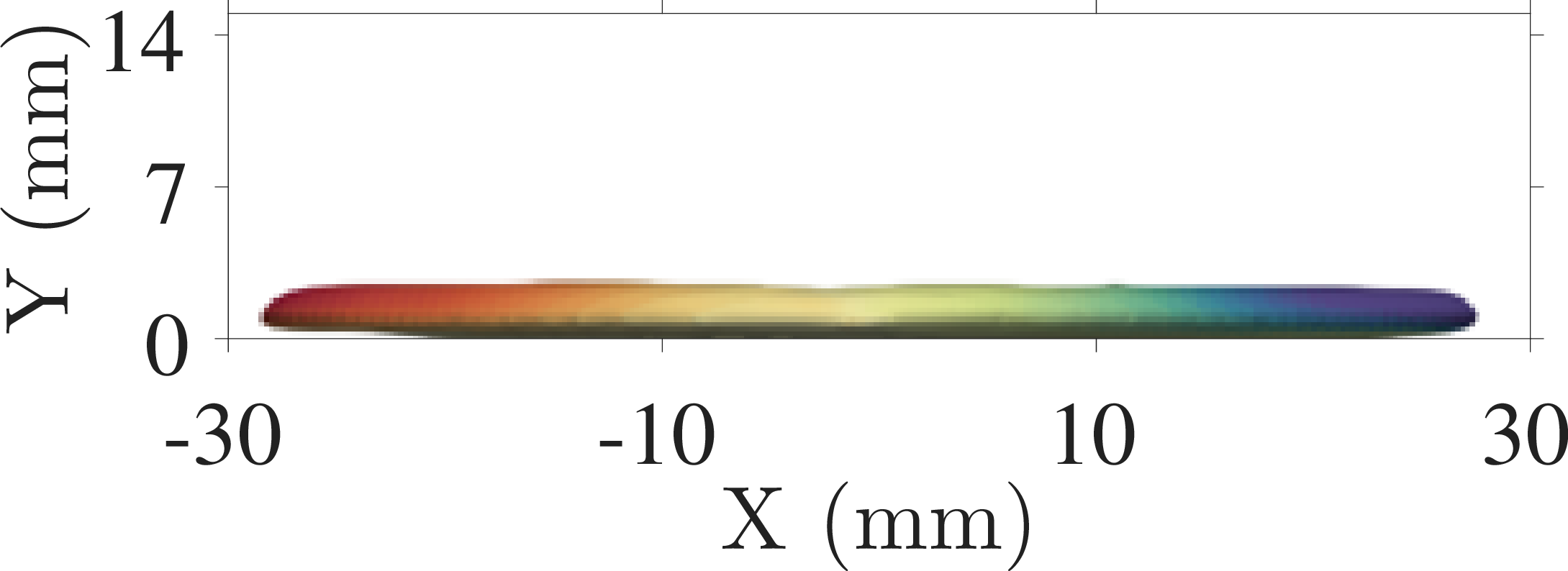}
\end{minipage}\hfill
\begin{minipage}{0.22\linewidth}
    \centering
    {\scriptsize $t=0.03\,\mathrm{s}$}\\[2pt]
    \includegraphics[width=\linewidth]{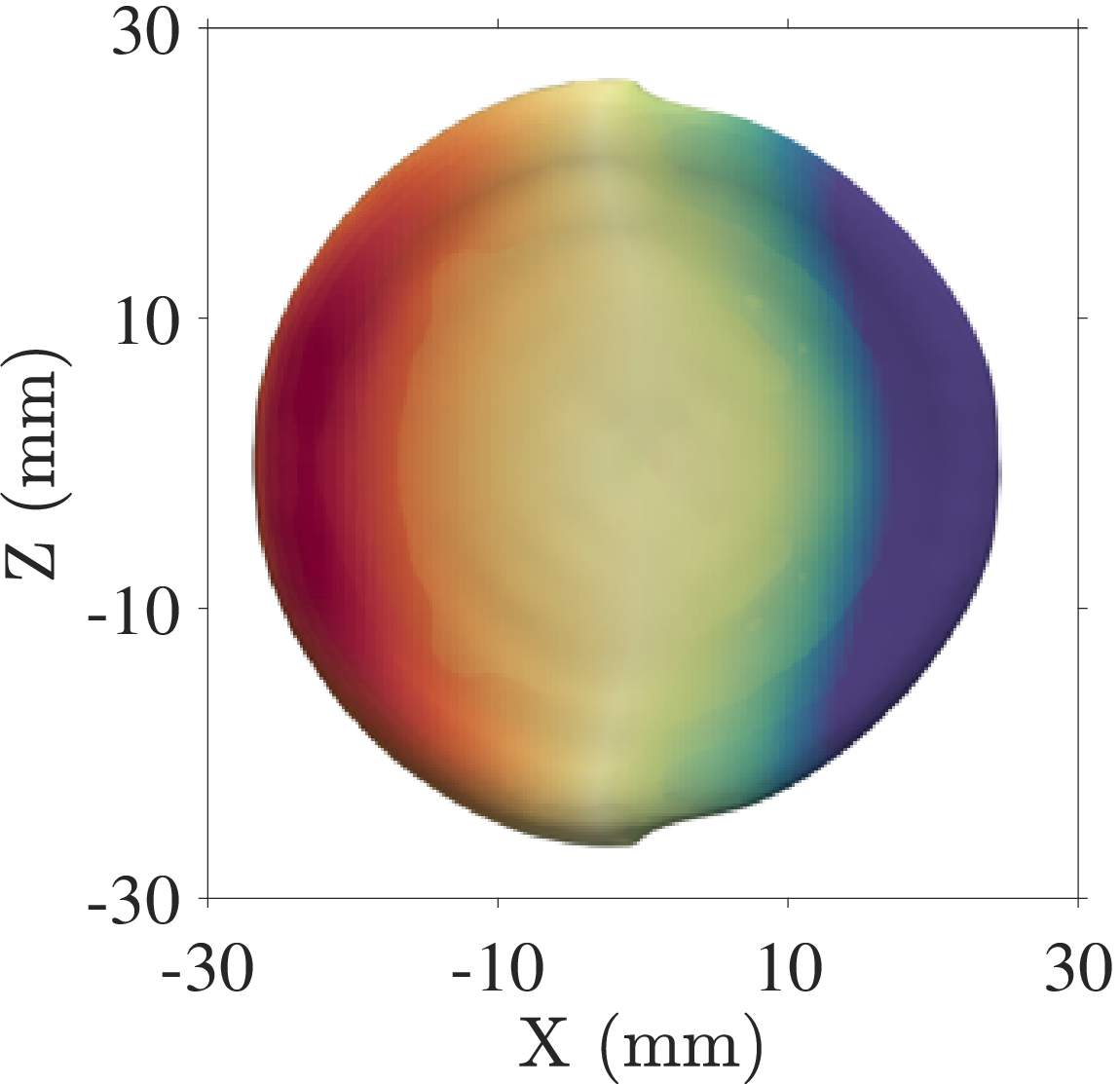}\\[3pt]
    \includegraphics[width=\linewidth]{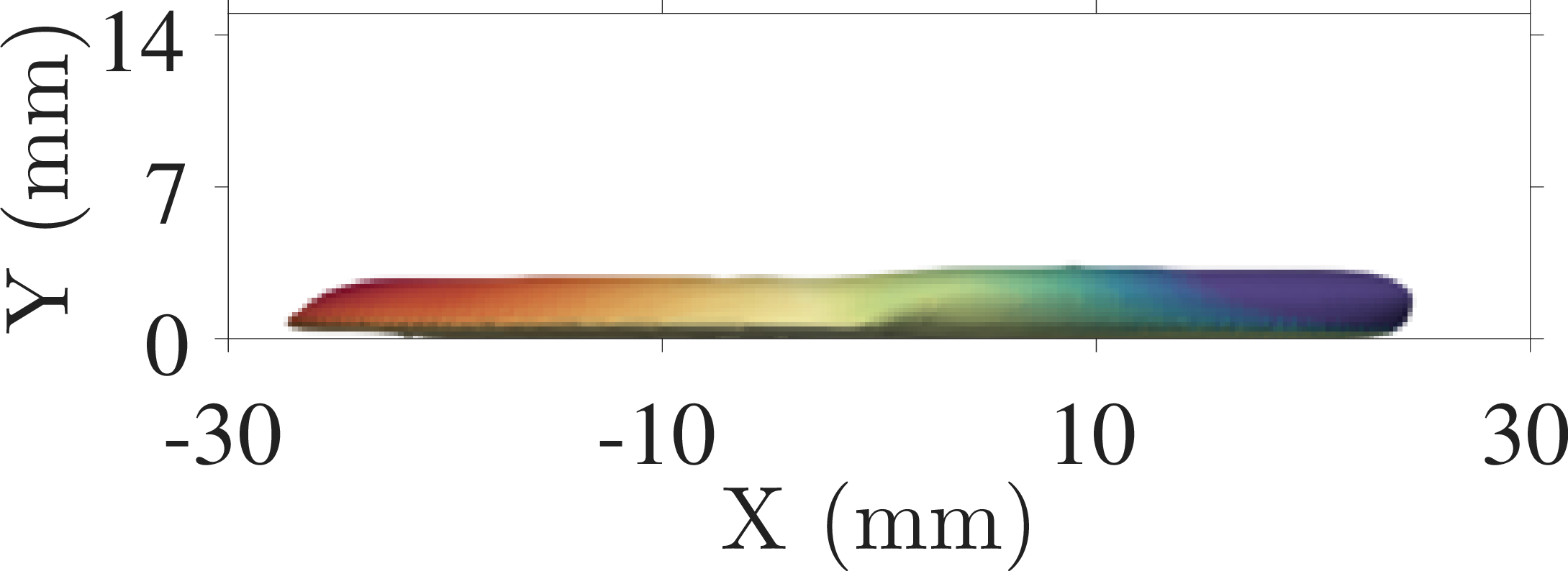}
\end{minipage}\hfill
\begin{minipage}{0.22\linewidth}
    \centering
    {\scriptsize $t=0.04\,\mathrm{s}$}\\[2pt]
    \includegraphics[width=\linewidth]{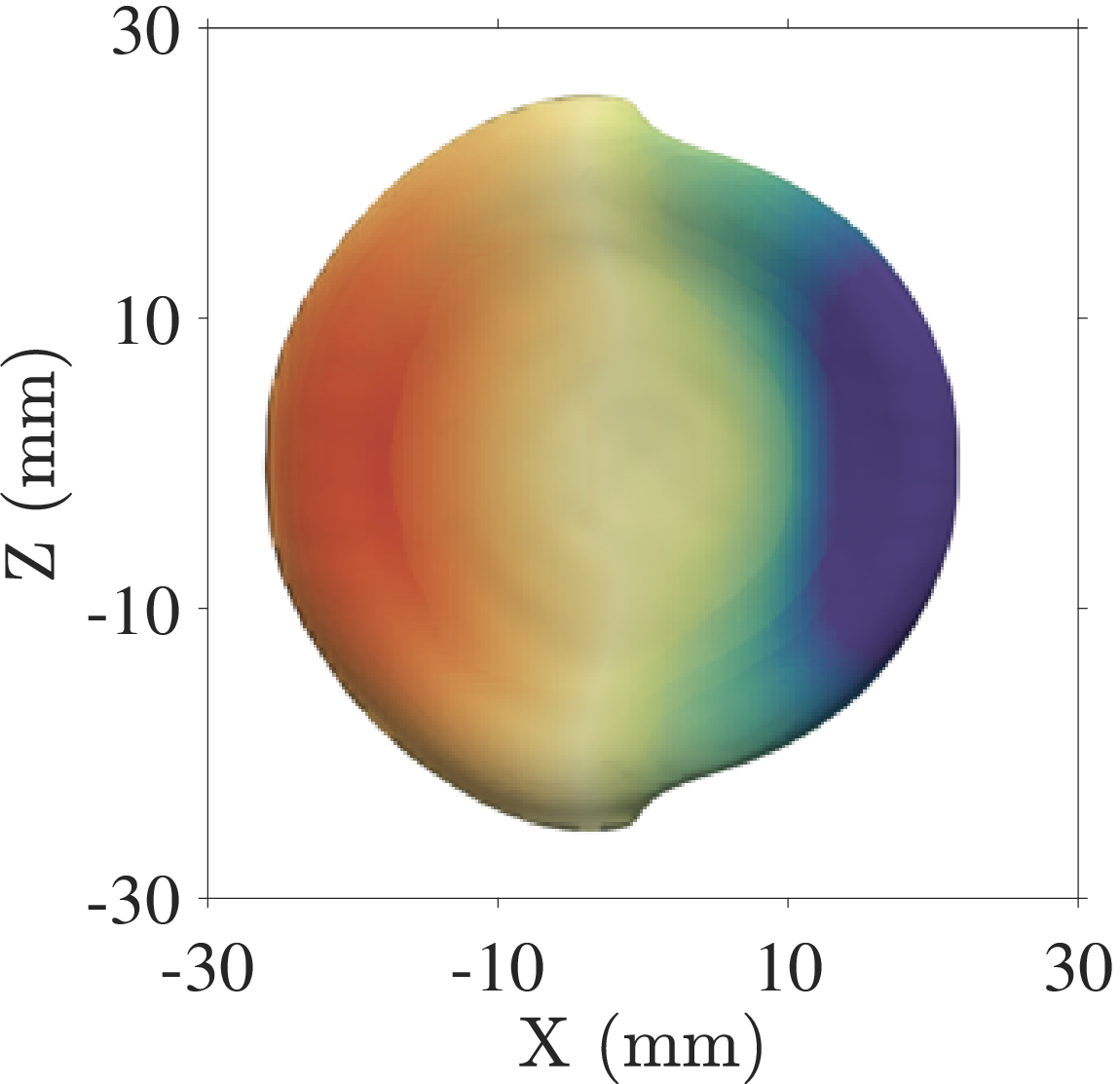}\\[3pt]
    \includegraphics[width=\linewidth]{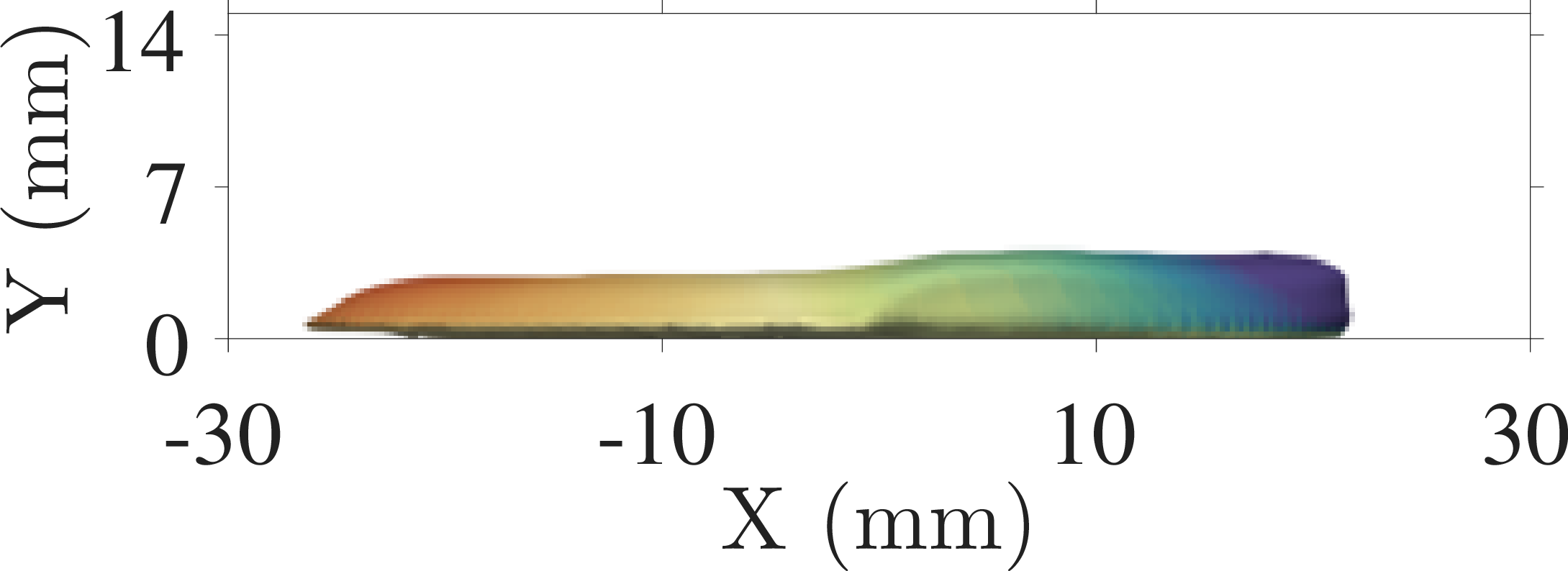}
\end{minipage}

\vspace{2pt}
\begin{center}\text{WCA = Hybrid $0^\circ$--$160^\circ$}\end{center}

\vspace{2pt}

% ================= Row 3: WCA = 160° =================
\begin{minipage}{0.22\linewidth}
    \centering
    {\scriptsize $t=0.01\,\mathrm{s}$}\\[2pt]
    \includegraphics[width=\linewidth]{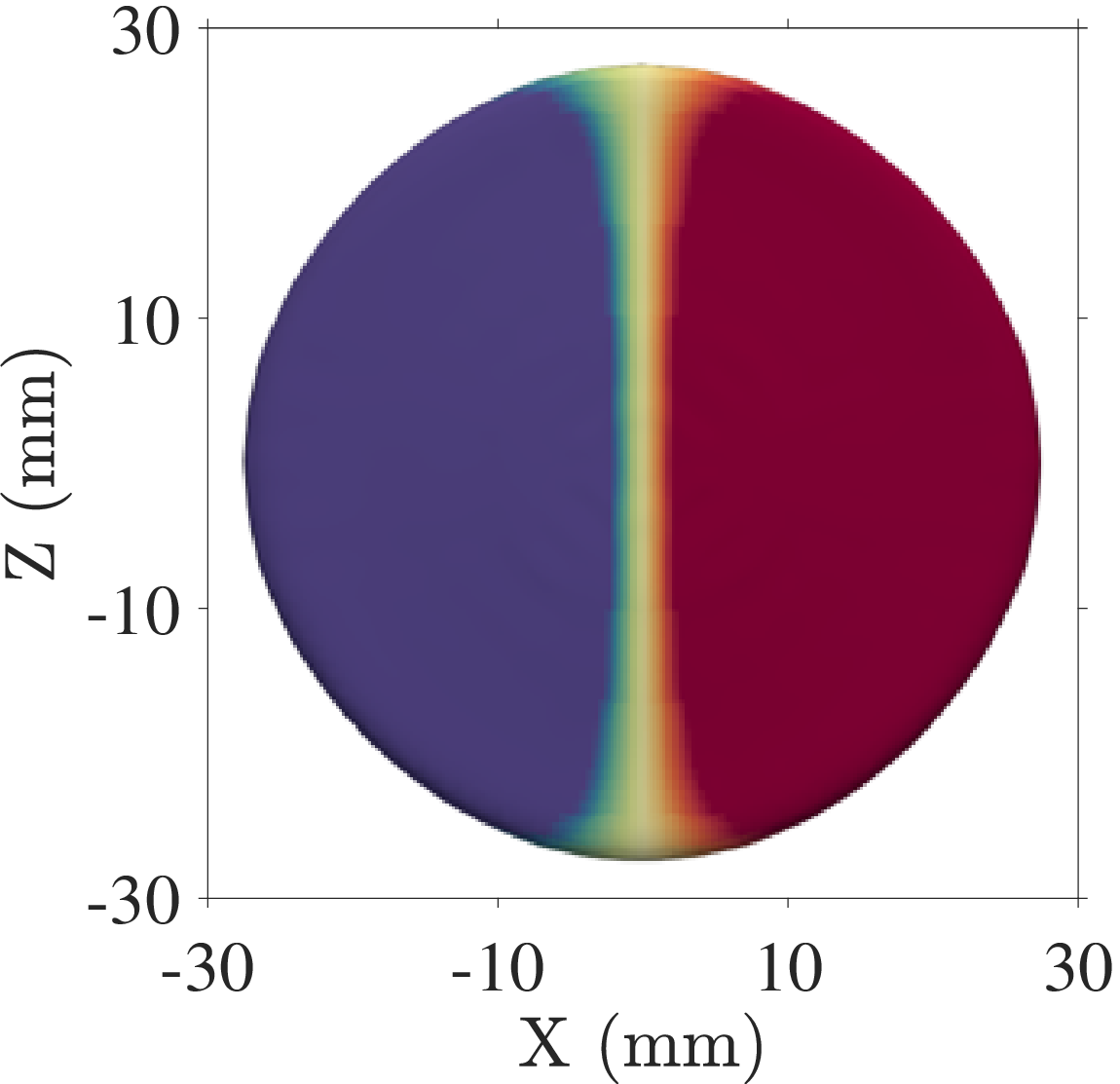}\\[3pt]
    \includegraphics[width=\linewidth]{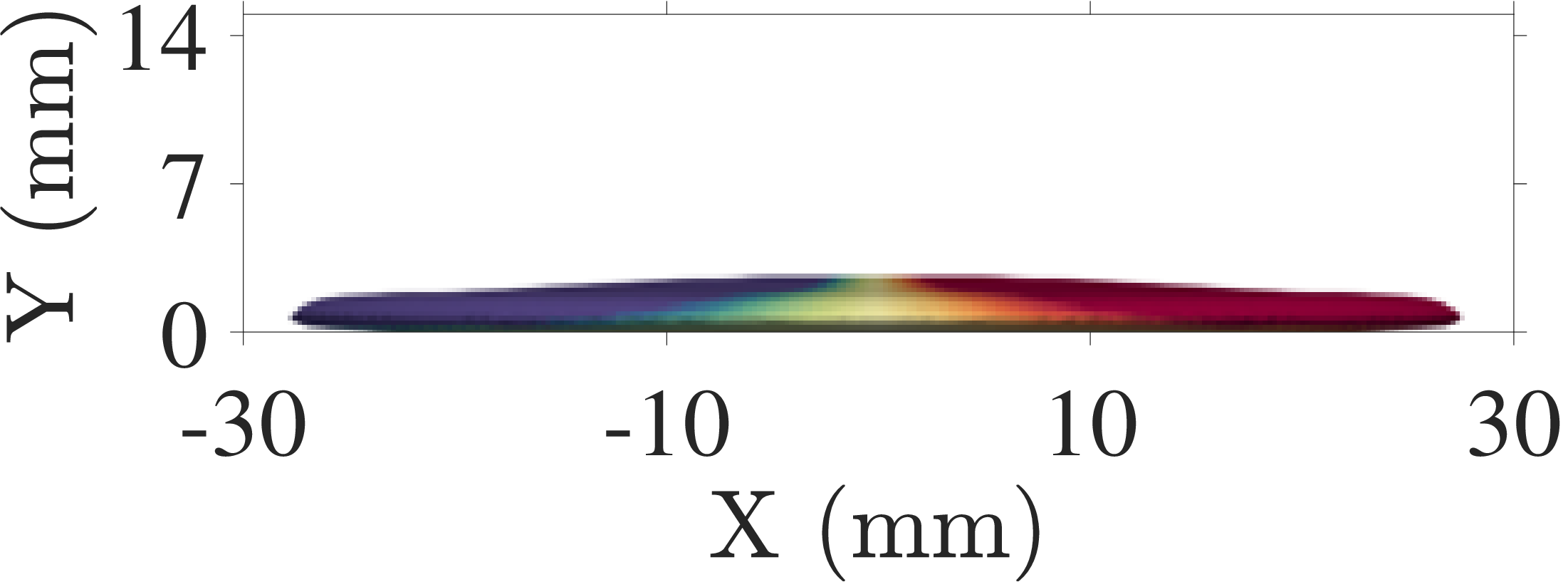}
\end{minipage}\hfill
\begin{minipage}{0.22\linewidth}
    \centering
    {\scriptsize $t=0.02\,\mathrm{s}$}\\[2pt]
    \includegraphics[width=\linewidth]{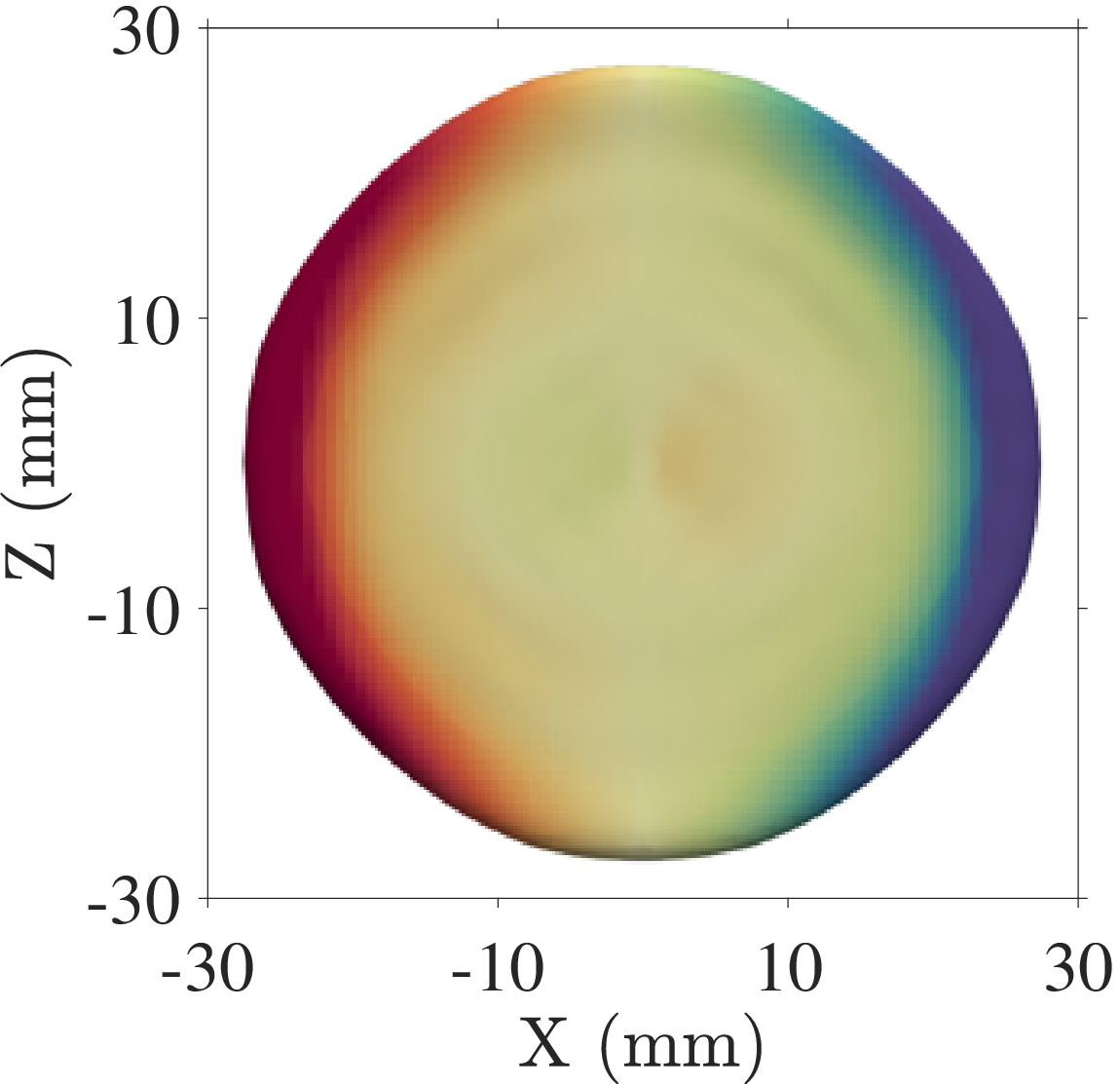}\\[3pt]
    \includegraphics[width=\linewidth]{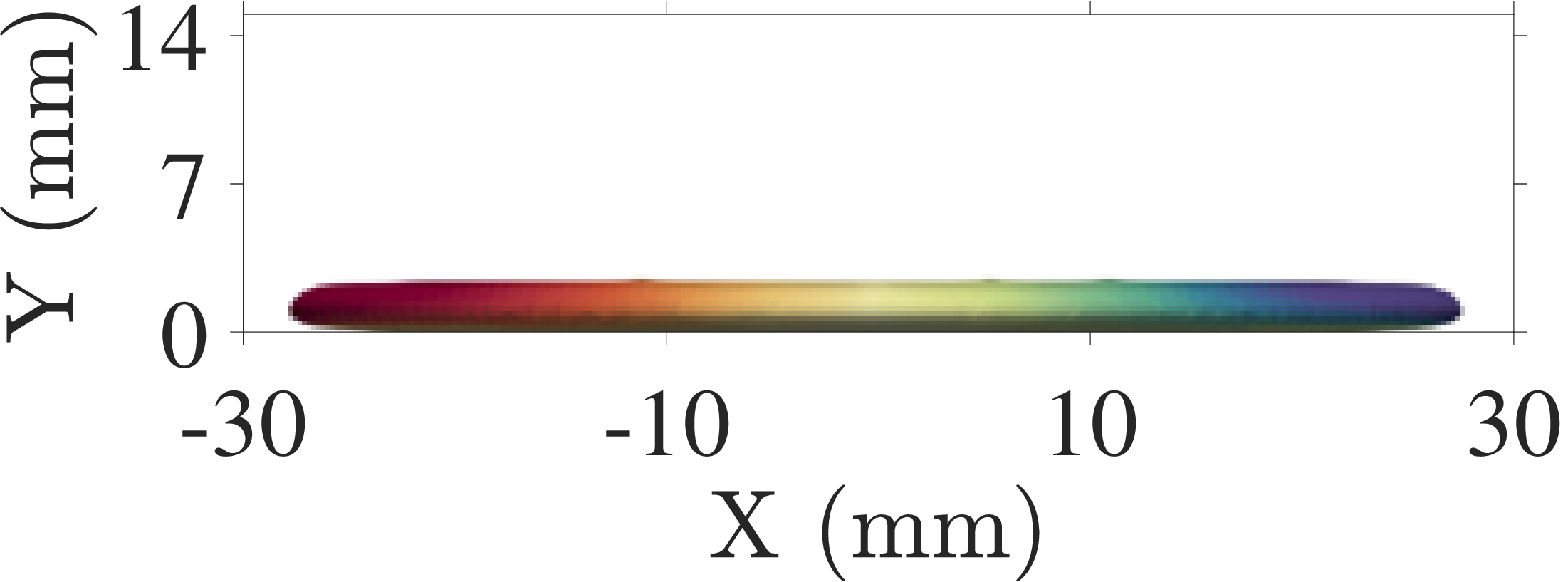}
\end{minipage}\hfill
\begin{minipage}{0.22\linewidth}
    \centering
    {\scriptsize $t=0.03\,\mathrm{s}$}\\[2pt]
    \includegraphics[width=\linewidth]{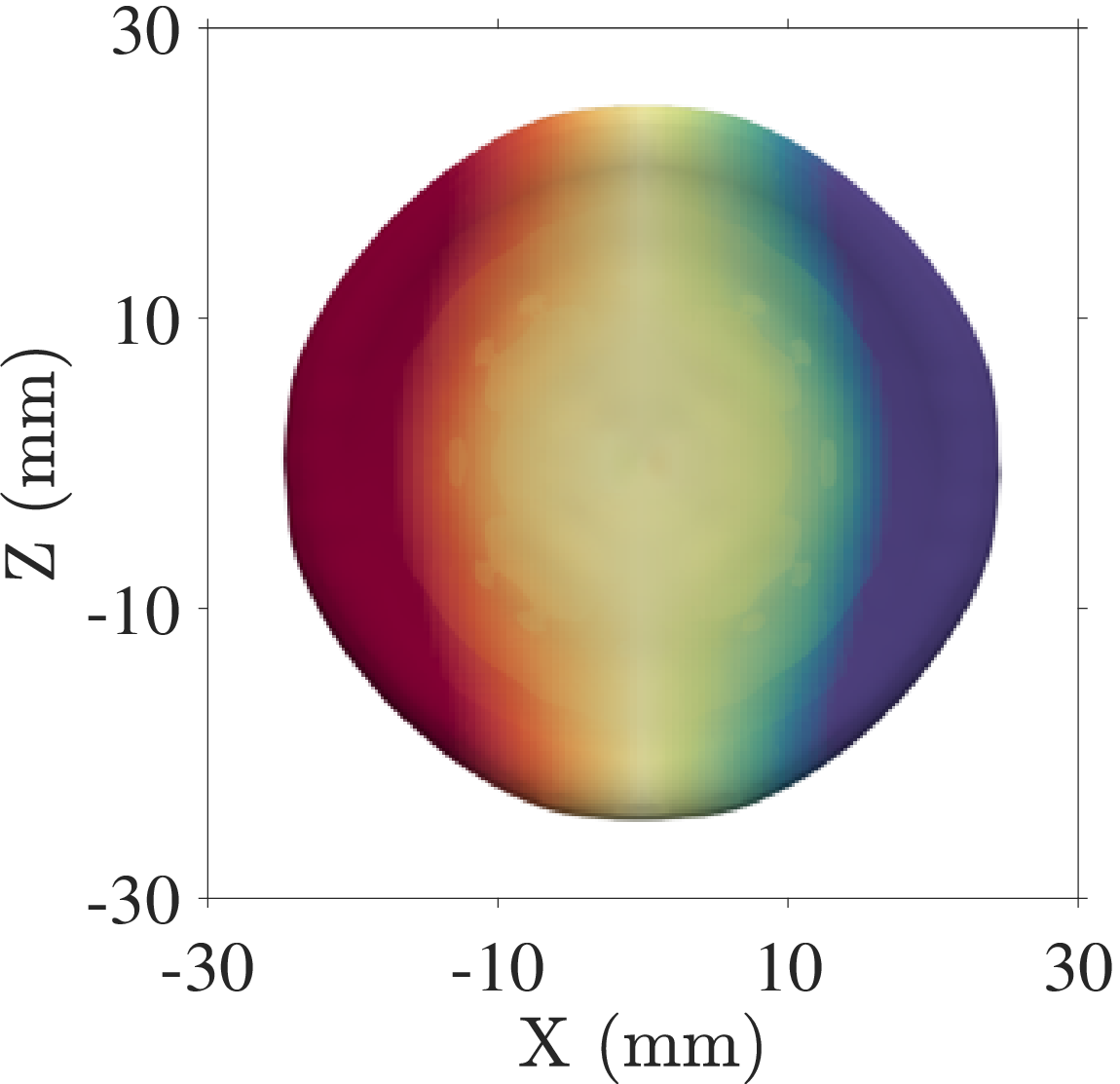}\\[3pt]
    \includegraphics[width=\linewidth]{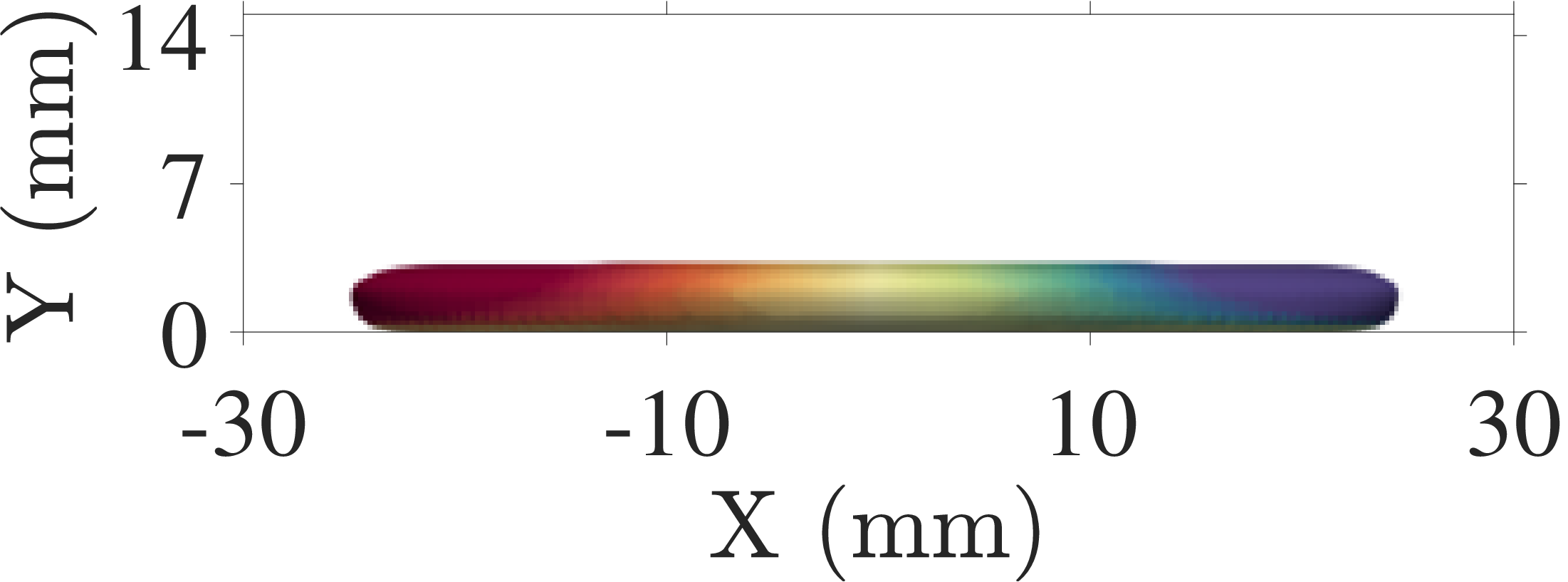}
\end{minipage}\hfill
\begin{minipage}{0.22\linewidth}
    \centering
    {\scriptsize $t=0.04\,\mathrm{s}$}\\[2pt]
    \includegraphics[width=\linewidth]{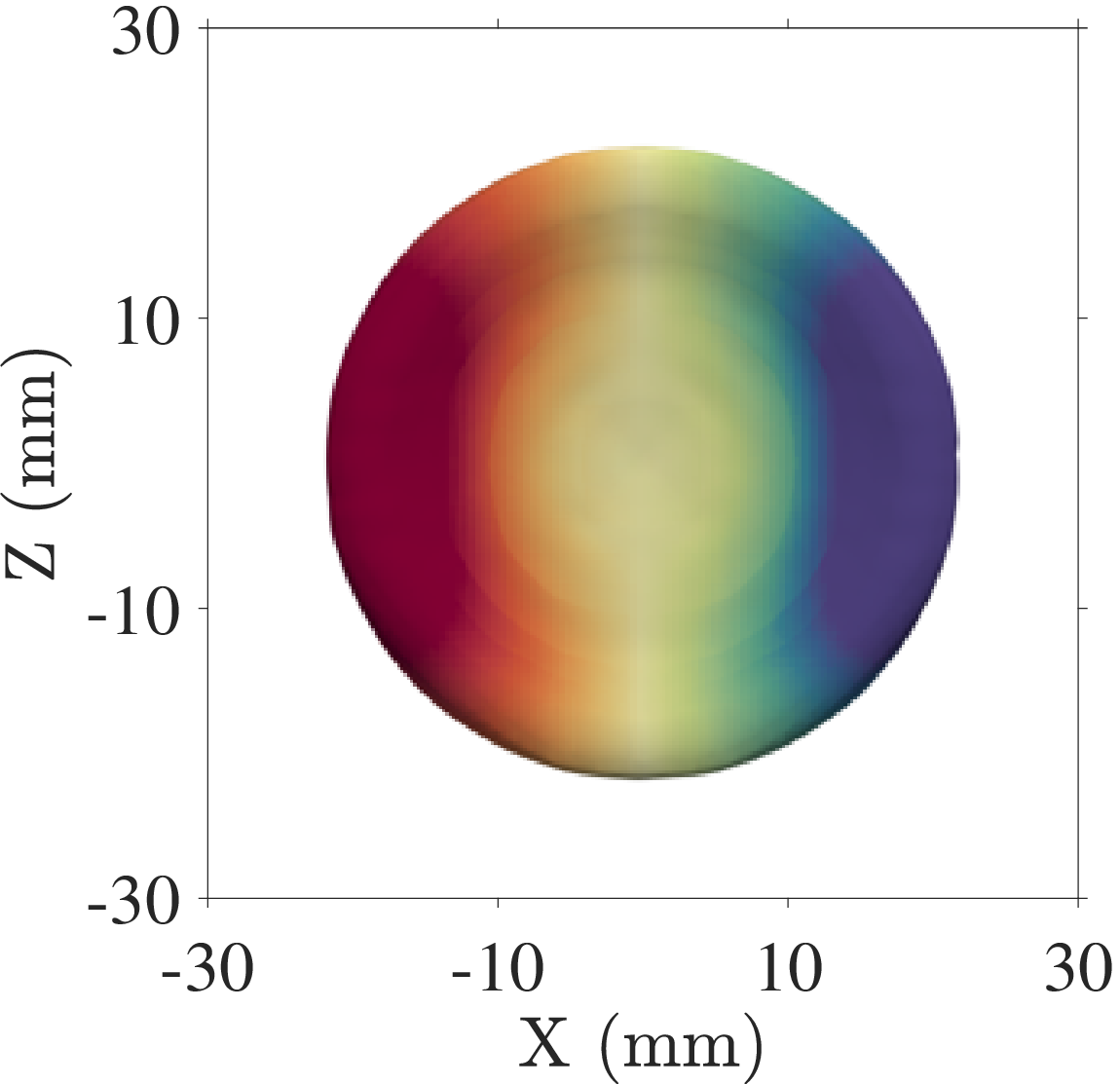}\\[3pt]
    \includegraphics[width=\linewidth]{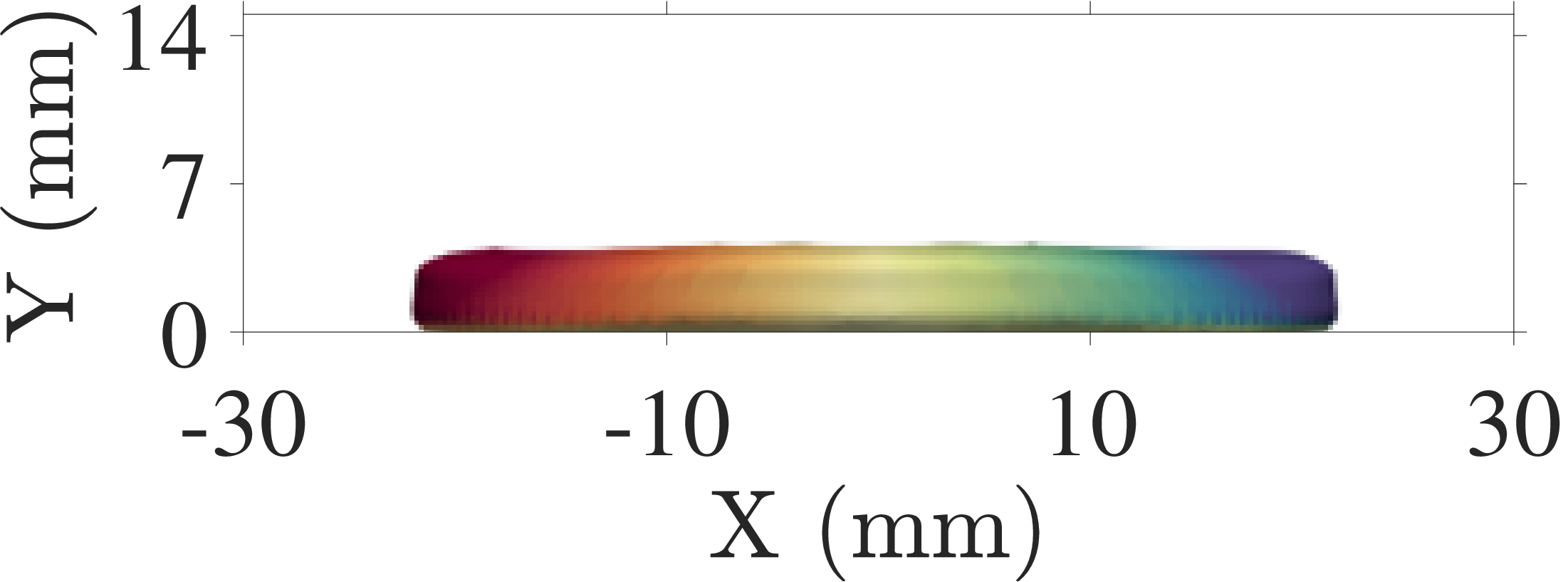}
\end{minipage}
\begin{center}\text{WCA = $160^\circ$}\end{center}

% -------- Figure for lambda = 0.12 --------
\caption{Temporal evolution of the axial velocity ($u_x$) distribution on the iso-surface of $\alpha = 0.5$ for $\lambda = 0.12$, from the instant of maximum spreading toward the final equilibrium state. The columns correspond to increasing time ($t = 0.01$ s to $t = 0.04$ s), while the rows represent different surface wettability conditions: hydrophilic (WCA = $0^\circ$), hybrid (WCA = $0^\circ$--$160^\circ$), and hydrophobic (WCA = $160^\circ$). Compared to lower relaxation time, the velocity field persists longer and exhibits stronger spatial asymmetry, reflecting enhanced elastic effects during recoil.}
\label{fig:ux_lambda_012}

\end{figure}

Figures~\ref{fig:ux_lambda_004} and \ref{fig:ux_lambda_012} show the temporal snapshots of droplet impact on solid surfaces with different wettability conditions. These snapshots capture the transition from maximum spreading to recoil-driven motion and eventually to the gradual decay of internal flow as the droplet approaches equilibrium. For both relaxation times, it is evident that the superhydrophilic surface allows the droplet to reach its equilibrium state significantly faster than the superhydrophobic surface, as strong liquid--solid affinity enhances energy dissipation through wetting and suppresses recoil, whereas weak adhesion on hydrophobic surfaces promotes prolonged oscillations and delayed relaxation \cite{josserand2016drop,bonn2009wetting}. 
For hybrid surfaces, the droplet consistently evolves into a dustpan-like morphology, characterized by a flattened concave base, a raised curved upper region, and an asymmetric leading edge. This shape emerges due to the coexistence of opposing capillary forces: the hydrophilic region draws liquid outward and anchors the contact line, while the hydrophobic region resists wetting and promotes recoil, thereby inducing a lateral capillary pressure gradient that redistributes fluid toward the more wettable side \cite{darhuber2005principles,quere2008wetting}. The presence of viscoelastic stresses further enhances this asymmetry, as polymer stretching near the wettability transition generates normal stress differences that sustain deformation and delay shape relaxation \cite{wang2017impact,yue2012phase}. 

A comparison between $\lambda = 0.04$ and $\lambda = 0.12$ reveals that for the hydrophilic surface (WCA = $0^\circ$), the droplet with lower relaxation time reaches equilibrium sooner. This behavior arises because smaller $\lambda$ corresponds to faster stress relaxation, allowing elastic energy stored during spreading to dissipate more rapidly, thereby accelerating the decay of internal flow \cite{bird1987dynamics,crooks2000influence}. This trend is clearly reflected in the velocity fields: at $t = 0.04$ s, the magnitude of the axial velocity component for $\lambda = 0.04$ is significantly lower than that for $\lambda = 0.12$, indicating that the higher-$\lambda$ droplet retains stronger internal motion for a longer duration. Similarly, at $t = 0.02$ s, a large portion of the droplet exhibits high axial velocity for both cases; however, when interpreted alongside Fig.~\ref{fig:rest_shape_lambda}, this state corresponds to a later stage of relaxation for $\lambda = 0.04$, whereas for $\lambda = 0.12$ it still represents an early recoil phase. This distinction highlights the role of elastic memory in extending the duration of flow-driven deformation. Notably, this behavior is consistent across all wettability conditions, indicating that viscoelastic effects act independently of surface properties during the early recoil stage. Furthermore, the maximum wetted area is consistently larger for higher relaxation time across all WCAs, which can be attributed to the ability of viscoelastic fluids to store kinetic energy in stretched polymer chains during impact, thereby delaying viscous dissipation and allowing the droplet to spread further before recoiling \cite{bartolo2007dynamics,bergeron2000controlling,wang2017impact}. Consequently, the combined influence of wettability and viscoelasticity governs not only the transient flow evolution but also the pathway through which the droplet reaches its final equilibrium configuration.
%%%%%%%%%%%%%%%%%%%%%%

%############################################

%%%%%%%%%%%%%%%%%%%%%%%%

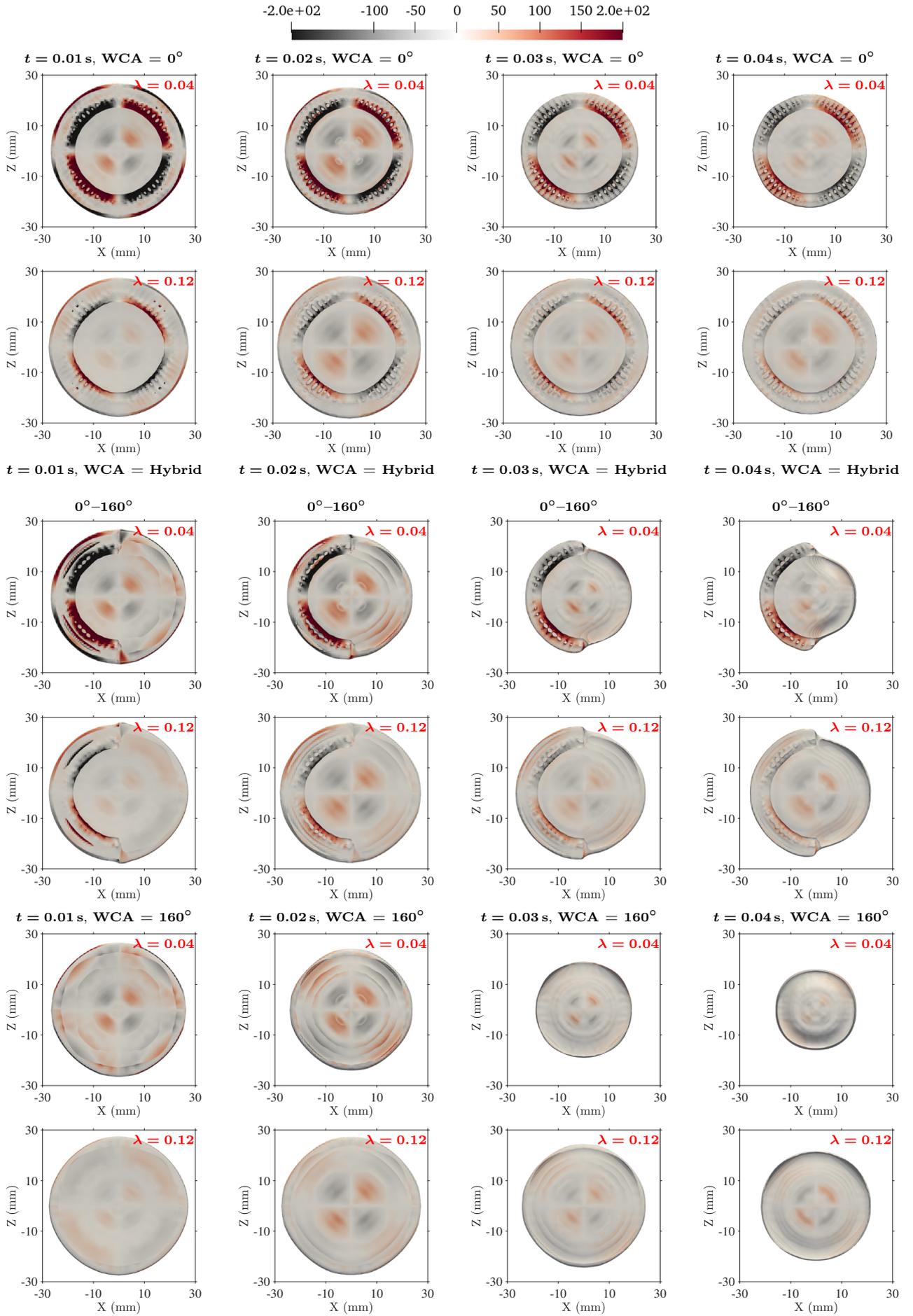
\begin{figure}[H]
\centering
    \includegraphics[width=0.45\linewidth]{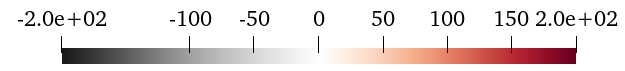}\\[3pt]
% ================= Row 1: WCA = 10° =================
\begin{minipage}{0.22\linewidth}
    \centering
{\scriptsize $\bm{t=0.01\,\mathrm{s}}$, \textbf{WCA = $\bm{0^\circ}$}}\\[2pt]
    \begin{tikzpicture}
    \node[anchor=south west, inner sep=0] (img) at (0,0)
        {\includegraphics[width=\linewidth]{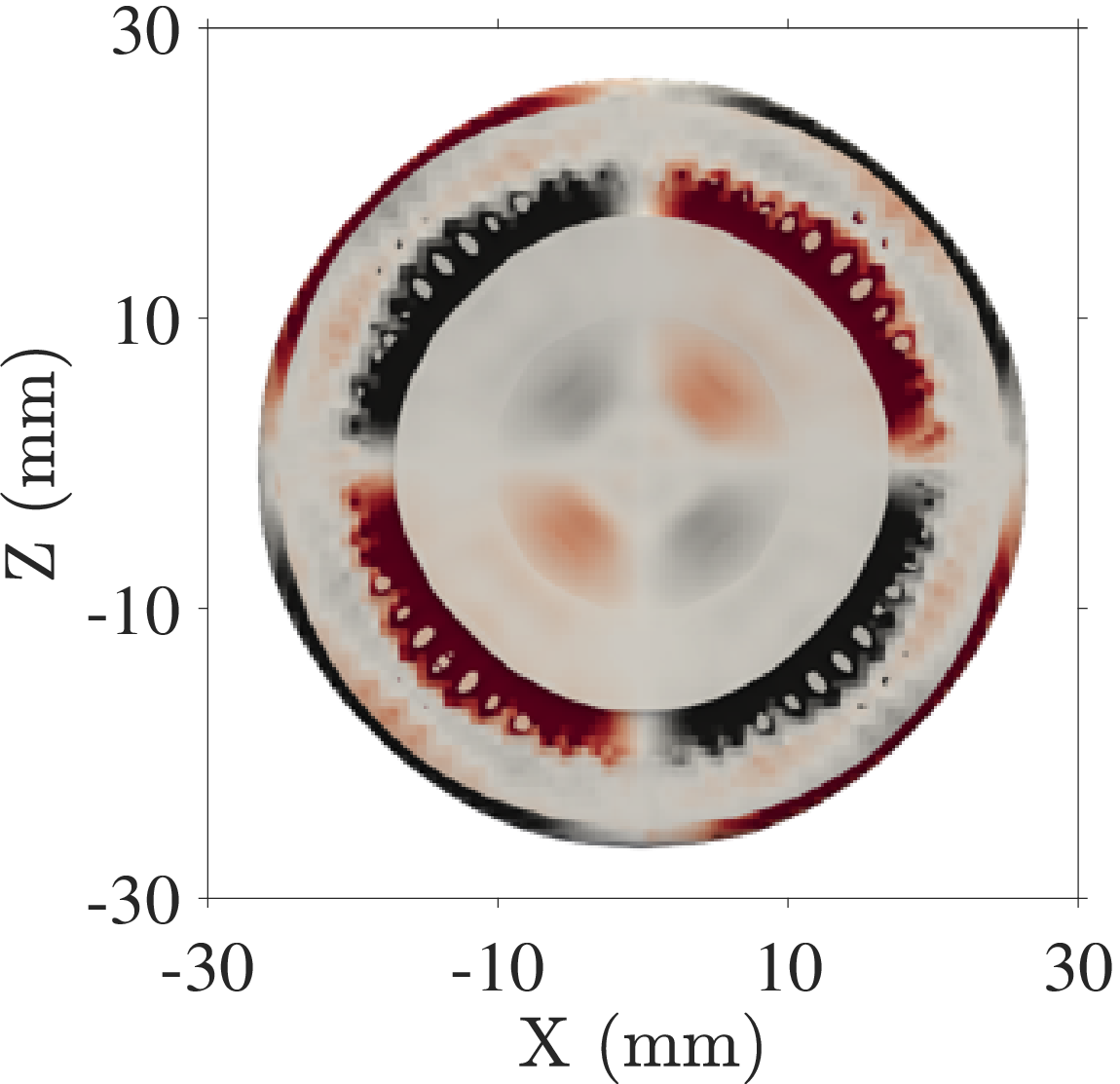}};
    \begin{scope}[x={(img.south east)},y={(img.north west)}]
        \node[anchor=north east] at (0.99,0.98) {\scriptsize \textcolor{red}{$\bm{\lambda=0.04}$}};
    \end{scope}
    \end{tikzpicture}\\[3pt]
    \begin{tikzpicture}
    \node[anchor=south west, inner sep=0] (img) at (0,0)
        {\includegraphics[width=\linewidth]{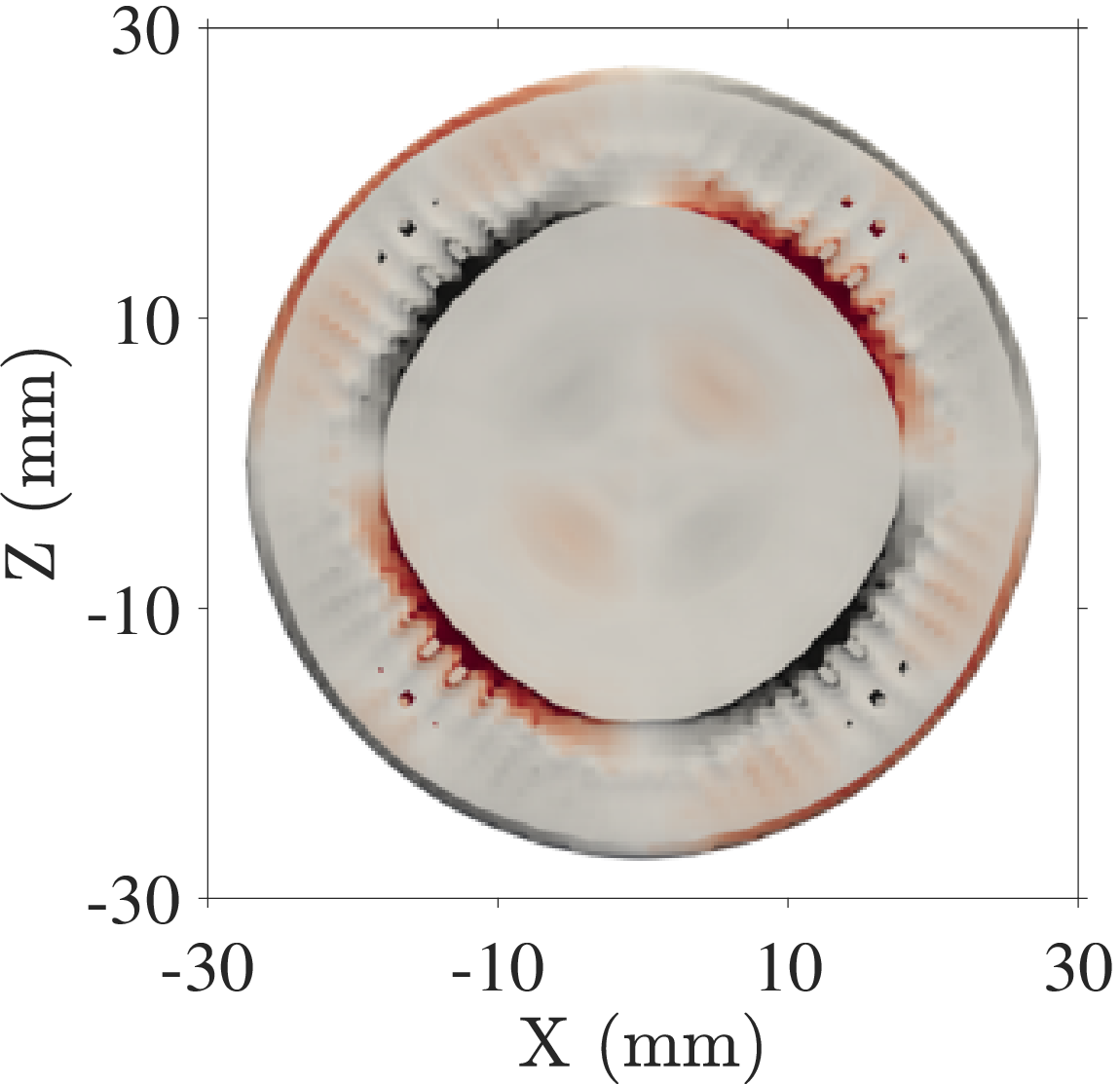}};
    \begin{scope}[x={(img.south east)},y={(img.north west)}]
        \node[anchor=north east] at (0.99,0.98) {\scriptsize \textcolor{red}{$\bm{\lambda=0.12}$}};
    \end{scope}
    \end{tikzpicture}
\end{minipage}\hfill
\begin{minipage}{0.22\linewidth}
    \centering
{\scriptsize $\bm{t=0.02\,\mathrm{s}}$, \textbf{WCA = $\bm{0^\circ}$}}\\[2pt]
    \begin{tikzpicture}
    \node[anchor=south west, inner sep=0] (img) at (0,0)
        {\includegraphics[width=\linewidth]{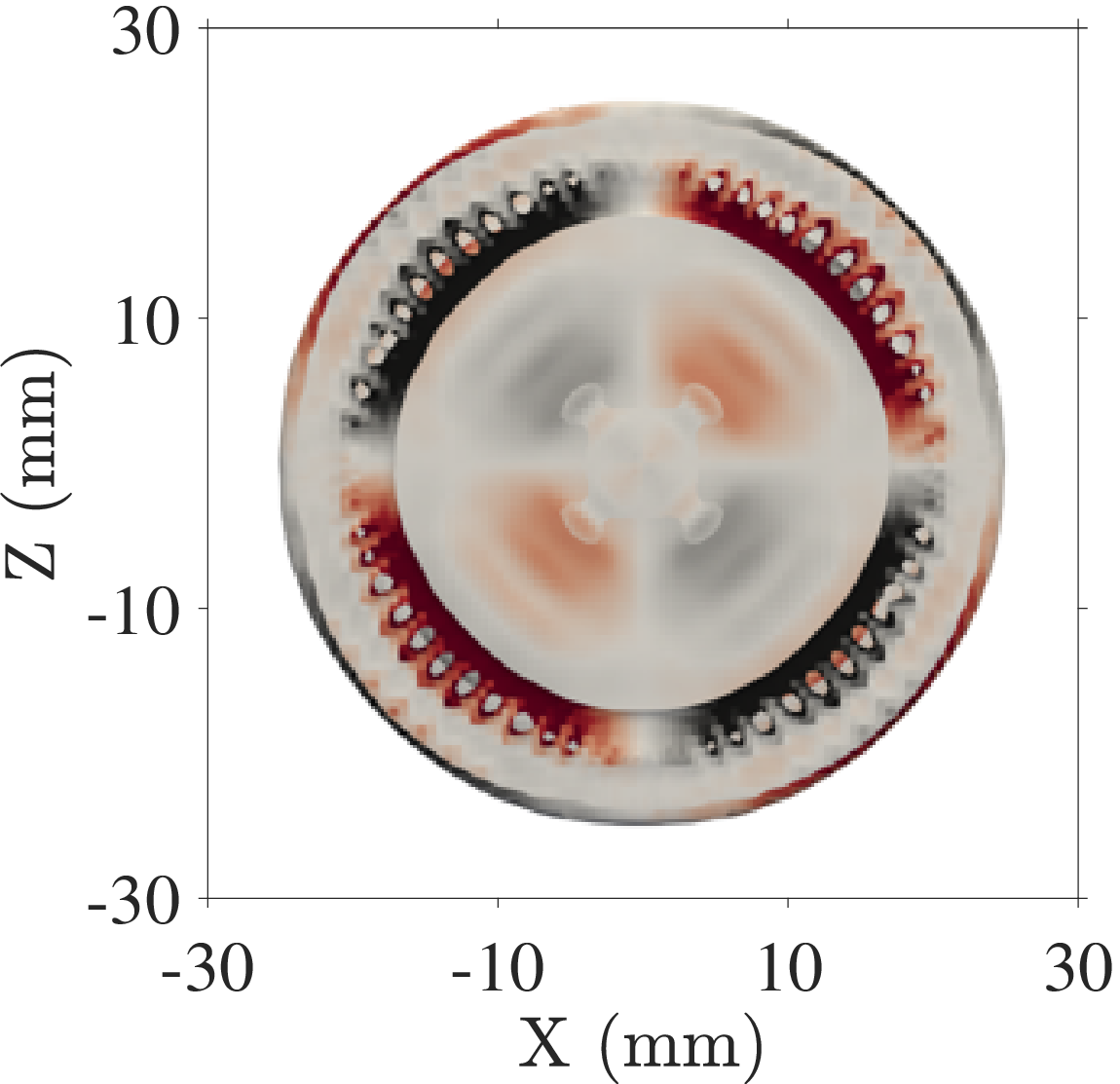}};
    \begin{scope}[x={(img.south east)},y={(img.north west)}]
        \node[anchor=north east] at (0.99,0.98) {\scriptsize \textcolor{red}{$\bm{\lambda=0.04}$}};
    \end{scope}
    \end{tikzpicture}\\[3pt]
    \begin{tikzpicture}
    \node[anchor=south west, inner sep=0] (img) at (0,0)
        {\includegraphics[width=\linewidth]{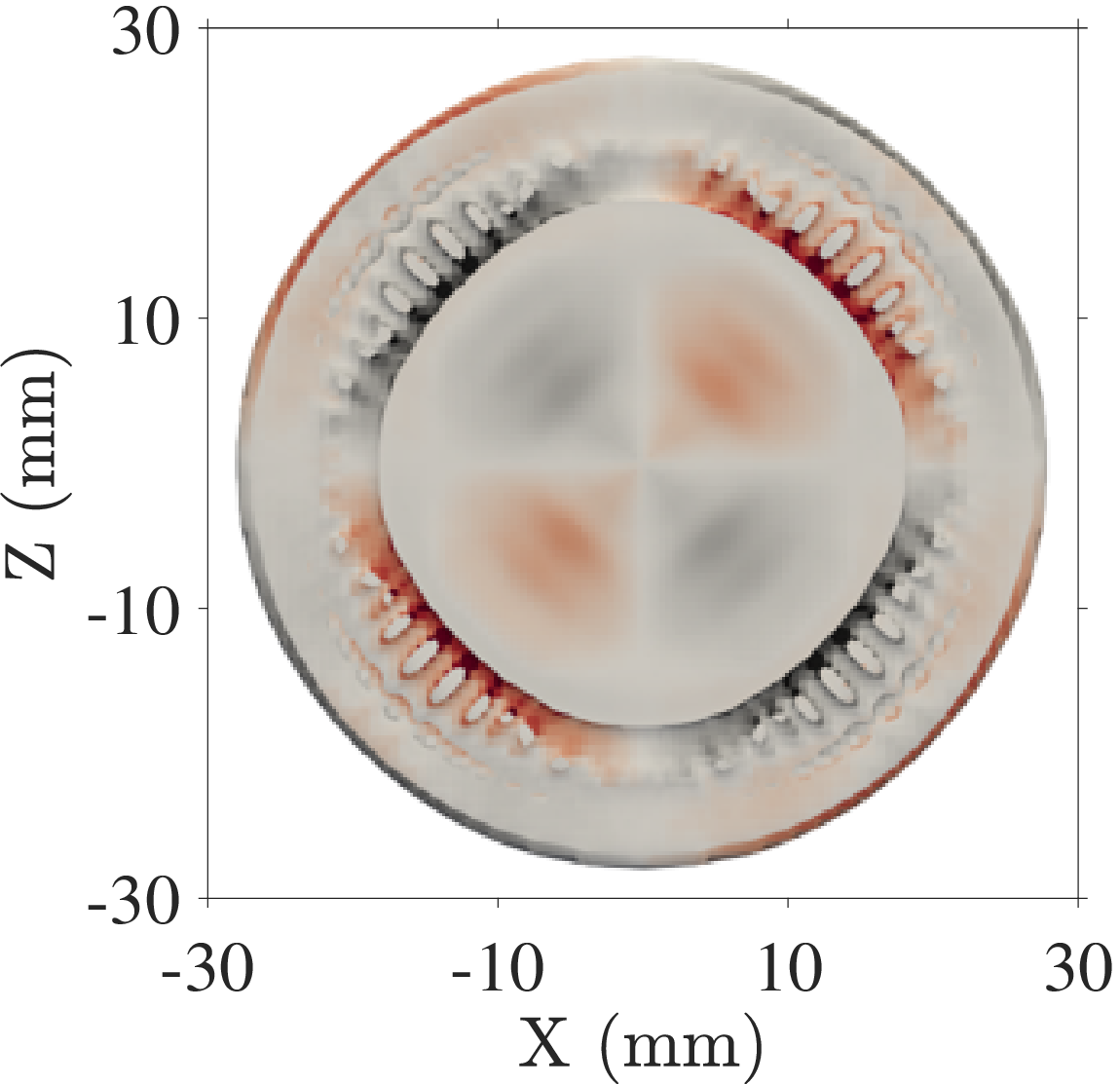}};
    \begin{scope}[x={(img.south east)},y={(img.north west)}]
        \node[anchor=north east] at (0.99,0.98) {\scriptsize \textcolor{red}{$\bm{\lambda=0.12}$}};
    \end{scope}
    \end{tikzpicture}
\end{minipage}\hfill
\begin{minipage}{0.22\linewidth}
    \centering
{\scriptsize $\bm{t=0.03\,\mathrm{s}}$, \textbf{WCA = $\bm{0^\circ}$}}\\[2pt]
    \begin{tikzpicture}
    \node[anchor=south west, inner sep=0] (img) at (0,0)
        {\includegraphics[width=\linewidth]{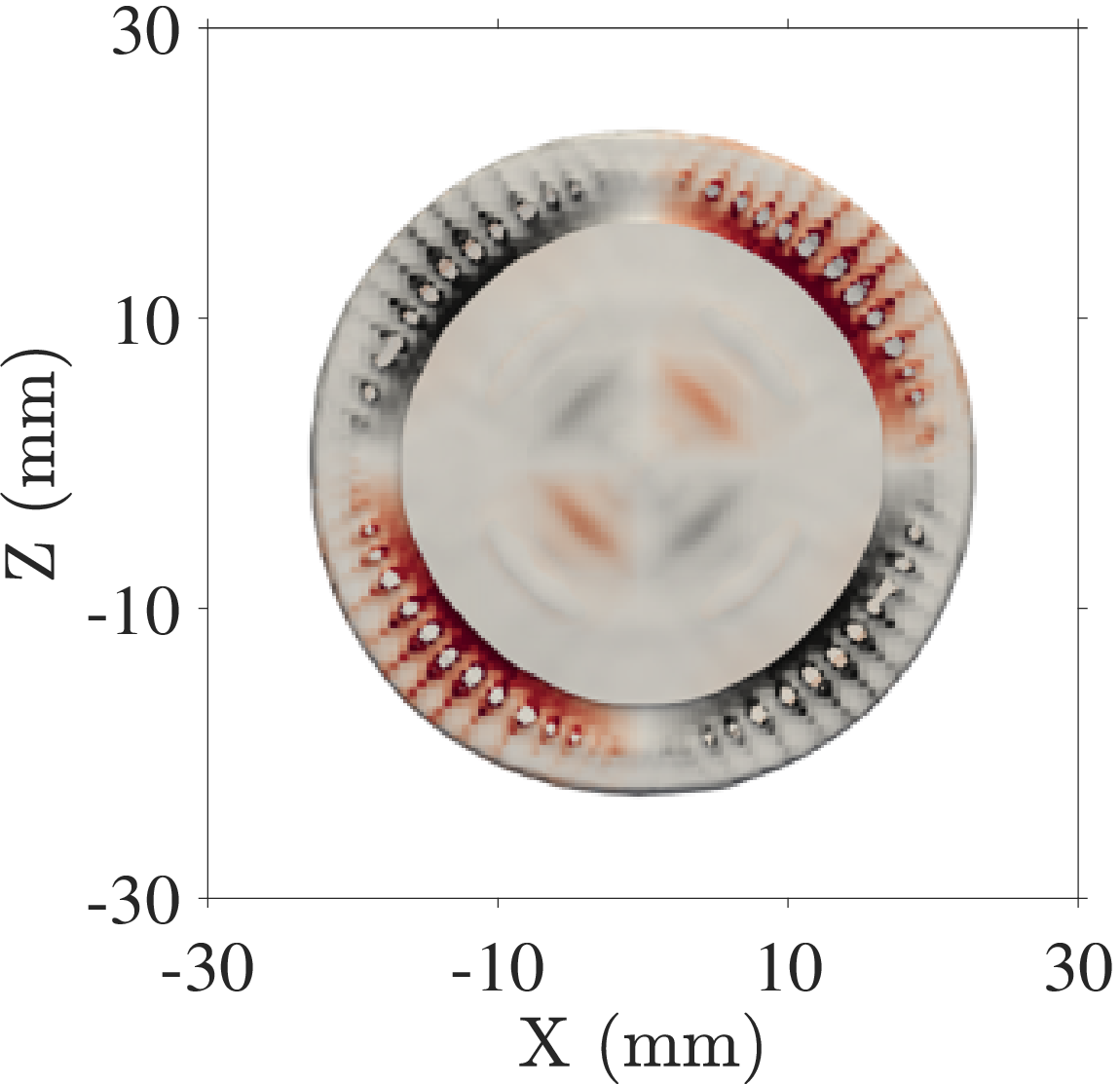}};
    \begin{scope}[x={(img.south east)},y={(img.north west)}]
        \node[anchor=north east] at (0.99,0.98) {\scriptsize \textcolor{red}{$\bm{\lambda=0.04}$}};
    \end{scope}
    \end{tikzpicture}\\[3pt]
    \begin{tikzpicture}
    \node[anchor=south west, inner sep=0] (img) at (0,0)
        {\includegraphics[width=\linewidth]{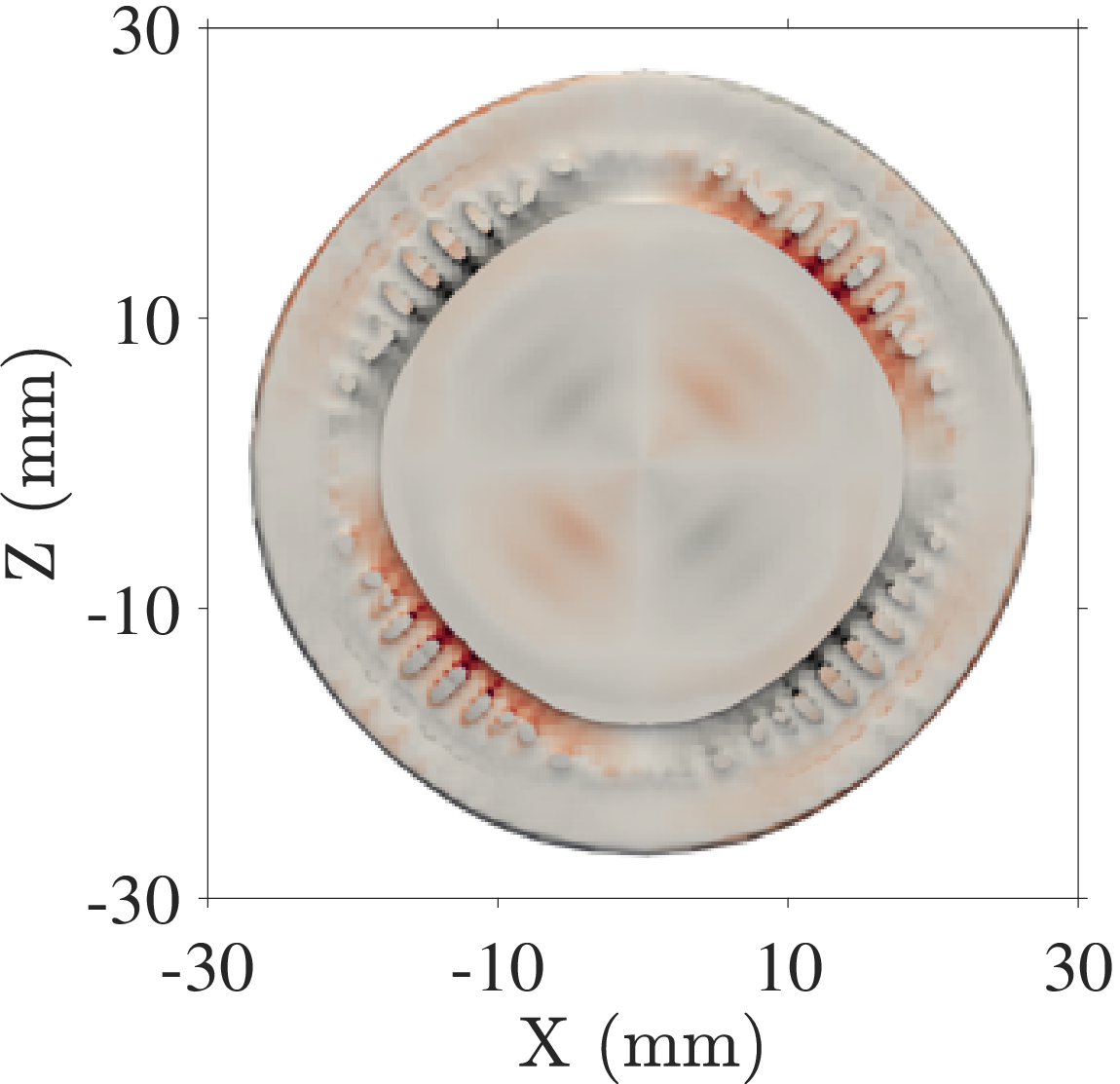}};
    \begin{scope}[x={(img.south east)},y={(img.north west)}]
        \node[anchor=north east] at (0.99,0.98) {\scriptsize \textcolor{red}{$\bm{\lambda=0.12}$}};
    \end{scope}
    \end{tikzpicture}\\[3pt]
\end{minipage}\hfill
\begin{minipage}{0.22\linewidth}
    \centering
{\scriptsize $\bm{t=0.04\,\mathrm{s}}$, \textbf{WCA = $\bm{0^\circ}$}}\\[2pt]
    \begin{tikzpicture}
    \node[anchor=south west, inner sep=0] (img) at (0,0)
        {\includegraphics[width=\linewidth]{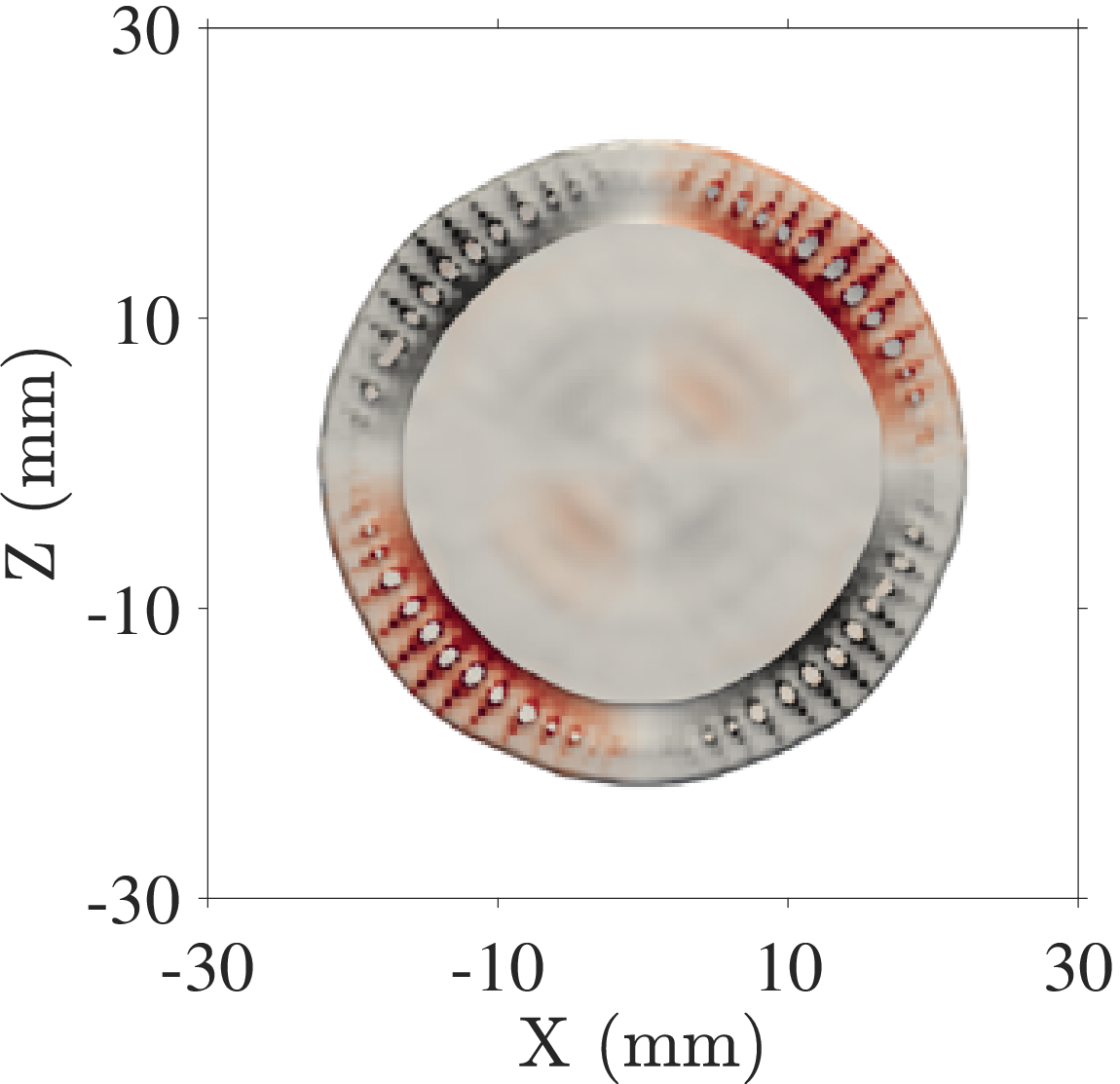}};
    \begin{scope}[x={(img.south east)},y={(img.north west)}]
        \node[anchor=north east] at (0.99,0.98) {\scriptsize \textcolor{red}{$\bm{\lambda=0.04}$}};
    \end{scope}
    \end{tikzpicture}\\[3pt]
    \begin{tikzpicture}
    \node[anchor=south west, inner sep=0] (img) at (0,0)
        {\includegraphics[width=\linewidth]{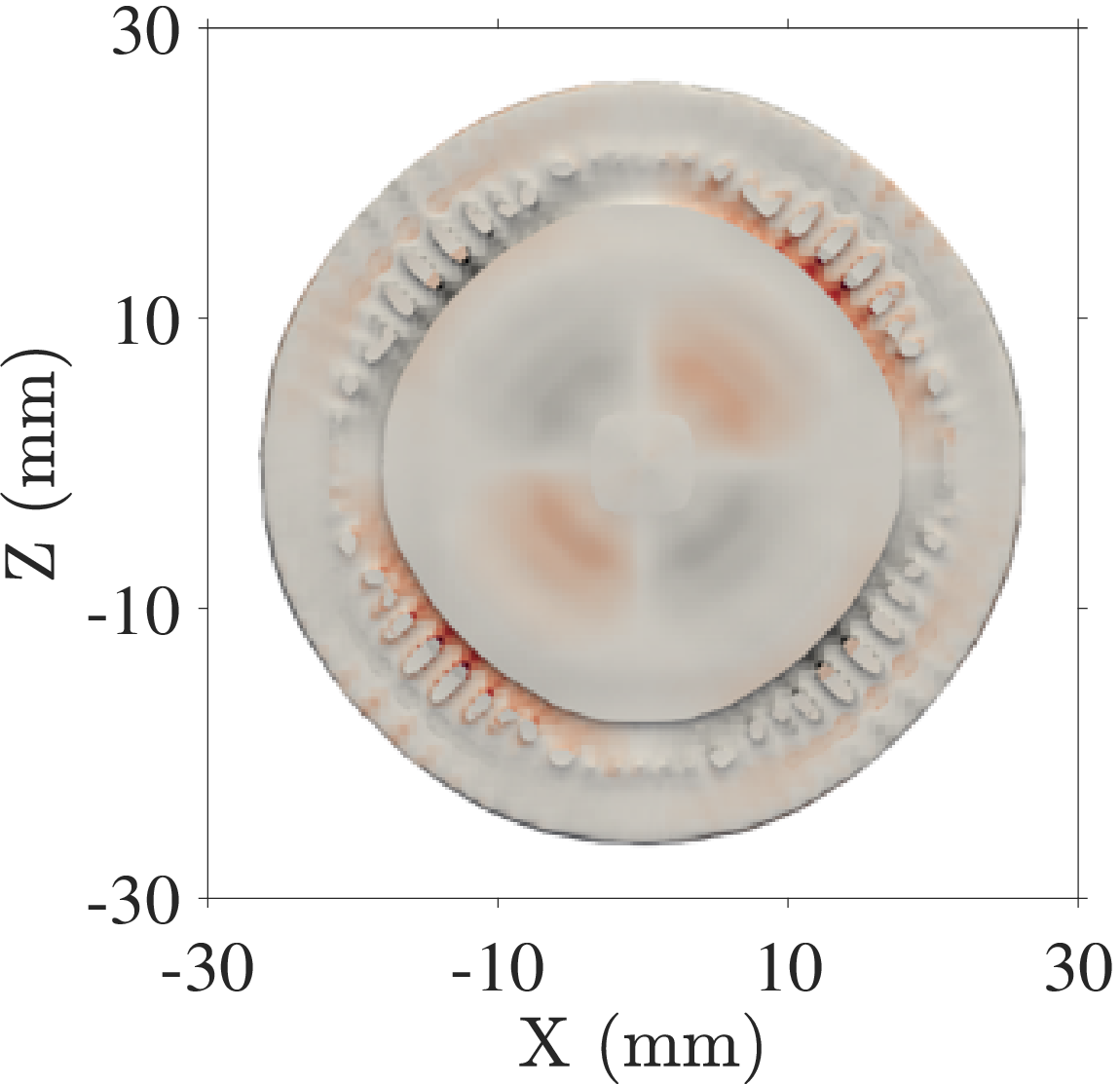}};
    \begin{scope}[x={(img.south east)},y={(img.north west)}]
        \node[anchor=north east] at (0.99,0.98) {\scriptsize \textcolor{red}{$\bm{\lambda=0.12}$}};
    \end{scope}
    \end{tikzpicture}\\[3pt]
\end{minipage}
\vspace{4pt}

% ================= Row 2: Hybrid =================
\begin{minipage}{0.22\linewidth}
    \centering
{\scriptsize $\bm{t=0.01\,\mathrm{s}}$, \textbf{WCA = Hybrid $\bm{0^\circ}$--$\bm{160^\circ}$}}\\[2pt]
    \begin{tikzpicture}
    \node[anchor=south west, inner sep=0] (img) at (0,0)
        {\includegraphics[width=\linewidth]{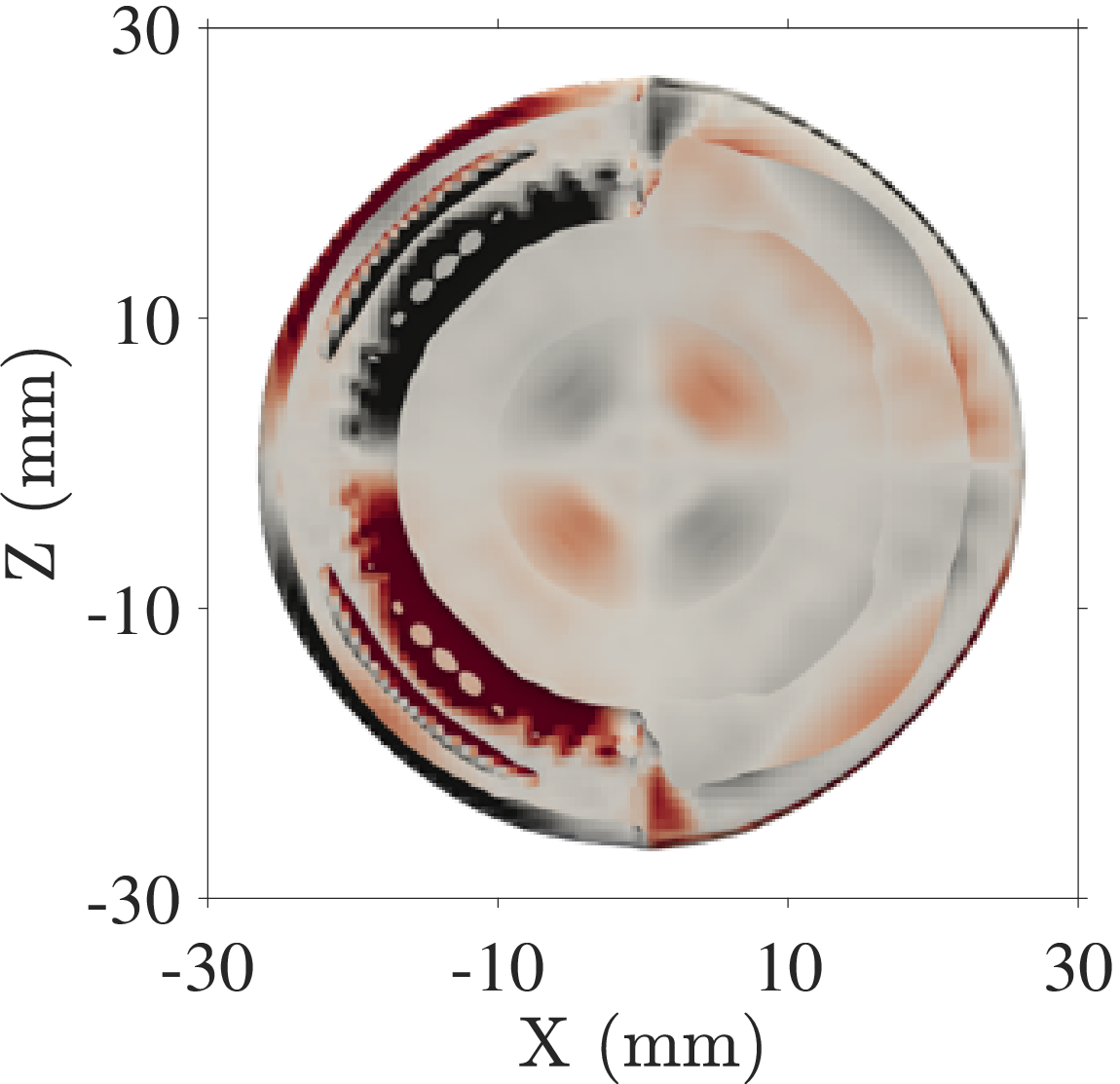}};
    \begin{scope}[x={(img.south east)},y={(img.north west)}]
        \node[anchor=north east] at (0.99,0.98) {\scriptsize \textcolor{red}{$\bm{\lambda=0.04}$}};
    \end{scope}
    \end{tikzpicture}\\[3pt]
    \begin{tikzpicture}
    \node[anchor=south west, inner sep=0] (img) at (0,0)
        {\includegraphics[width=\linewidth]{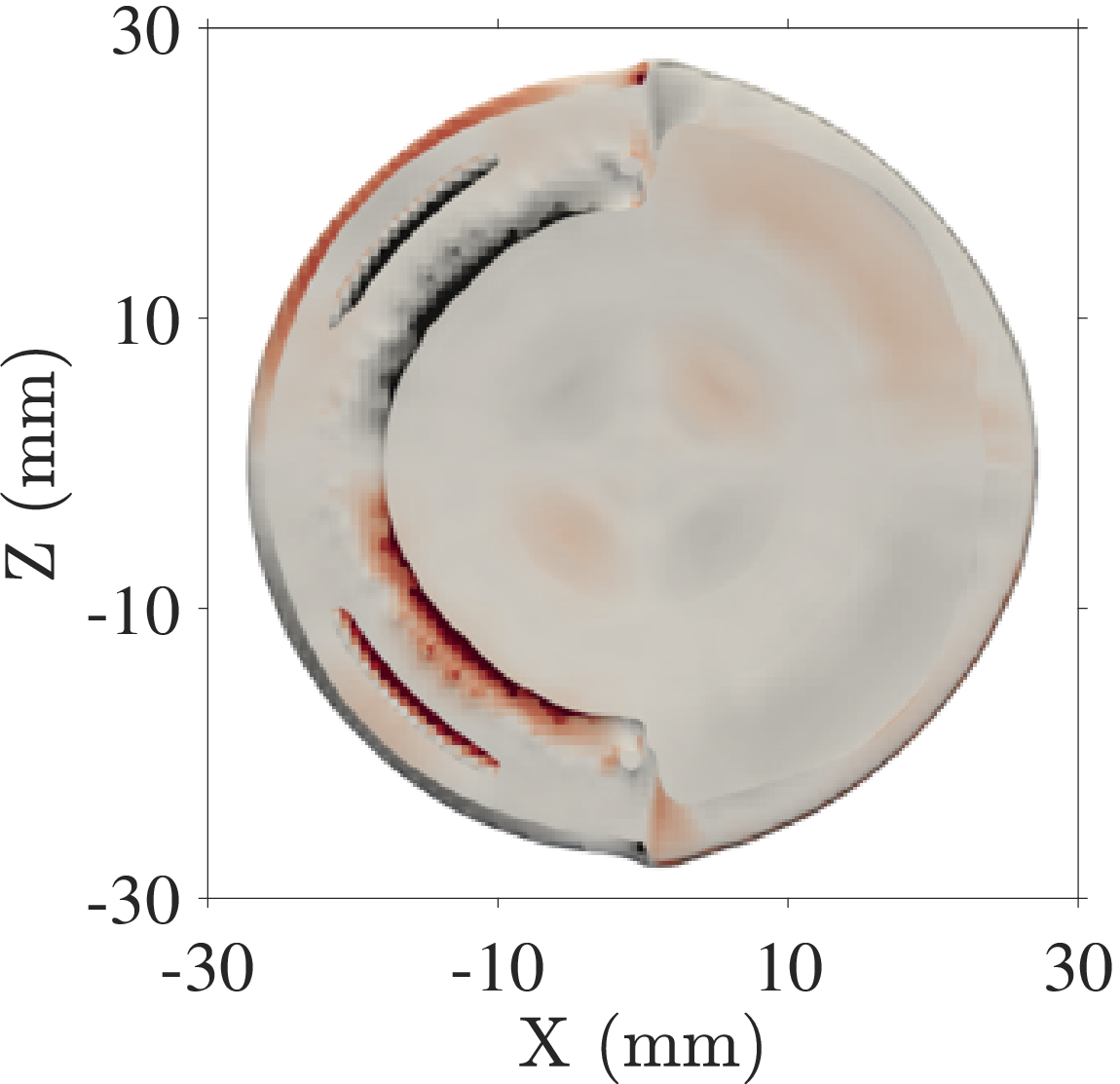}};
    \begin{scope}[x={(img.south east)},y={(img.north west)}]
        \node[anchor=north east] at (0.99,0.98) {\scriptsize \textcolor{red}{$\bm{\lambda=0.12}$}};
    \end{scope}
    \end{tikzpicture}
\end{minipage}\hfill
\begin{minipage}{0.22\linewidth}
    \centering
{\scriptsize $\bm{t=0.02\,\mathrm{s}}$, \textbf{WCA = Hybrid $\bm{0^\circ}$--$\bm{160^\circ}$}}\\[2pt]
    \begin{tikzpicture}
    \node[anchor=south west, inner sep=0] (img) at (0,0)
        {\includegraphics[width=\linewidth]{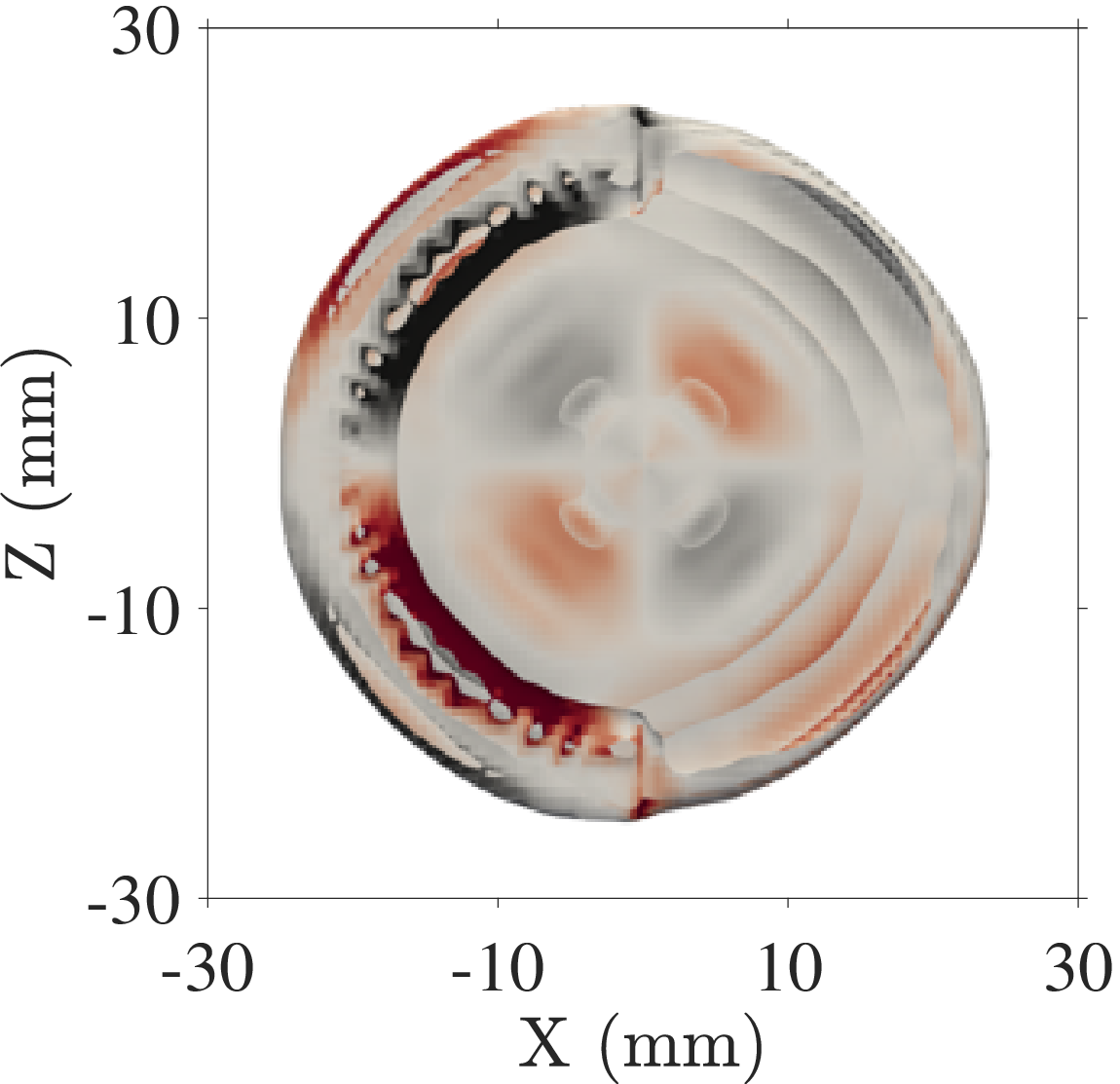}};
    \begin{scope}[x={(img.south east)},y={(img.north west)}]
        \node[anchor=north east] at (0.99,0.98) {\scriptsize \textcolor{red}{$\bm{\lambda=0.04}$}};
    \end{scope}
    \end{tikzpicture}\\[3pt]
    \begin{tikzpicture}
    \node[anchor=south west, inner sep=0] (img) at (0,0)
        {\includegraphics[width=\linewidth]{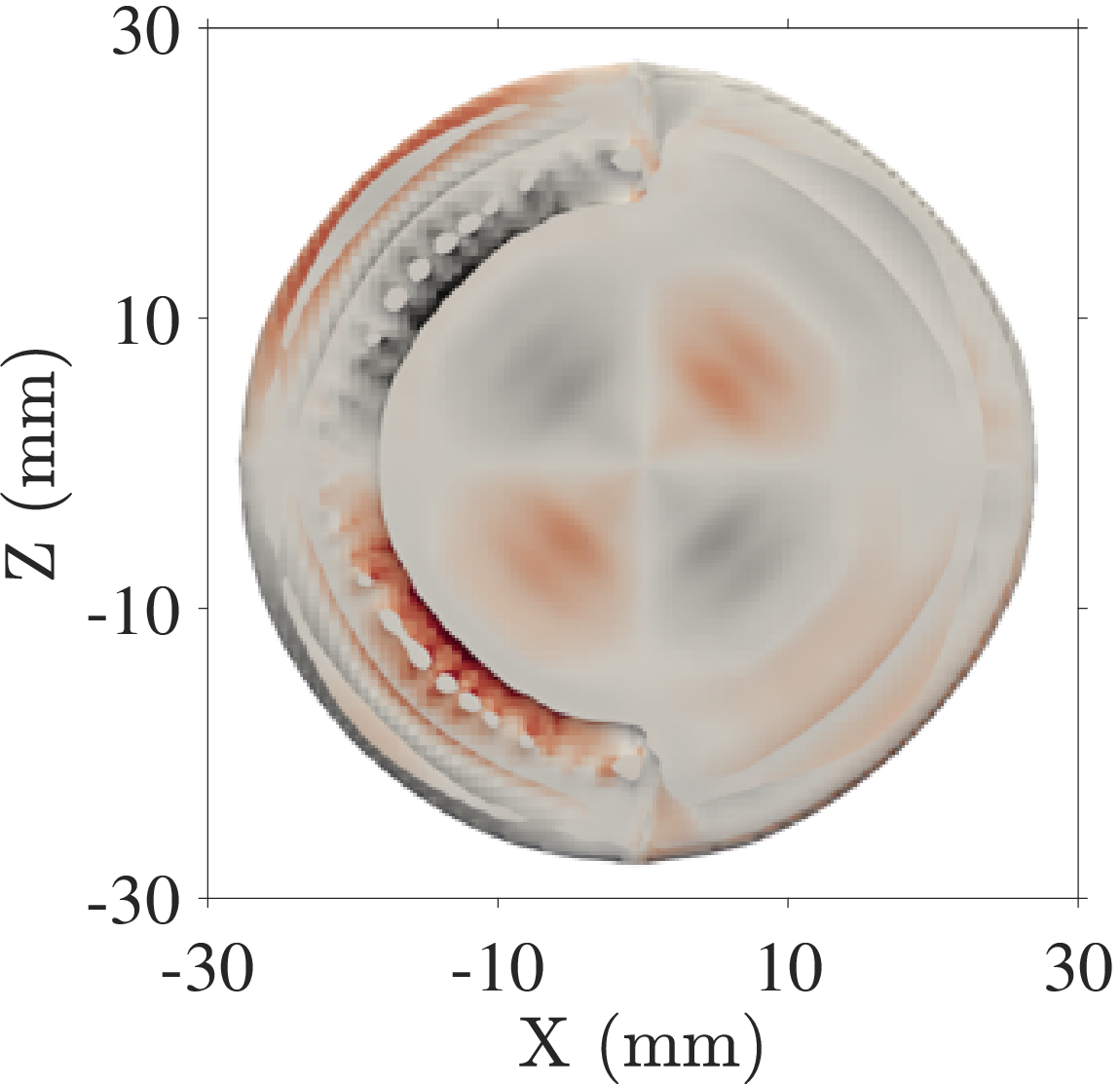}};
    \begin{scope}[x={(img.south east)},y={(img.north west)}]
        \node[anchor=north east] at (0.99,0.98) {\scriptsize \textcolor{red}{$\bm{\lambda=0.12}$}};
    \end{scope}
    \end{tikzpicture}
\end{minipage}\hfill
\begin{minipage}{0.22\linewidth}
    \centering
{\scriptsize $\bm{t=0.03\,\mathrm{s}}$, \textbf{WCA = Hybrid $\bm{0^\circ}$--$\bm{160^\circ}$}}\\[2pt]
    \begin{tikzpicture}
    \node[anchor=south west, inner sep=0] (img) at (0,0)
        {\includegraphics[width=\linewidth]{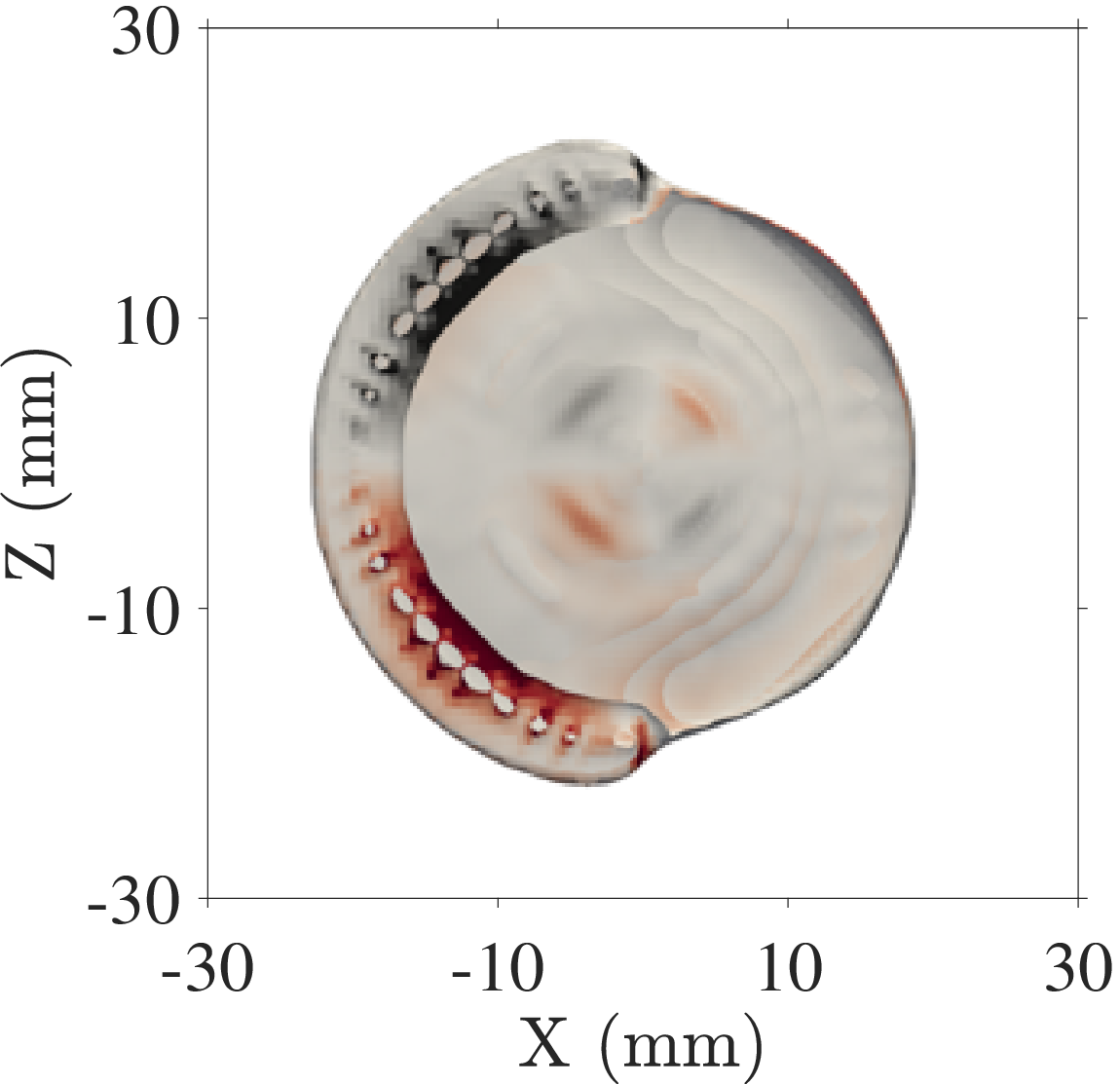}};
    \begin{scope}[x={(img.south east)},y={(img.north west)}]
        \node[anchor=north east] at (0.99,0.98) {\scriptsize \textcolor{red}{$\bm{\lambda=0.04}$}};
    \end{scope}
    \end{tikzpicture}\\[3pt]
    \begin{tikzpicture}
    \node[anchor=south west, inner sep=0] (img) at (0,0)
        {\includegraphics[width=\linewidth]{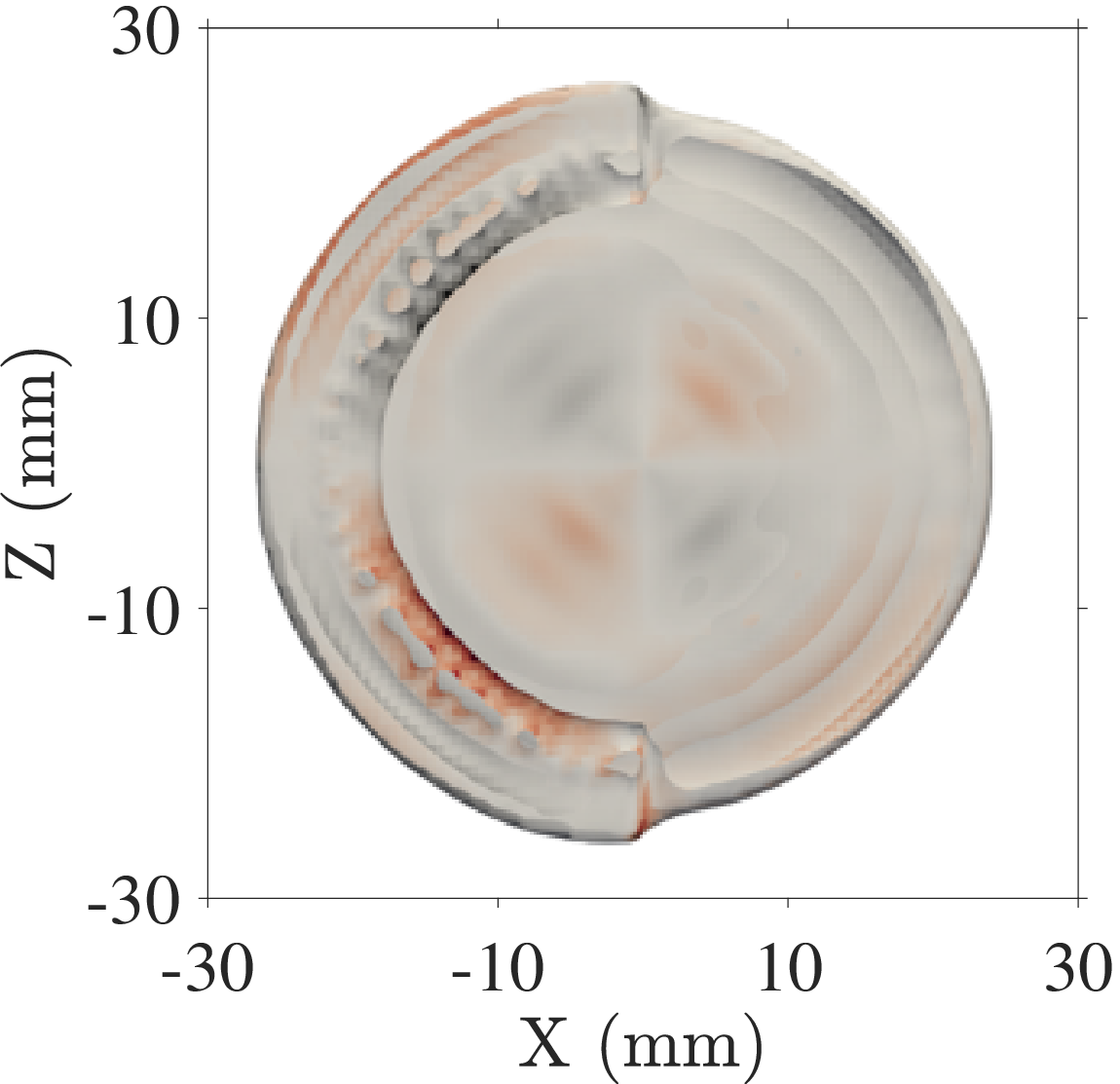}};
    \begin{scope}[x={(img.south east)},y={(img.north west)}]
        \node[anchor=north east] at (0.99,0.98) {\scriptsize \textcolor{red}{$\bm{\lambda=0.12}$}};
    \end{scope}
    \end{tikzpicture}
\end{minipage}\hfill
\begin{minipage}{0.22\linewidth}
    \centering
{\scriptsize $\bm{t=0.04\,\mathrm{s}}$, \textbf{WCA = Hybrid $\bm{0^\circ}$--$\bm{160^\circ}$}}\\[2pt]
    \begin{tikzpicture}
    \node[anchor=south west, inner sep=0] (img) at (0,0)
        {\includegraphics[width=\linewidth]{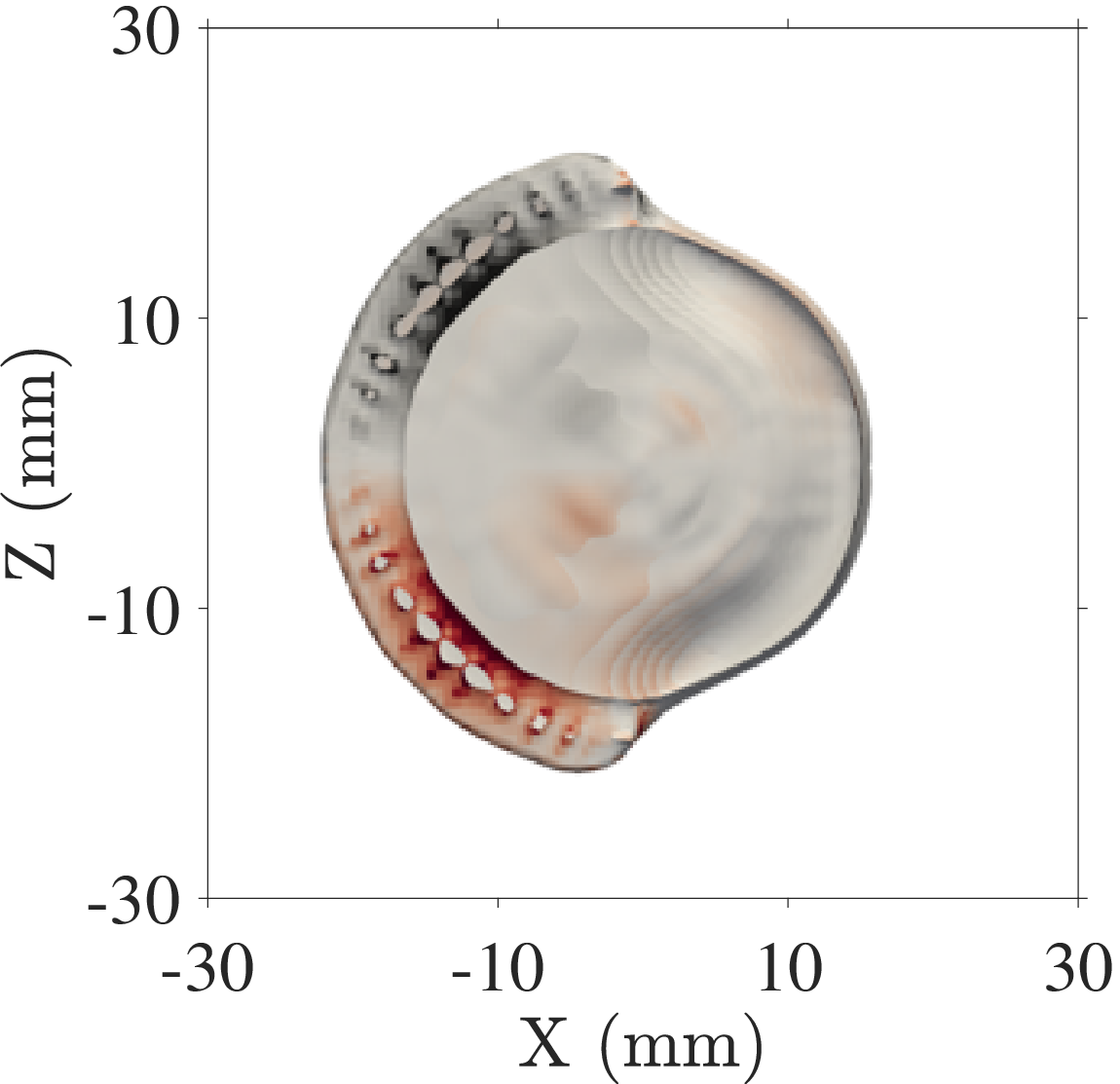}};
    \begin{scope}[x={(img.south east)},y={(img.north west)}]
        \node[anchor=north east] at (0.99,0.98) {\scriptsize \textcolor{red}{$\bm{\lambda=0.04}$}};
    \end{scope}
    \end{tikzpicture}\\[3pt]
    \begin{tikzpicture}
    \node[anchor=south west, inner sep=0] (img) at (0,0)
        {\includegraphics[width=\linewidth]{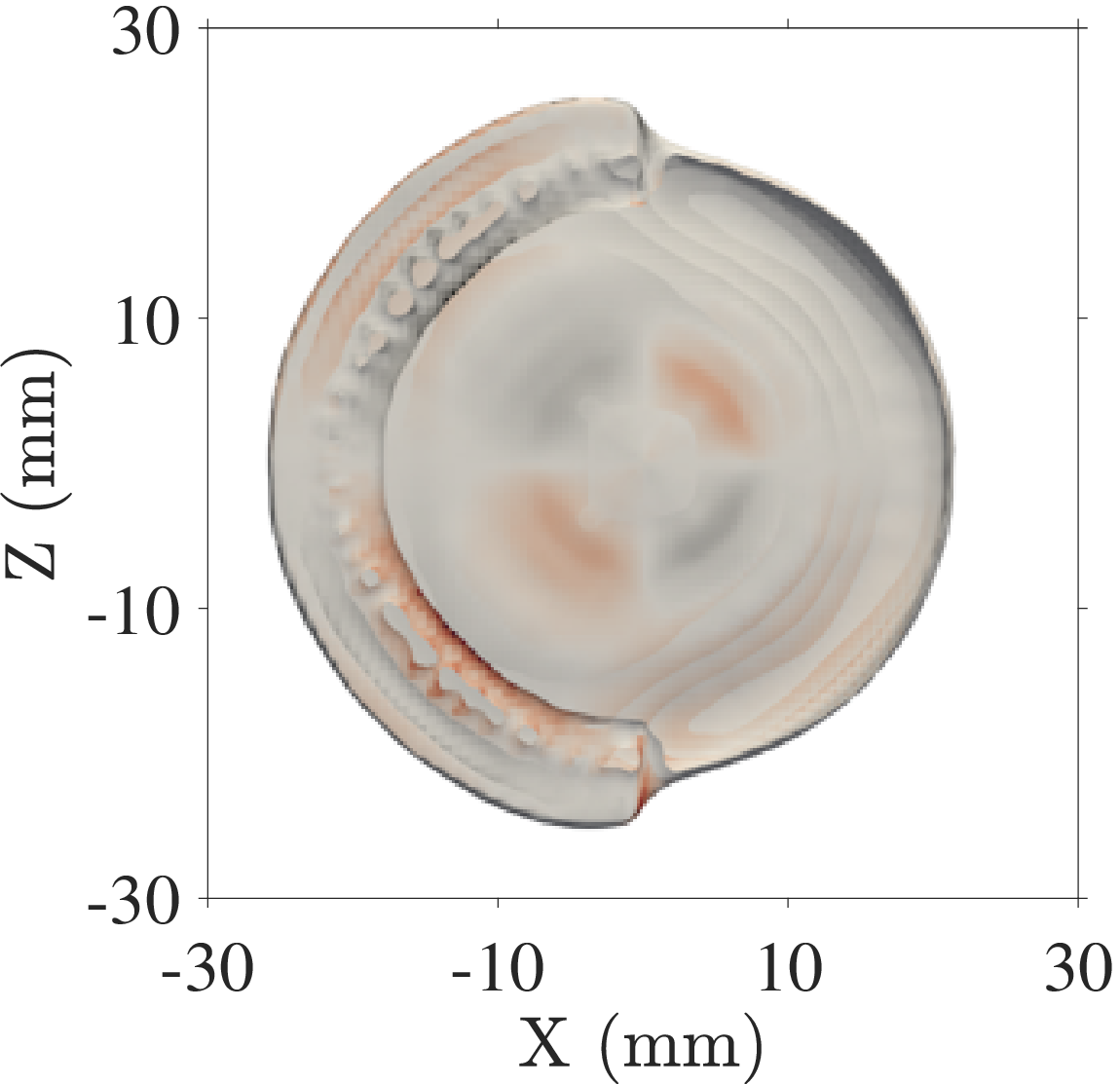}};
    \begin{scope}[x={(img.south east)},y={(img.north west)}]
        \node[anchor=north east] at (0.99,0.98) {\scriptsize \textcolor{red}{$\bm{\lambda=0.12}$}};
    \end{scope}
    \end{tikzpicture}\\[3pt]
\end{minipage}

\vspace{4pt}

% ================= Row 3: WCA = 160° =================
\begin{minipage}{0.22\linewidth}
    \centering
{\scriptsize $\bm{t=0.01\,\mathrm{s}}$, \textbf{WCA = $\bm{160^\circ}$}}\\[2pt]
    \begin{tikzpicture}
    \node[anchor=south west, inner sep=0] (img) at (0,0)
        {\includegraphics[width=\linewidth]{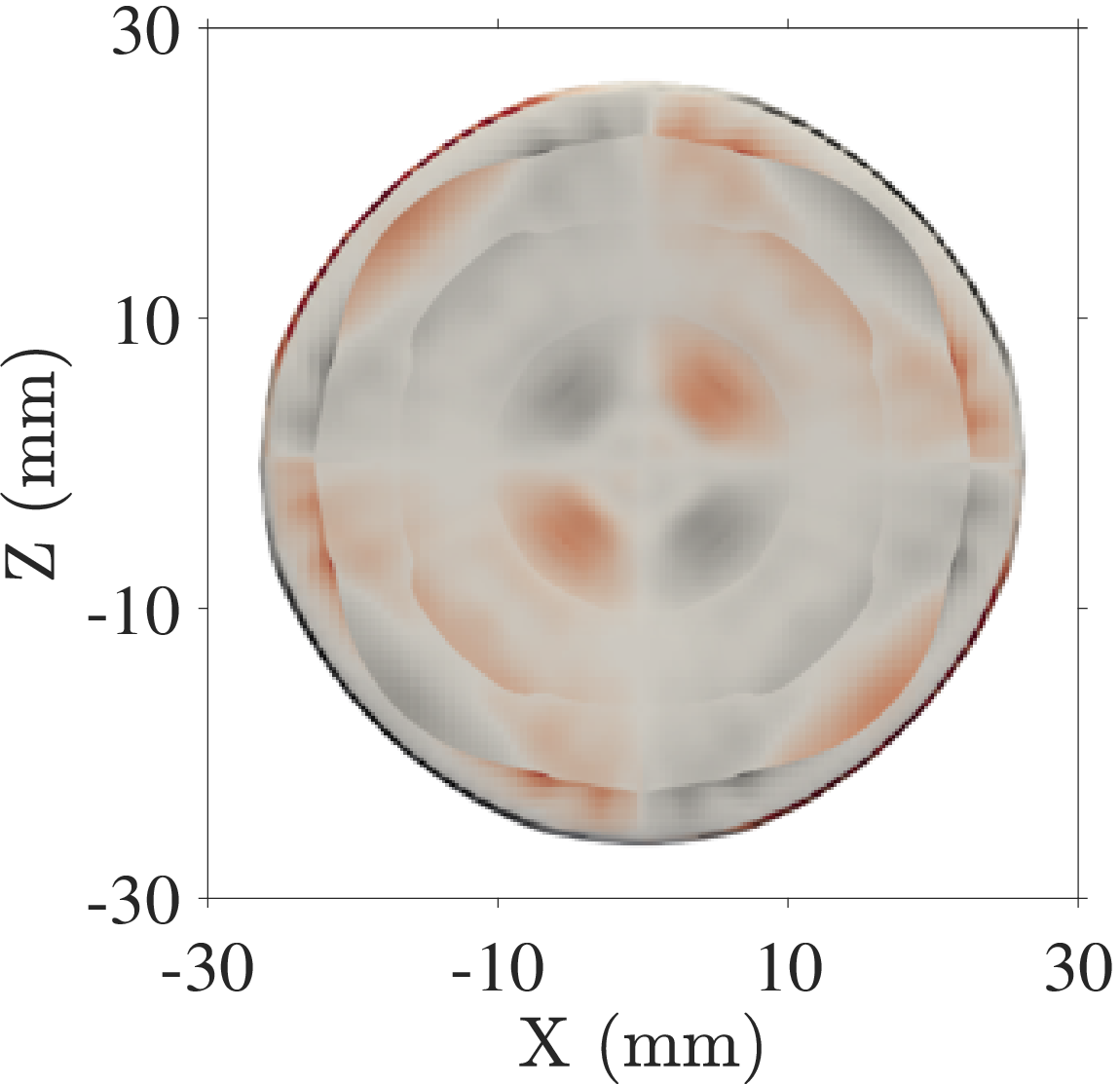}};
    \begin{scope}[x={(img.south east)},y={(img.north west)}]
        \node[anchor=north east] at (0.99,0.98) {\scriptsize \textcolor{red}{$\bm{\lambda=0.04}$}};
    \end{scope}
    \end{tikzpicture}\\[3pt]
    \begin{tikzpicture}
    \node[anchor=south west, inner sep=0] (img) at (0,0)
        {\includegraphics[width=\linewidth]{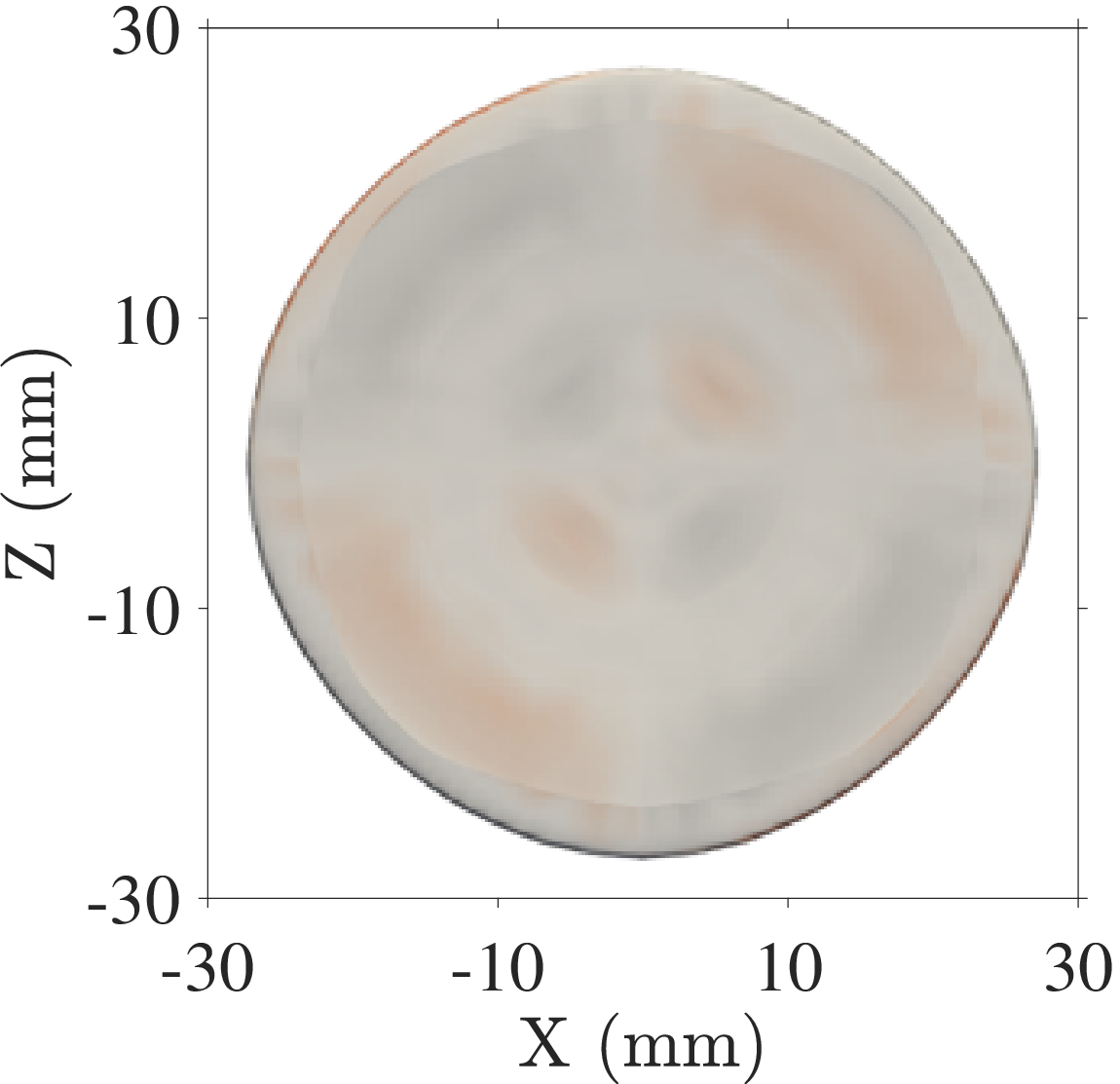}};
    \begin{scope}[x={(img.south east)},y={(img.north west)}]
        \node[anchor=north east] at (0.99,0.98) {\scriptsize \textcolor{red}{$\bm{\lambda=0.12}$}};
    \end{scope}
    \end{tikzpicture}
\end{minipage}\hfill
\begin{minipage}{0.22\linewidth}
    \centering
{\scriptsize $\bm{t=0.02\,\mathrm{s}}$, \textbf{WCA = $\bm{160^\circ}$}}\\[2pt]
    \begin{tikzpicture}
    \node[anchor=south west, inner sep=0] (img) at (0,0)
        {\includegraphics[width=\linewidth]{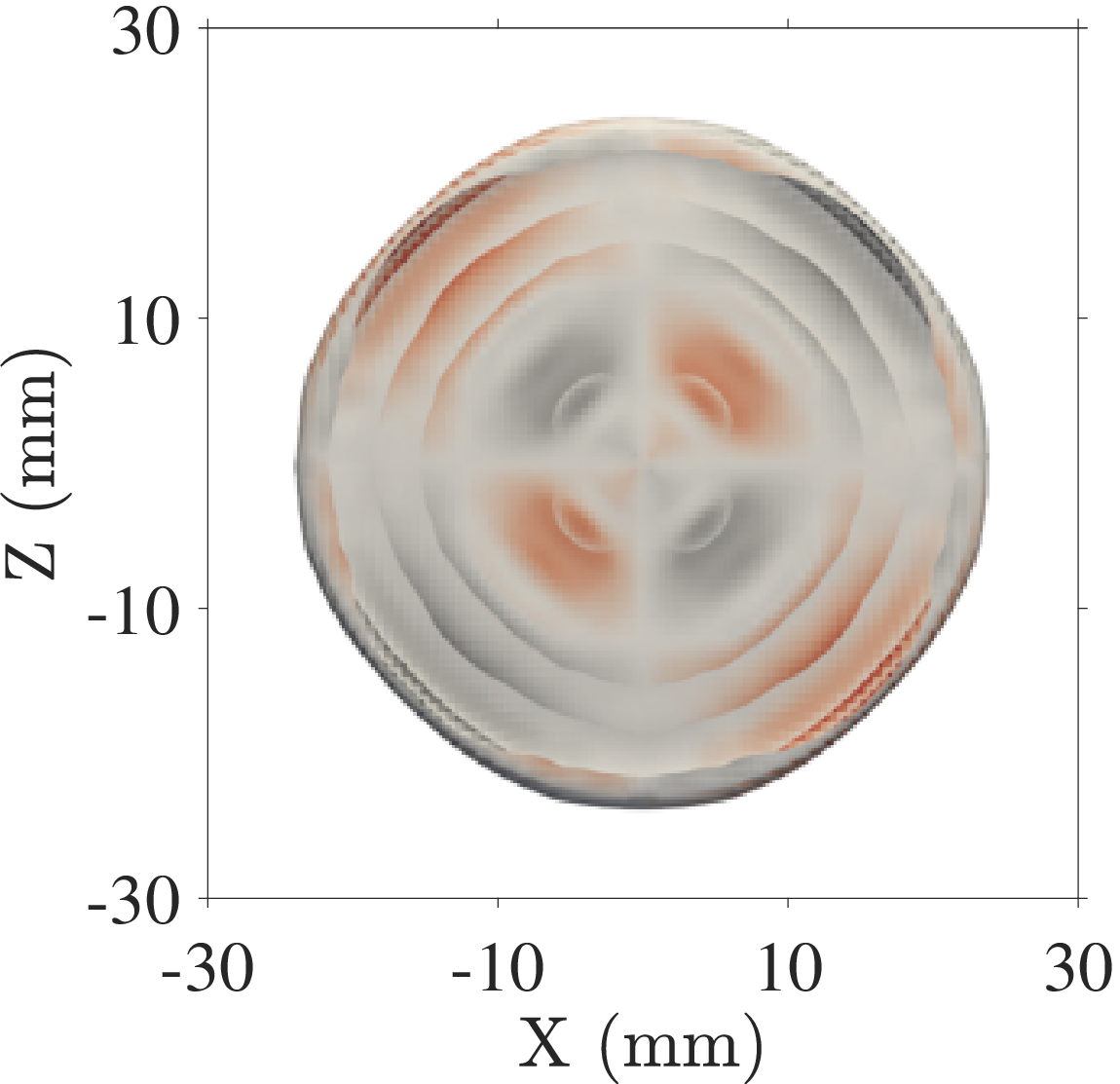}};
    \begin{scope}[x={(img.south east)},y={(img.north west)}]
        \node[anchor=north east] at (0.99,0.98) {\scriptsize \textcolor{red}{$\bm{\lambda=0.04}$}};
    \end{scope}
    \end{tikzpicture}\\[3pt]
    \begin{tikzpicture}
    \node[anchor=south west, inner sep=0] (img) at (0,0)
        {\includegraphics[width=\linewidth]{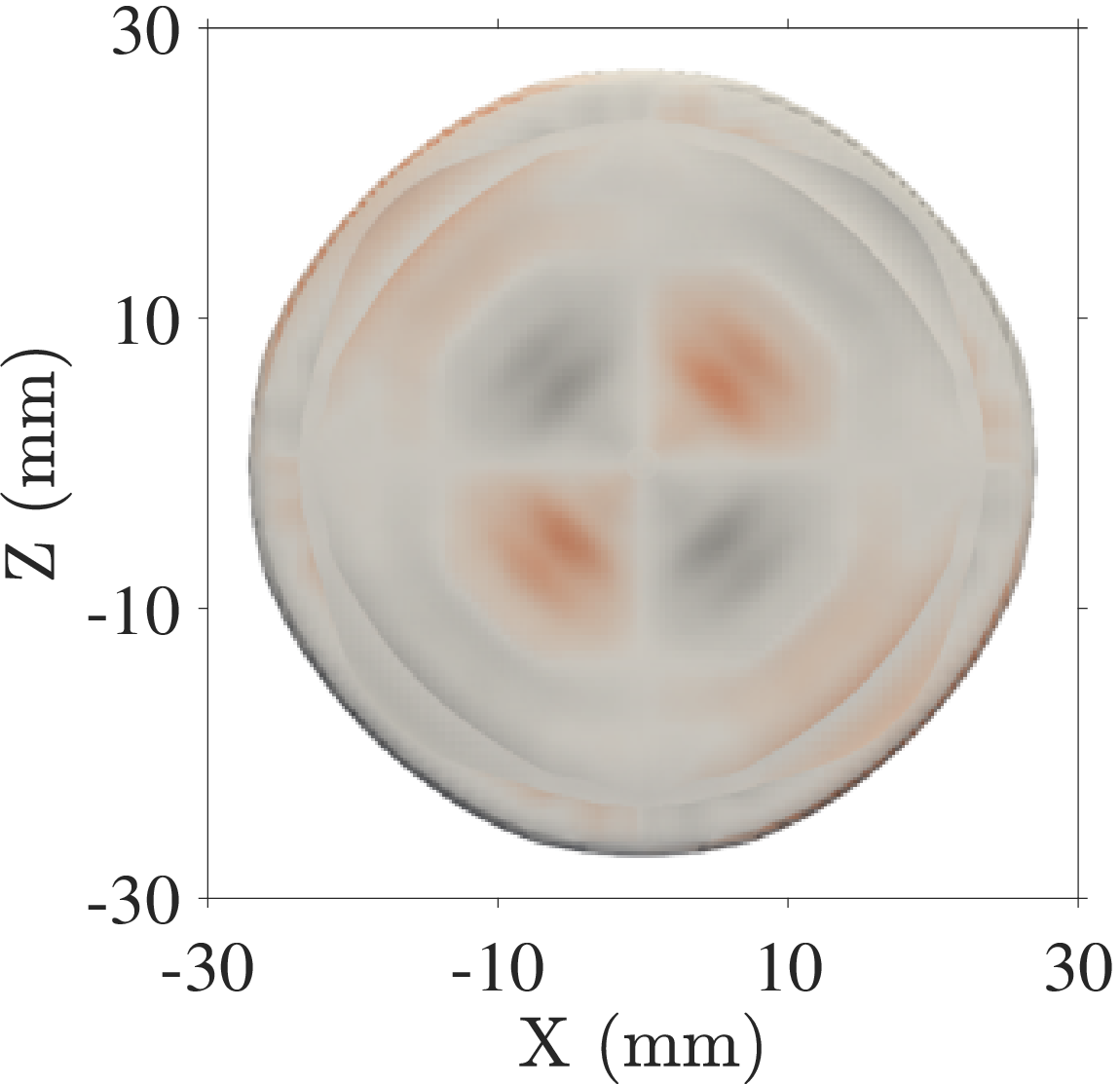}};
    \begin{scope}[x={(img.south east)},y={(img.north west)}]
        \node[anchor=north east] at (0.99,0.98) {\scriptsize \textcolor{red}{$\bm{\lambda=0.12}$}};
    \end{scope}
    \end{tikzpicture}\\[3pt]
\end{minipage}\hfill
\begin{minipage}{0.22\linewidth}
    \centering
{\scriptsize $\bm{t=0.03\,\mathrm{s}}$, \textbf{WCA = $\bm{160^\circ}$}}\\[2pt]
    \begin{tikzpicture}
    \node[anchor=south west, inner sep=0] (img) at (0,0)
        {\includegraphics[width=\linewidth]{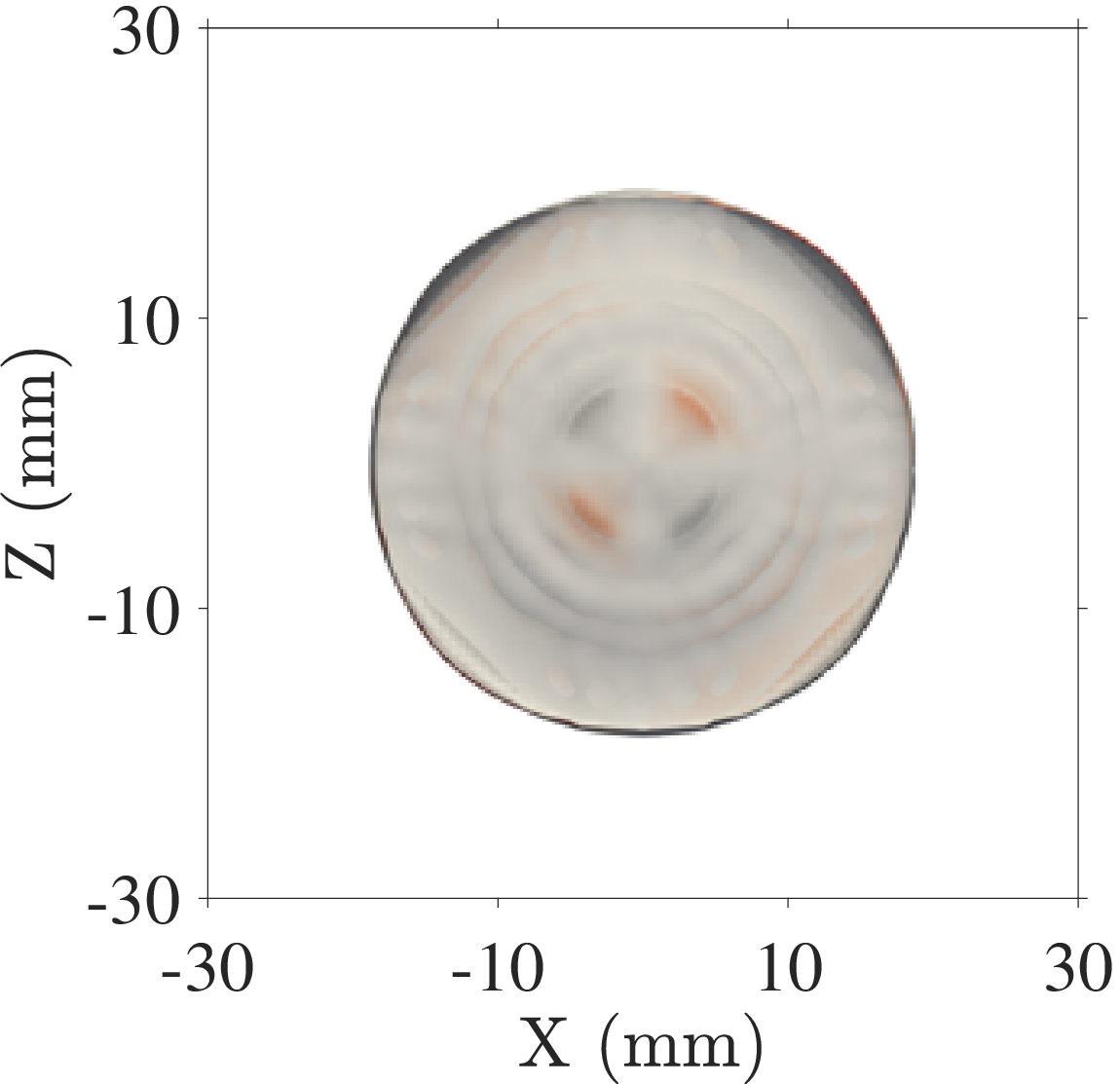}};
    \begin{scope}[x={(img.south east)},y={(img.north west)}]
        \node[anchor=north east] at (0.99,0.98) {\scriptsize \textcolor{red}{$\bm{\lambda=0.04}$}};
    \end{scope}
    \end{tikzpicture}\\[3pt]
    \begin{tikzpicture}
    \node[anchor=south west, inner sep=0] (img) at (0,0)
        {\includegraphics[width=\linewidth]{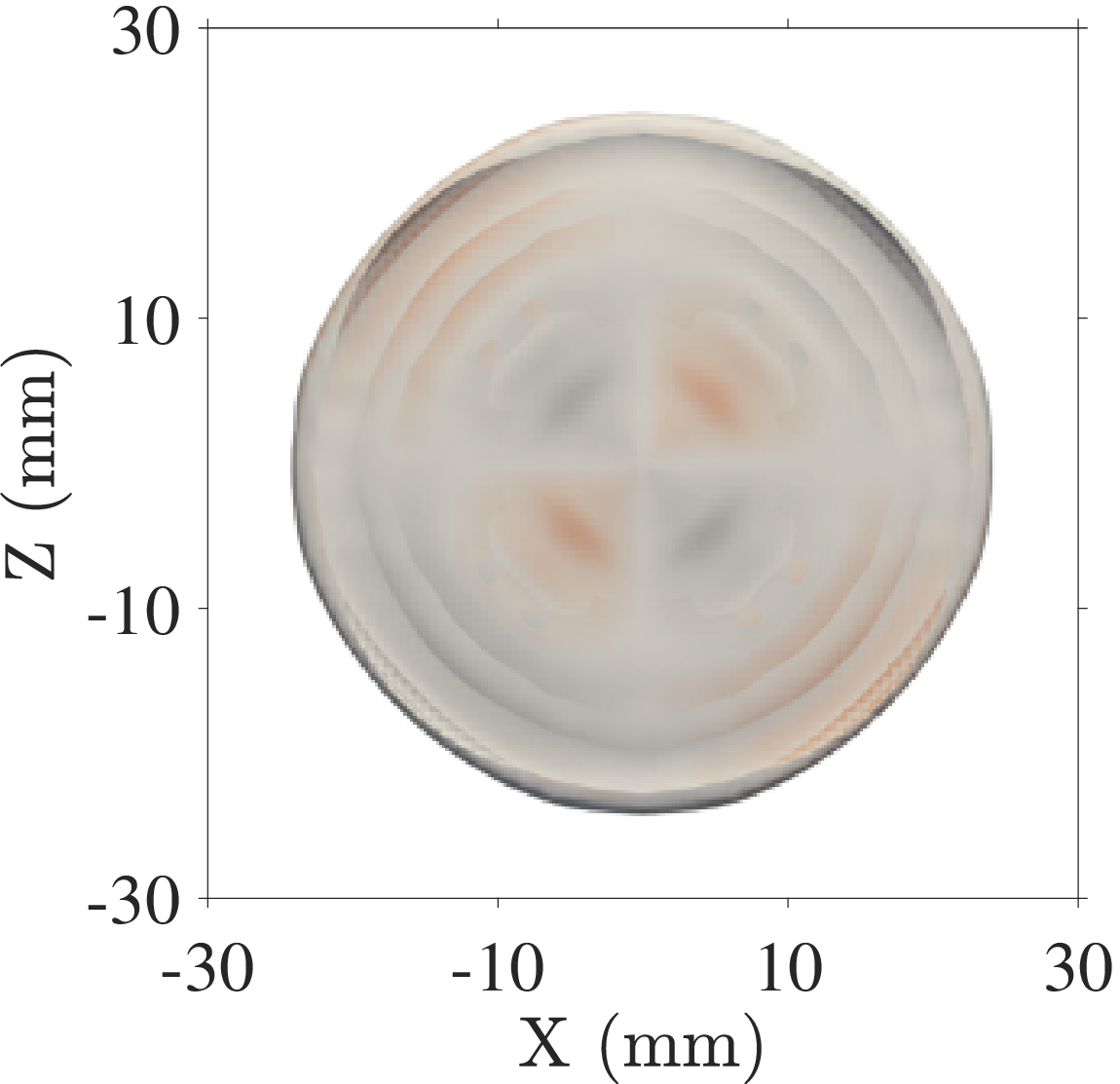}};
    \begin{scope}[x={(img.south east)},y={(img.north west)}]
        \node[anchor=north east] at (0.99,0.98) {\scriptsize \textcolor{red}{$\bm{\lambda=0.12}$}};
    \end{scope}
    \end{tikzpicture}\\[3pt]
\end{minipage}\hfill
\begin{minipage}{0.22\linewidth}
    \centering
{\scriptsize $\bm{t=0.04\,\mathrm{s}}$, \textbf{WCA = $\bm{160^\circ}$}}\\[2pt]
    \begin{tikzpicture}
    \node[anchor=south west, inner sep=0] (img) at (0,0)
        {\includegraphics[width=\linewidth]{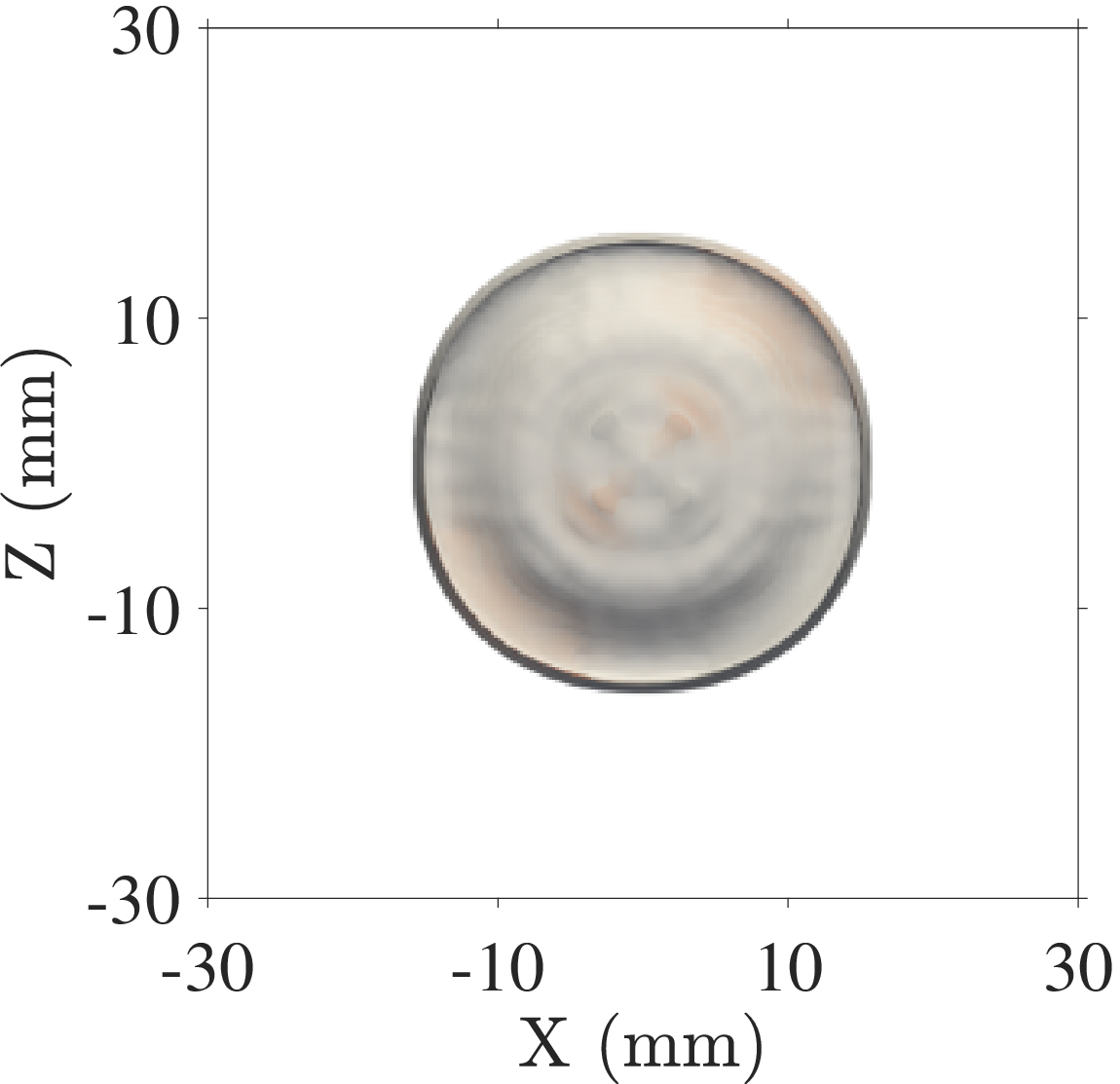}};
    \begin{scope}[x={(img.south east)},y={(img.north west)}]
        \node[anchor=north east] at (0.99,0.98) {\scriptsize \textcolor{red}{$\bm{\lambda=0.04}$}};
    \end{scope}
    \end{tikzpicture}\\[3pt]
    \begin{tikzpicture}
    \node[anchor=south west, inner sep=0] (img) at (0,0)
        {\includegraphics[width=\linewidth]{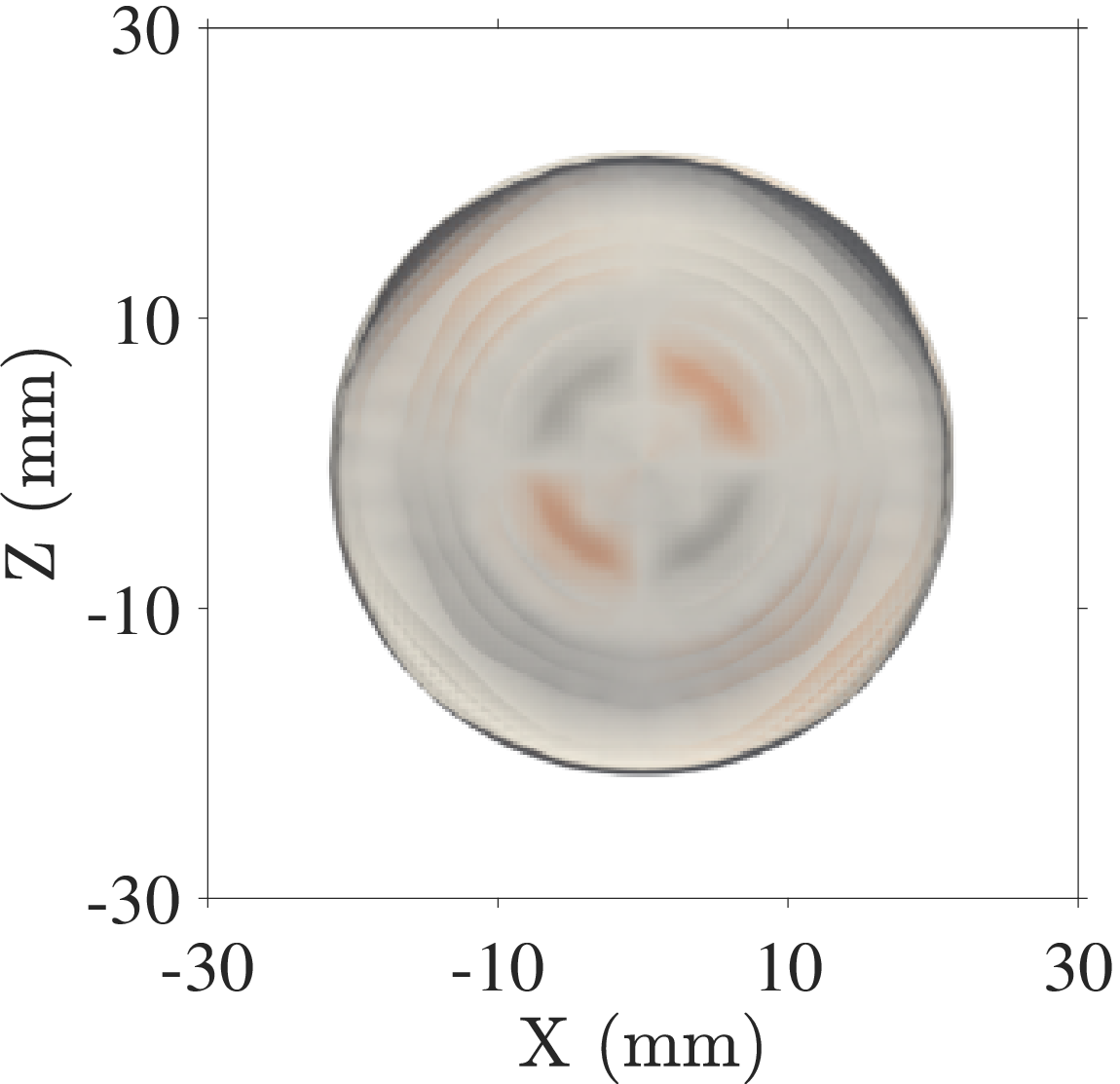}};
    \begin{scope}[x={(img.south east)},y={(img.north west)}]
        \node[anchor=north east] at (0.99,0.98) {\scriptsize \textcolor{red}{$\bm{\lambda=0.12}$}};
    \end{scope}
    \end{tikzpicture}\\[3pt]
\end{minipage}

\caption{Distribution of the $XY$--viscoelastic stress ($\tau_{MF}$) on the iso-surface of $\alpha = 0.5$ from maximum receding to the static states of the droplet. The snapshots are presented from the bottom view (contact region with the solid surface) for two relaxation times, $\lambda = 0.04$ and $\lambda = 0.12$, under different wettability conditions.}
\label{fig:tauMF_distribution}

\end{figure}

The distribution of viscoelastic stress in the $xy$-plane shown in Fig.~\ref{fig:tauMF_distribution} provides direct insight into how elastic stresses evolve during recoil and how they influence the final droplet configuration. A clear trend across all wettability conditions is that the stress intensity is significantly higher for $\lambda = 0.04$ than for $\lambda = 0.12$, which can be attributed to the shorter relaxation time associated with lower $\lambda$, leading to faster deformation rates and reduced ability of polymer chains to relax, thereby generating larger instantaneous elastic stresses \cite{bird1987dynamics,crooks2000influence}. In contrast, for higher $\lambda$, the fluid exhibits stronger elastic memory and slower stress relaxation, which distributes the deformation over a longer time and reduces peak stress levels \cite{wang2017impact,yue2012phase}. 

For the hydrophilic surface (WCA = $0^\circ$), Fig.~\ref{fig:tauMF_distribution} shows that the wetted region is noticeably larger for $\lambda = 0.12$ compared to $\lambda = 0.04$, indicating that higher relaxation time promotes sustained spreading before recoil. This behavior is consistent with the ability of viscoelastic fluids to store kinetic energy in stretched polymer chains, delaying dissipation and allowing a larger contact area to develop \cite{bartolo2007dynamics,bergeron2000controlling}. The stress distribution in this case appears more diffuse for higher $\lambda$, reflecting the gradual release of stored elastic energy.  On the hybrid surface, the stress field becomes strongly asymmetric, with higher stress concentrations localized near the hydrophobic side where the contact line experiences greater resistance. This asymmetry aligns with the previously observed directional flow and deformation, confirming that the wettability gradient induces a lateral redistribution of stresses and drives fluid migration toward the hydrophilic region \cite{darhuber2005principles,quere2008wetting}. The contrast between $\lambda = 0.04$ and $\lambda = 0.12$ remains evident, with lower $\lambda$ producing sharper and more localized stress peaks, while higher $\lambda$ results in smoother and more distributed stress patterns due to prolonged elastic relaxation. For the hydrophobic surface (WCA = $160^\circ$), the stress is highly concentrated near the contact line at early times and rapidly diminishes as the droplet retracts, reflecting the reduced wetted area and strong recoil tendency. Overall, Fig.~\ref{fig:tauMF_distribution} demonstrates that viscoelastic stress is not only sensitive to relaxation time but also strongly coupled with surface wettability, governing both the transient flow behavior and the extent of droplet attachment to the solid surface, which ultimately determines the final equilibrium morphology observed in Fig.~\ref{fig:rest_shape_lambda}.

%%%%%%%%%%%%^^^^^^^^^^^^^^^^^^^^^^^^^^^^^^^^with labale inside the figures

%%%%%%%%%%%%%%%%%%%%%%
%%%%%%%%%%%%%%%%%%%%%%

\begin{figure}[H]
\centering

% -------- Row 1: KE --------
\begin{subfigure}[b]{0.475\linewidth}
    \centering
    \includegraphics[width=\linewidth]{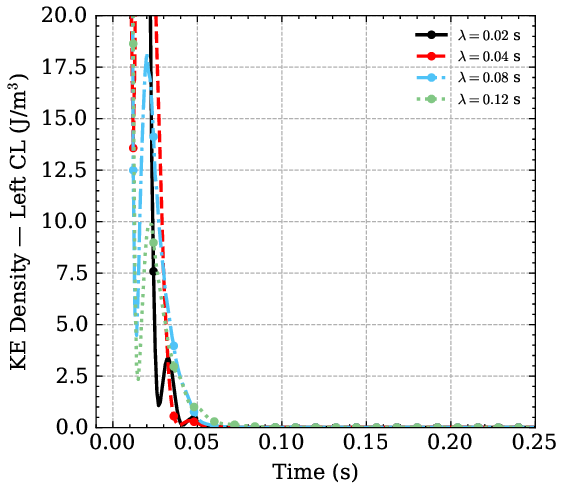}
    \subcaption{Kinetic energy (left)}
\end{subfigure}
\hfill
\begin{subfigure}[b]{0.475\linewidth}
    \centering
    \includegraphics[width=\linewidth]{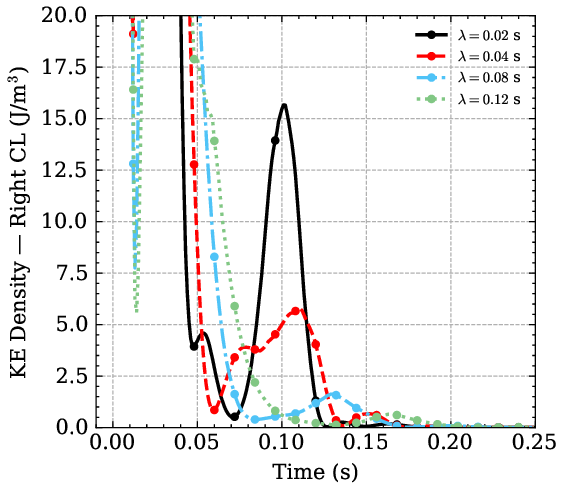}
    \subcaption{Kinetic energy (right)}
\end{subfigure}

\vspace{3pt}

% -------- Row 2: N1 --------
\begin{subfigure}[b]{0.475\linewidth}
    \centering
    \includegraphics[width=\linewidth]{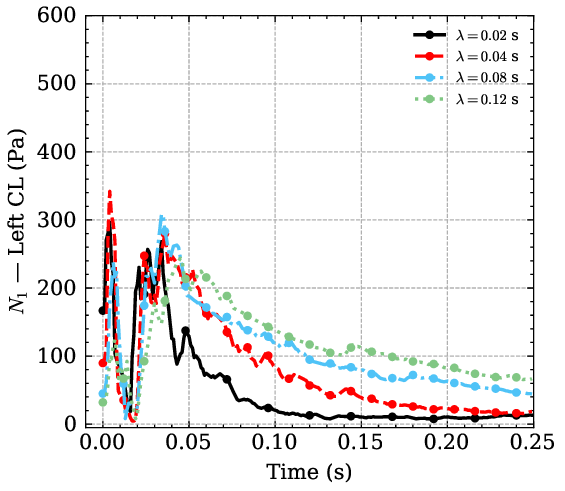}
    \subcaption{First normal stress difference $N_1$ (left)}
\end{subfigure}
\hfill
\begin{subfigure}[b]{0.475\linewidth}
    \centering
    \includegraphics[width=\linewidth]{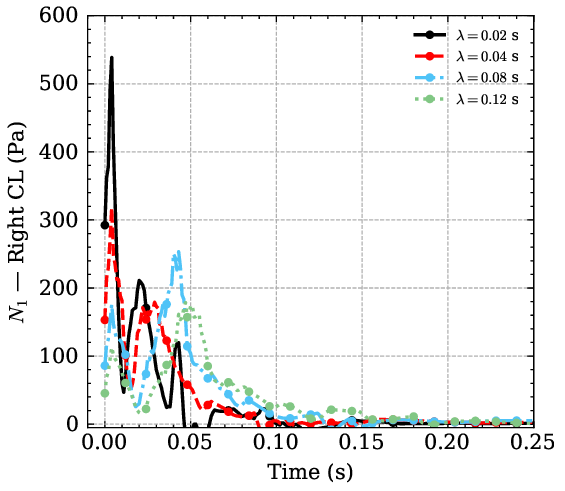}
    \subcaption{First normal stress difference $N_1$ (right)}
\end{subfigure}

\caption{Comparison of kinetic energy (top row) and first normal stress difference $N_1$ (bottom row) for left and right configurations.}
\label{fig:ke_n1_comparison}

\end{figure}

%%%%%%%%%%%%^^^^^^^^^^^^^^^^^^^^^^^^^^^^^^^^with labale inside the figures
Figure~\ref{fig:ke_n1_comparison} presents the temporal evolution of kinetic energy and first normal stress difference $N_1$ for the hybrid surface, providing a quantitative description of the asymmetric axisymmetric bubble-like dynamics. The kinetic energy plots in Fig.~\ref{fig:ke_n1_comparison}a,b show a rapid post-impact decay, but the decay is clearly delayed on the hydrophobic side, where secondary peaks persist longer and reflect stronger recoil under weaker solid--liquid affinity. By contrast, the hydrophilic side loses kinetic energy more quickly, consistent with stronger wetting and more efficient damping of motion \cite{josserand2016drop,bonn2009wetting}. The comparison across relaxation times shows that increasing $\lambda$ prolongs the lifetime of kinetic-energy fluctuations on both sides, with the longest persistence appearing for $\lambda = 0.12$, whereas $\lambda = 0.02$ relaxes the fastest; this indicates that a larger relaxation time sustains internal motion and delays the approach to equilibrium. This imbalance in energy decay contributes directly to the asymmetric droplet morphology, since fluid is retained longer on the hydrophilic side while the hydrophobic side undergoes stronger rebound.

The evolution of $N_1$ in Fig.~\ref{fig:ke_n1_comparison}c,d further clarifies the role of viscoelasticity in this process. The initial sharp peaks correspond to rapid polymer stretching during impact and early spreading, followed by a gradual decay as elastic stresses relax. Here again, the effect of $\lambda$ is pronounced: larger values produce higher and more persistent $N_1$ levels, indicating stronger elastic memory and slower stress relaxation \cite{bird1987dynamics,wang2017impact}. The stress response is also more oscillatory on the hydrophobic side, where the contact line experiences stronger localized deformation during recoil, whereas the hydrophilic side relaxes more smoothly. Taken together, the graphs show that increasing $\lambda$ amplifies both the duration and asymmetry of the kinetic and elastic responses, so the competition between rapid dissipation on the hydrophilic side and sustained recoil on the hydrophobic side governs the axisymmetric bubble-like deformation and delays the transition to equilibrium.

\subsection{Effect of Surface Tension at Fixed Relaxation Time}

\begin{figure}[H]
\centering

% -------- Row 1: Three subfigures --------
\begin{subfigure}[b]{0.32\linewidth}
\centering
\includegraphics[width=\linewidth]{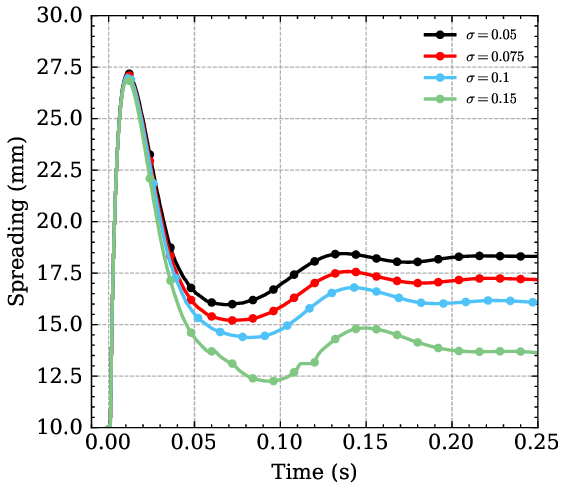}
\subcaption{Spreading}
\end{subfigure}
\hfill
\begin{subfigure}[b]{0.32\linewidth}
\centering
\includegraphics[width=\linewidth]{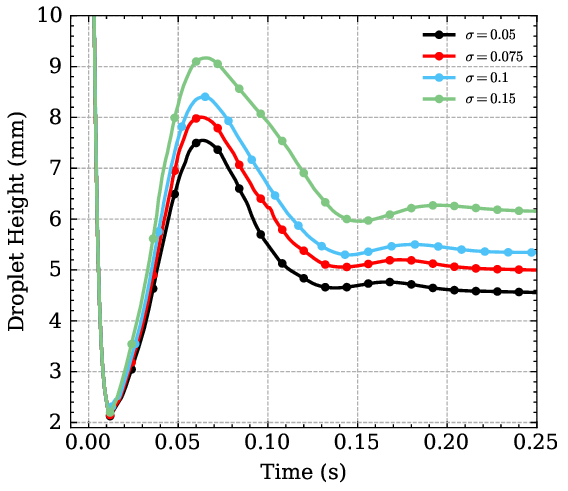}
\subcaption{Droplet Height}
\end{subfigure}
\hfill
\begin{subfigure}[b]{0.32\linewidth}
\centering
\includegraphics[width=\linewidth]{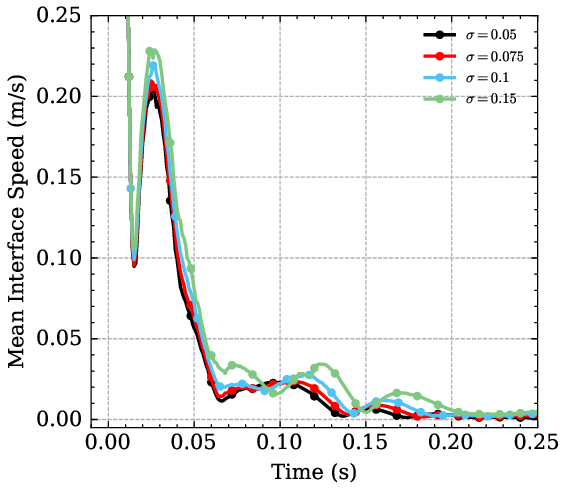}
\subcaption{Mean Interface Speed}
\end{subfigure}

\vspace{10pt}

% -------- Row 2: Two subfigures nicely centered --------
\centering
\begin{subfigure}[b]{0.32\linewidth}
\centering
\includegraphics[width=\linewidth]{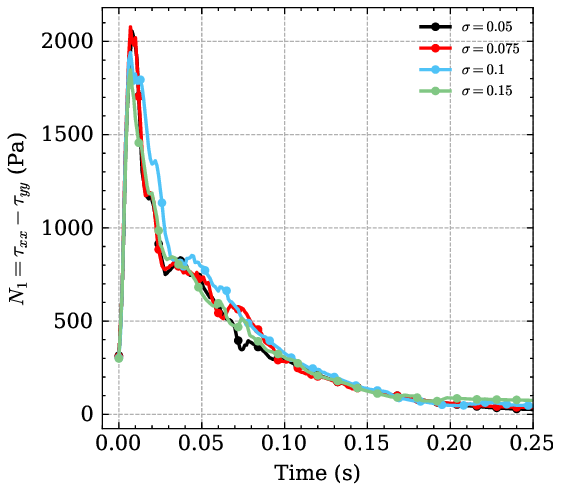}
\subcaption{Normal Stress Difference $N_1$}
\end{subfigure}
\hspace{0.04\linewidth}
\begin{subfigure}[b]{0.32\linewidth}
\centering
\includegraphics[width=\linewidth]{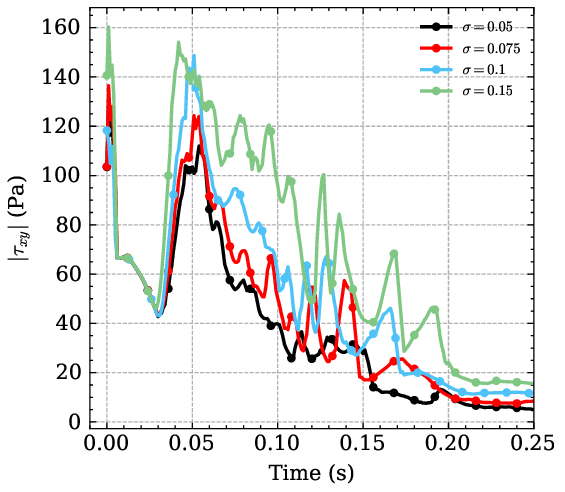}
\subcaption{Shear Stress $|\tau_{xy}|$}
\end{subfigure}

\caption{Time evolution of key interface quantities for viscoelastic droplet impact at $\lambda = 0.05~\text{s}$, showing the influence of surface tension ($\sigma = 0.05$, 0.075, 0.1, and 0.15~N/m): (a) spreading, (b) droplet height, (c) mean interface speed, (d) normal stress difference $N_1 = \tau_{xx} - \tau_{yy}$, and (e) interfacial shear stress $|\tau_{xy}|$.}
\label{fig:zigma_impact}
\end{figure}

To isolate the influence of capillary forces on viscoelastic droplet impact, the effect of surface tension \(\sigma\) is examined at a fixed relaxation time of \(\lambda = 0.05~\mathrm{s}\) on the hybrid wettability surface (\(\mathrm{WCA}=0^\circ{-}160^\circ\)). Varying \(\sigma\) from 0.05 to 0.15~N/m directly modifies the Weber and E\"otv\"os numbers while keeping other material properties and impact conditions constant, thereby shifting the balance among inertia, gravity, and capillarity. Figure~\ref{fig:zigma_impact} presents the time evolution of key interfacial quantities for \(\sigma = 0.05\), 0.075, 0.1, and 0.15~N/m. As shown in Fig.~\ref{fig:zigma_impact}(a), lower surface tension leads to larger maximum spreading due to a higher effective Weber number that allows inertia to dominate during the early impact stage; increasing \(\sigma\) suppresses radial expansion by 1.1\% (from 27.21~mm to 26.90~mm) and results in a more compact footprint \cite{josserand2016drop}. The corresponding droplet height evolution in Fig.~\ref{fig:zigma_impact}(b) reveals that higher \(\sigma\) maintains a thicker profile and accelerates vertical recovery, with the minimum height increasing by 3.3\% (from 2.12~mm to 2.20~mm), consistent with stronger capillary-driven recoil \cite{bonn2009wetting}. Higher \(\sigma\) also produces stronger rebound-driven motion in the mean interface speed (Fig.~\ref{fig:zigma_impact}(c)), indicating more efficient conversion of surface energy into kinetic energy after the spreading phase.

The viscoelastic stress responses further highlight the coupling between capillarity and elasticity. The normal stress difference \(N_1\) in Fig.~\ref{fig:zigma_impact}(d) exhibits slightly lower peaks at higher \(\sigma\) due to reduced deformation during spreading, whereas the shear stress \(|\tau_{xy}|\) in Fig.~\ref{fig:zigma_impact}(e) remains elevated for longer durations and increases by 37.6\% in its maximum value, reflecting prolonged interfacial deformation and slower but more controlled stress relaxation \cite{wang2017impact,bird1987dynamics}. These trends occur because stronger capillary forces limit the extent of polymer stretching early in the impact while sustaining shear at the interface over a longer period. On the hybrid surface the wettability gradient amplifies the overall effect, resulting in a coupled redistribution of mass, momentum, and elastic stresses that governs the droplet’s approach to equilibrium.
%%%%%%%%%%%%%%%%%%%%%%%%%%%%%%%%%%%%%%%%%%%%

\begin{figure}[htbp]
\centering
    \includegraphics[width=0.45\linewidth]{taulegend.png}\\[3pt]
% -------- Figure 1 --------
\begin{minipage}{0.22\linewidth}
    \centering
\begin{tikzpicture}
\node[anchor=south west, inner sep=0] (img) at (0,0)
    {\includegraphics[width=\linewidth]{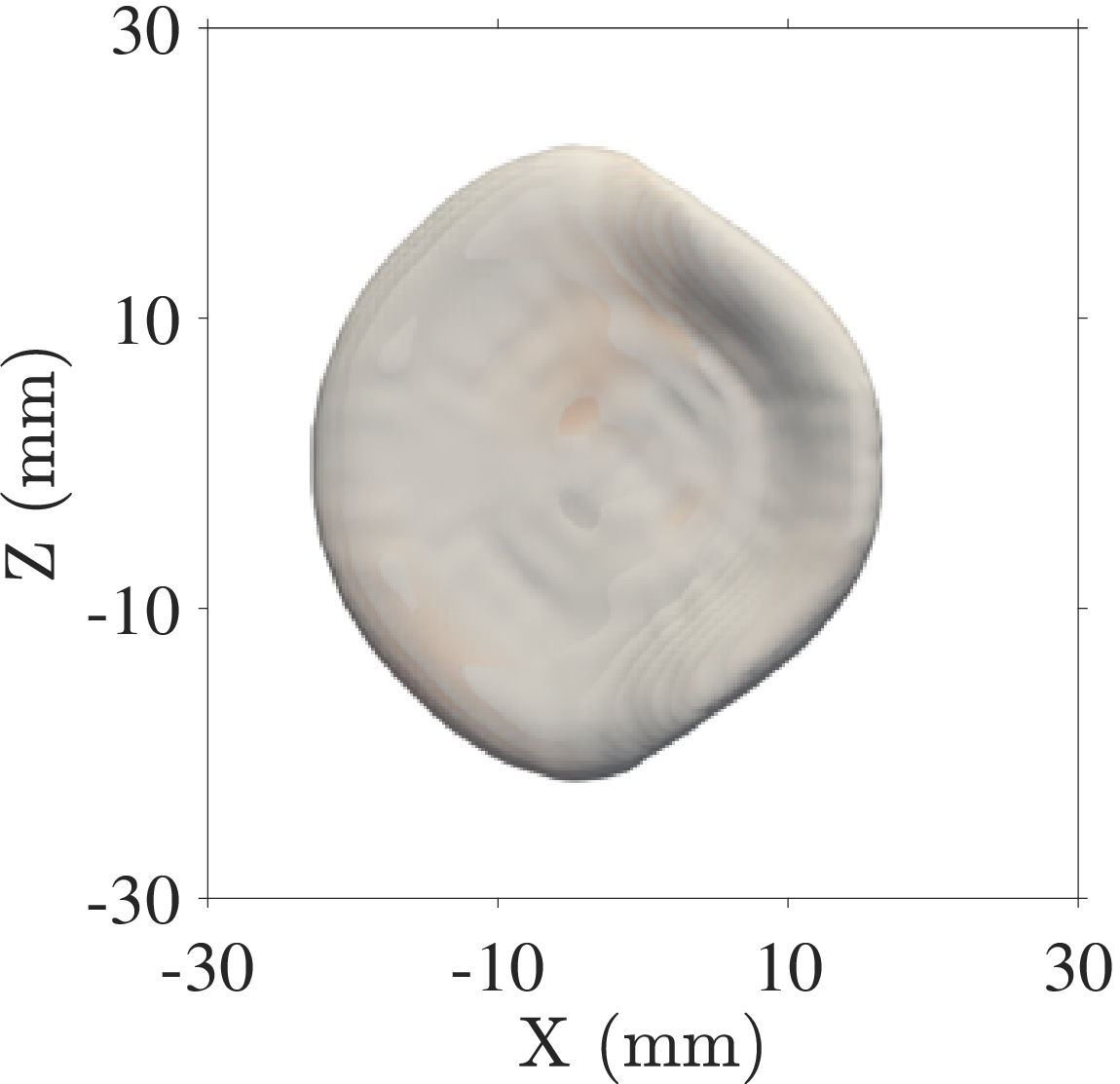}};
\begin{scope}[x={(img.south east)},y={(img.north west)}]
\draw[blue, dashed, thick] (0.6, 0.67) ellipse (0.125 and 0.175);
\end{scope}
\end{tikzpicture}
    
    \vspace{2pt}

    \includegraphics[width=\linewidth]{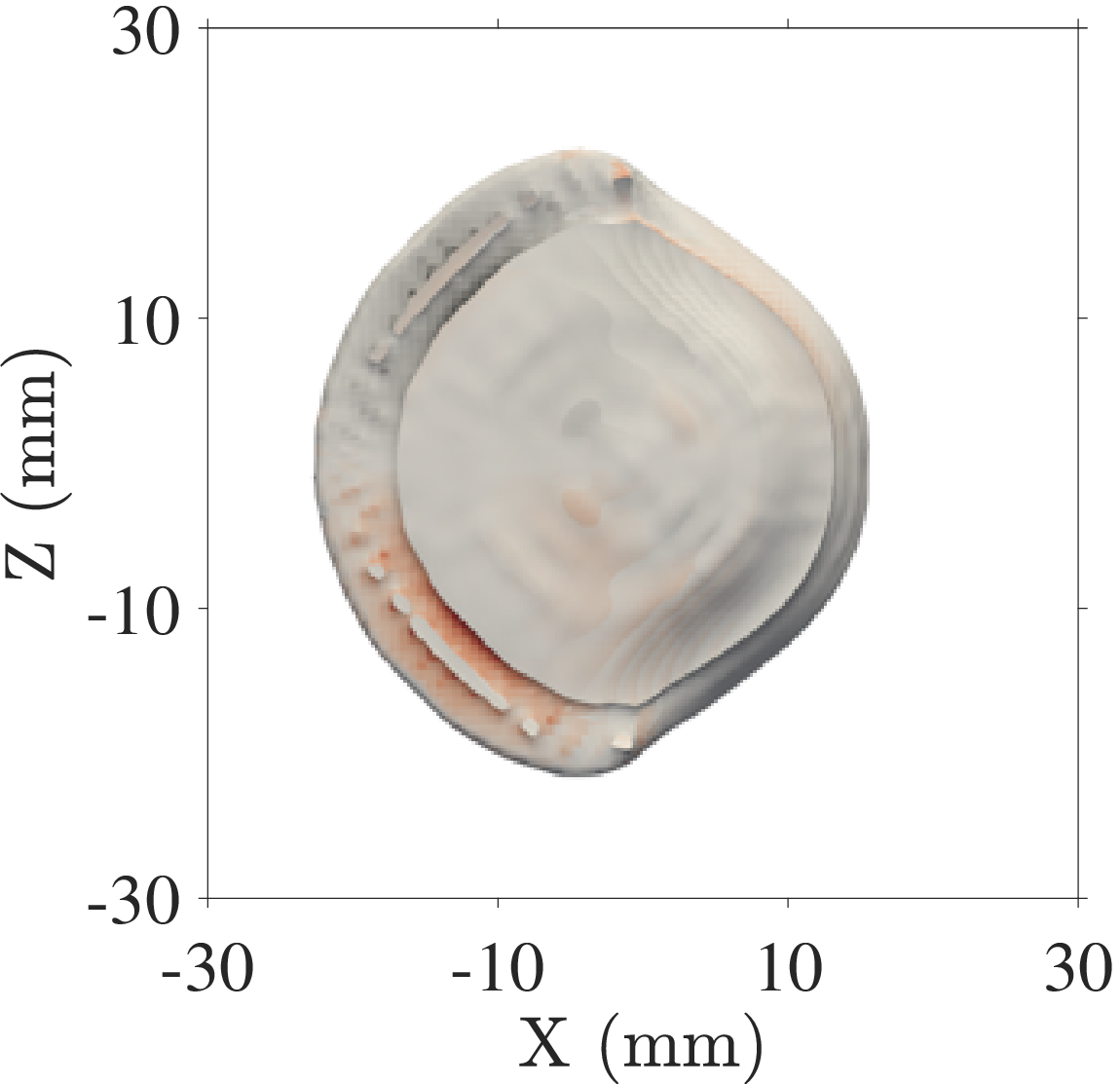}
    \vspace{2pt}
    \includegraphics[width=\linewidth]{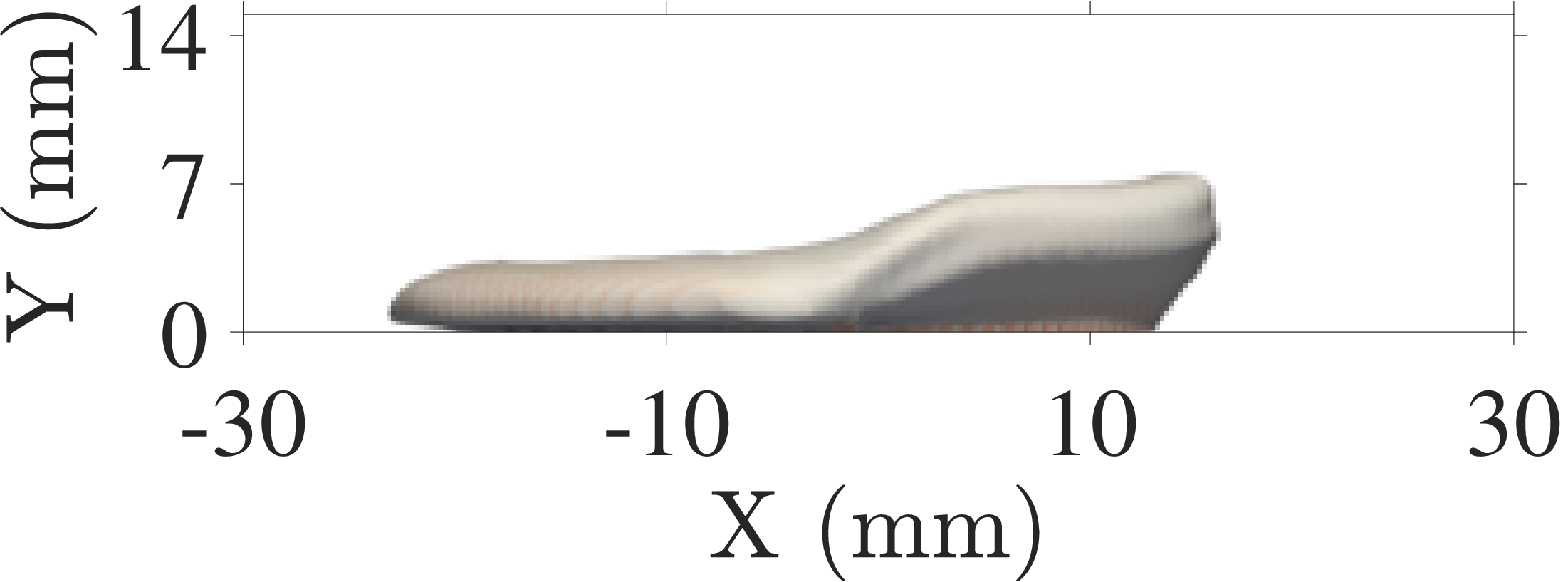}
    
    \vspace{2pt}
    {\small $\sigma = 0.05~\mathrm{N/m}$}
\end{minipage}
\hfill
% -------- Figure 2 --------
\begin{minipage}{0.22\linewidth}
    \centering
\begin{tikzpicture}
\node[anchor=south west, inner sep=0] (img) at (0,0)
    {    \includegraphics[width=\linewidth]{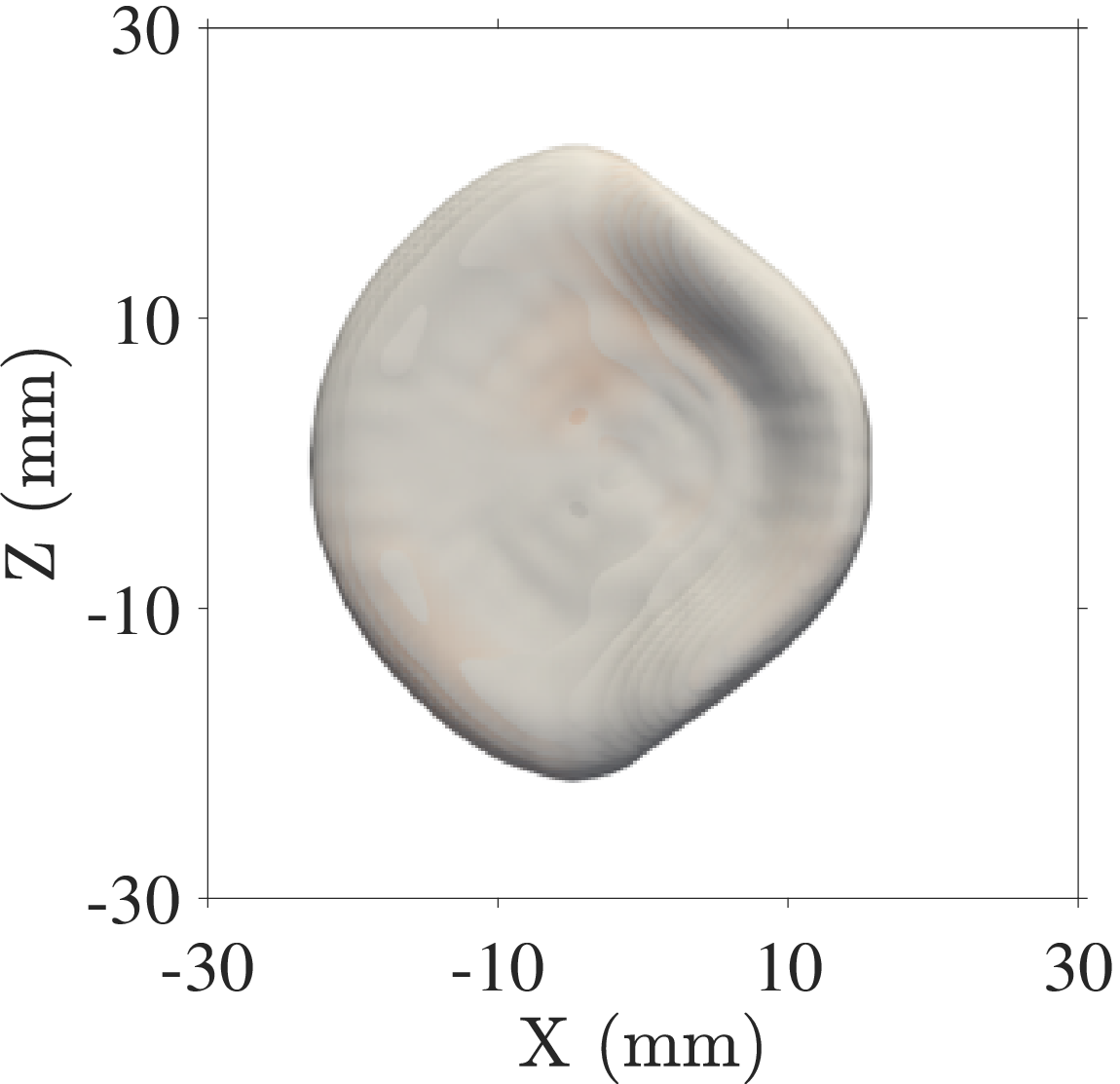}};
\begin{scope}[x={(img.south east)},y={(img.north west)}]
\draw[blue, dashed, thick] (0.6, 0.67) ellipse (0.125 and 0.175);
\end{scope}
\end{tikzpicture}
    \vspace{2pt}
    \includegraphics[width=\linewidth]{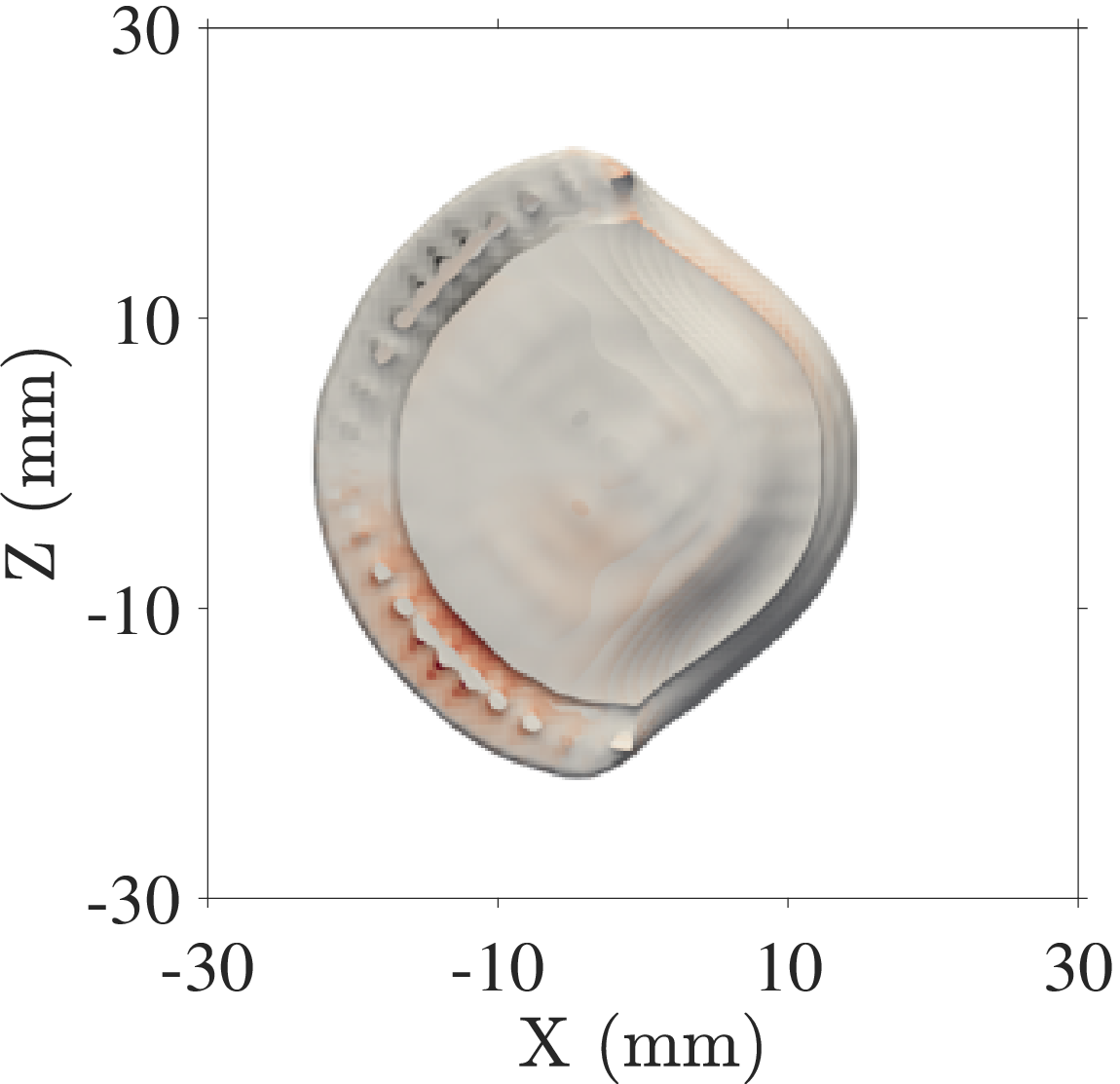}
    \vspace{2pt}
    \includegraphics[width=\linewidth]{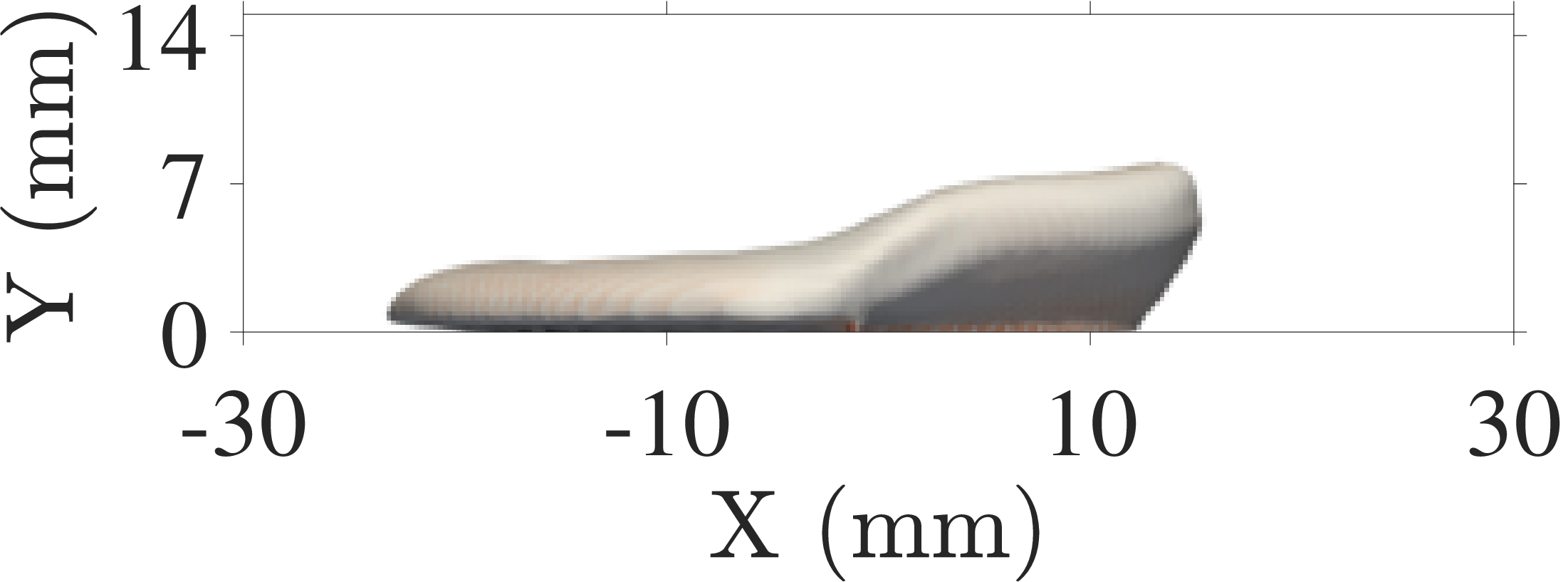}
    
    \vspace{2pt}
    {\small $\sigma = 0.075~\mathrm{N/m}$}
\end{minipage}
\hfill
% -------- Figure 3 --------
\begin{minipage}{0.22\linewidth}
    \centering
\begin{tikzpicture}
\node[anchor=south west, inner sep=0] (img) at (0,0)
    {    \includegraphics[width=\linewidth]{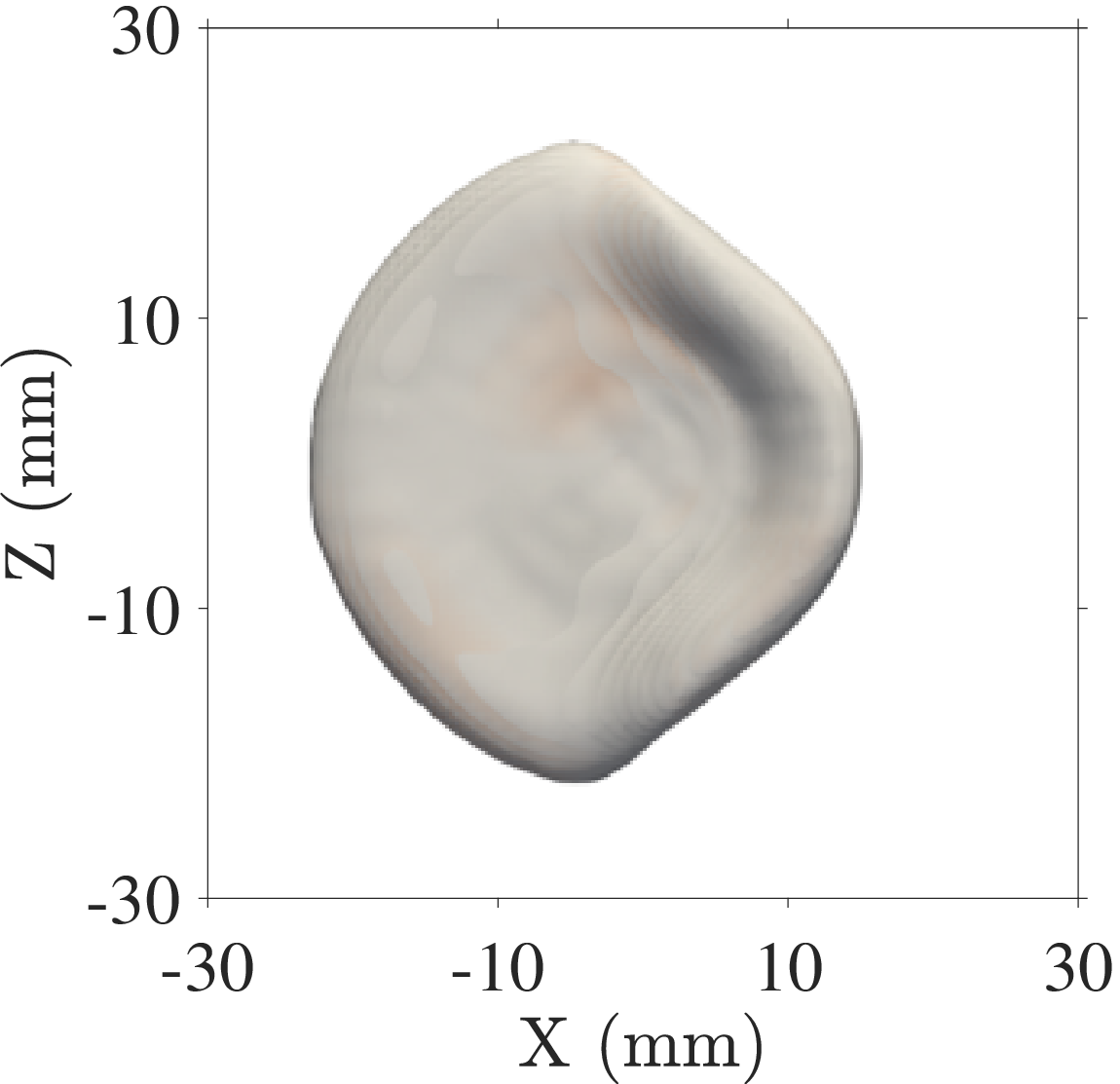}};
\begin{scope}[x={(img.south east)},y={(img.north west)}]
\draw[blue, dashed, thick] (0.6, 0.67) ellipse (0.125 and 0.175);
\end{scope}
\end{tikzpicture}
    \vspace{2pt}
    \includegraphics[width=\linewidth]{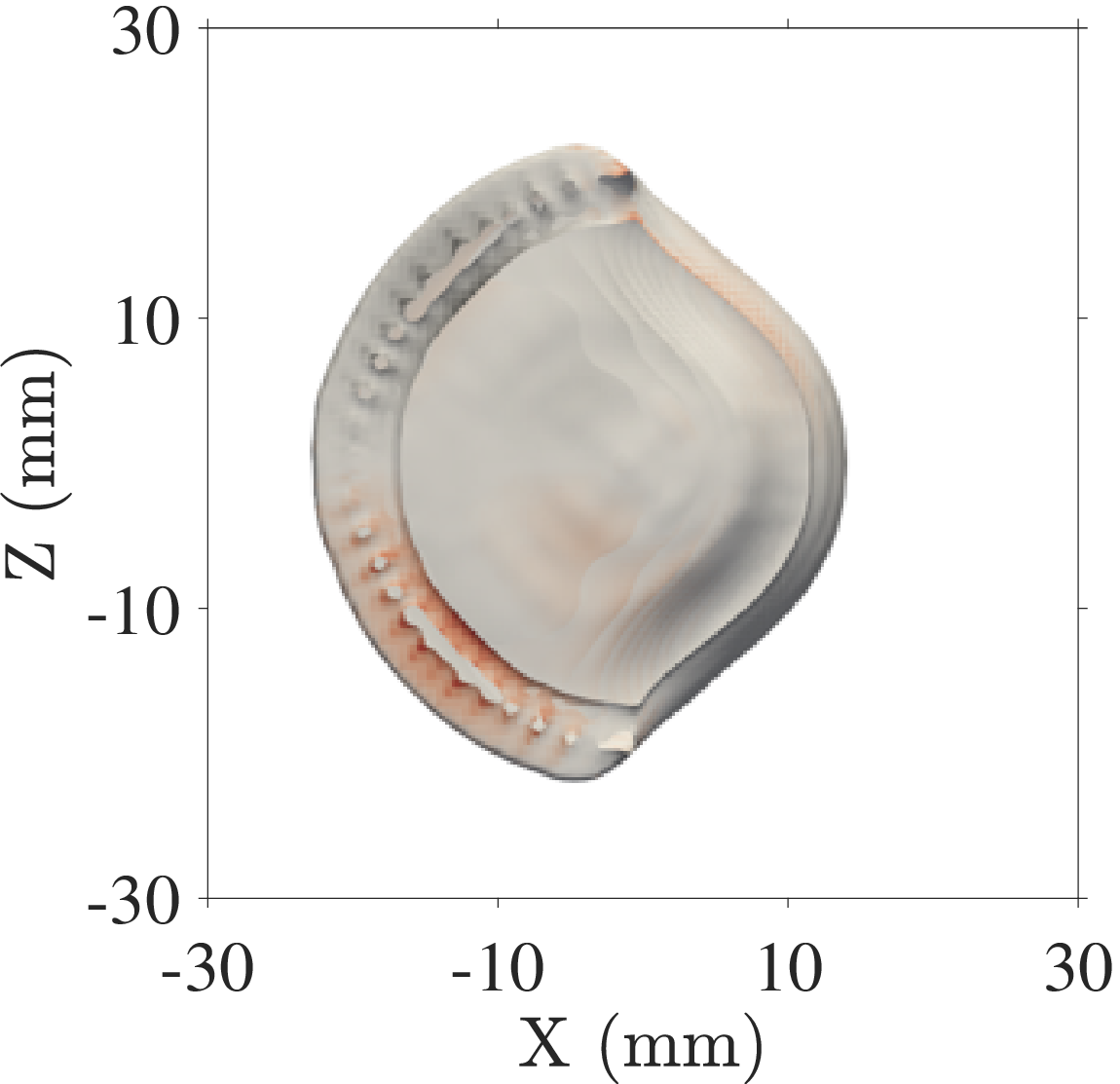}
    \vspace{2pt}
    \includegraphics[width=\linewidth]{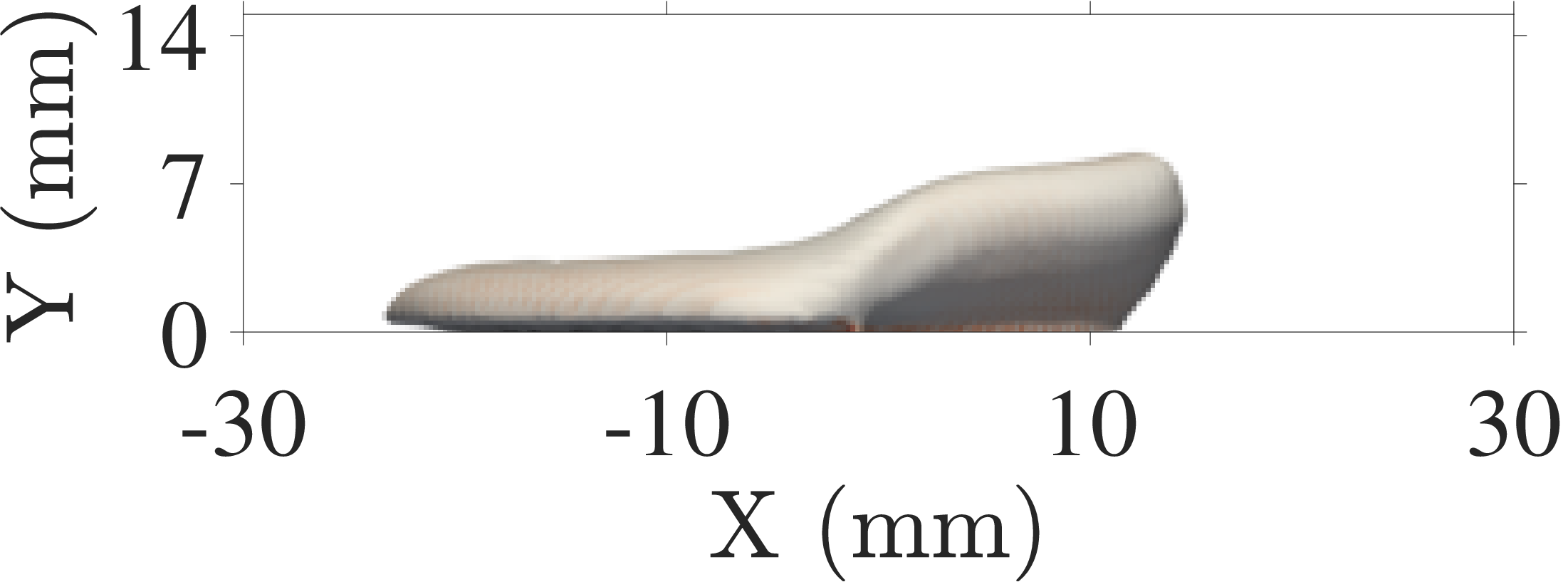}
    
    \vspace{2pt}
    {\small $\sigma = 0.10~\mathrm{N/m}$}
\end{minipage}
\hfill
% -------- Figure 4 --------
\begin{minipage}{0.22\linewidth}
    \centering
\begin{tikzpicture}
\node[anchor=south west, inner sep=0] (img) at (0,0)
    { \includegraphics[width=\linewidth]{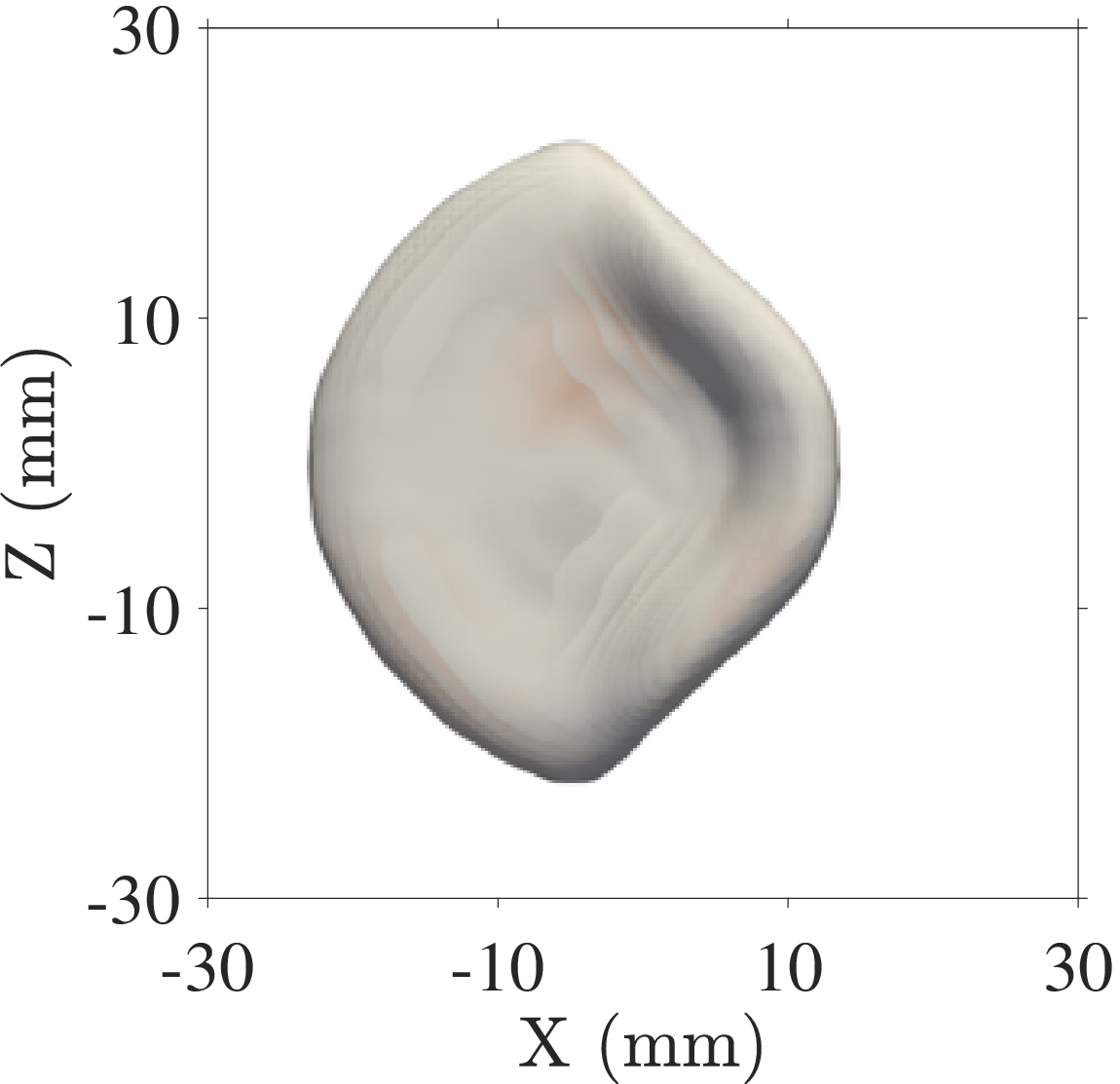}};
\begin{scope}[x={(img.south east)},y={(img.north west)}]
\draw[blue, dashed, thick] (0.6, 0.67) ellipse (0.125 and 0.175);
\end{scope}
\end{tikzpicture}
    \vspace{2pt}
    \includegraphics[width=\linewidth]{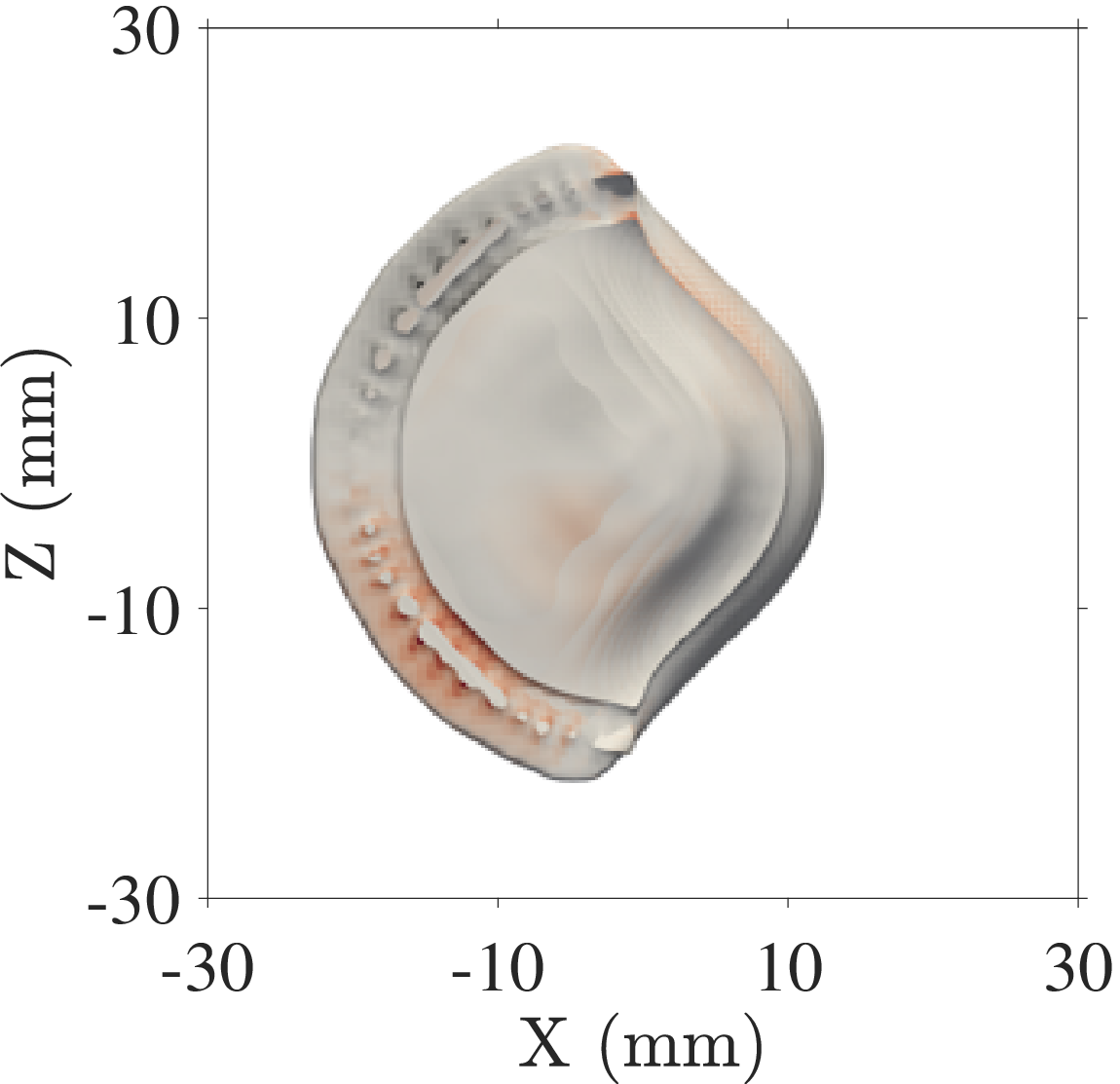}
    \vspace{2pt}
    \includegraphics[width=\linewidth]{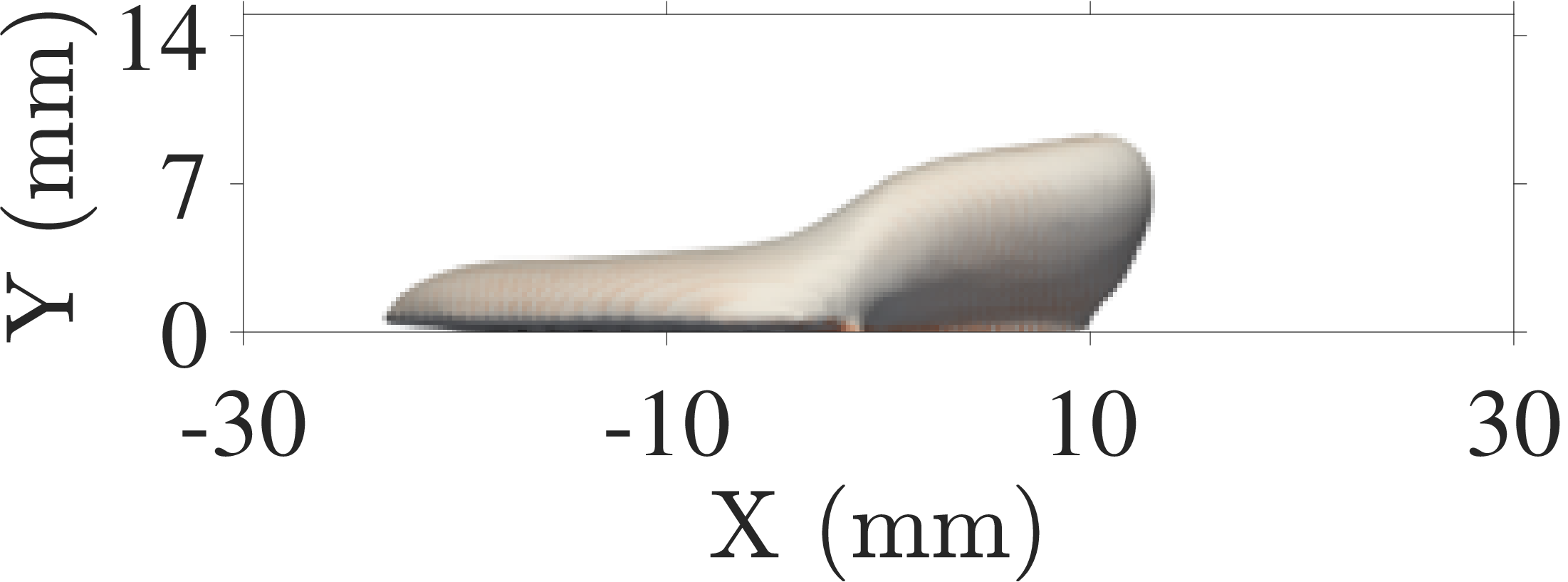}
    
    \vspace{2pt}
    {\small $\sigma = 0.15~\mathrm{N/m}$}
\end{minipage}

\caption{Distribution of the viscoelastic stress $\tau_{MF}$ on the iso-surface of $\alpha=0.5$ for the axisymmetric bubble-like droplet under different surface tensions. The top row shows the top view, the second row shows the bottom view in contact with the solid surface, and the third row shows the side view, which exhibits the shoe-like droplet morphology. The columns correspond to $\sigma = 0.05$, $0.075$, $0.10$, and $0.15~\mathrm{N/m}$.}
\label{fig:tauMF_sigma_views}
\end{figure}

Figure~\ref{fig:tauMF_sigma_views} illustrates the distribution of the viscoelastic stress $\tau_{MF}$ in the $xy$-plane for the axisymmetric bubble-like droplet at different surface tensions, revealing a clear modification of both stress localization and droplet morphology. In the top view (first row), the droplet exhibits a distinct dustpan-like shape, as highlighted by the dashed ellipse, where a scooped and asymmetric front develops along the advancing side. As the surface tension increases, this dustpan structure becomes deeper and more pronounced. This behavior arises because higher $\sigma$ strengthens capillary restoring forces, which suppress lateral spreading while promoting curvature-driven contraction toward the droplet interior. As a result, fluid accumulates more strongly in the advancing region, enhancing the concave geometry and concentrating viscoelastic stresses in a narrower zone \cite{josserand2016drop,bonn2009wetting}.  In contrast, at lower surface tension, the droplet spreads more extensively due to a higher effective Weber number, leading to a shallower and more diffuse dustpan shape, accompanied by broader stress distribution. The bottom view (second row) confirms this trend, where the contact region exhibits more irregular and extended stress patterns at low $\sigma$, while higher $\sigma$ produces smoother and more localized stress bands along the contact line. The side view (third row) consistently shows a shoe-like profile, where increasing $\sigma$ results in a thicker and more elevated front, reflecting enhanced capillary resistance to flattening. Therefore, Fig.~\ref{fig:tauMF_sigma_views} demonstrates that increasing surface tension not only limits spreading but also intensifies curvature-driven deformation, leading to a deeper dustpan morphology in the top view and a more pronounced shoe-like structure in the side view. These changes are directly linked to the redistribution of viscoelastic stresses, where higher $\sigma$ confines deformation and organizes stress along the interface, while lower $\sigma$ allows more widespread deformation and weaker stress localization \cite{wang2017impact,bird1987dynamics}.

%%%%%%%%%%%%%%%%%%%%%%%%%%%%end of with frame

\section{Conclusion}

This study presents a comprehensive three-dimensional numerical investigation of viscoelastic droplet impact on surfaces with uniform and hybrid wettability. Using a high-fidelity OpenFOAM-based solver that incorporates the Oldroyd-B constitutive model and a velocity-dependent dynamic contact angle formulation, we systematically examined the coupled influence of fluid relaxation time and surface tension under gravitational-capillary balance.

Key findings reveal that increasing the relaxation time from $0.02~\mathrm{s}$ to $0.12~\mathrm{s}$ enhances elastic energy storage, resulting in a $12.6\%$--$12.9\%$ increase in the maximum spreading diameter (from approximately $24.97~\mathrm{mm}$ to $28.09$--$28.17~\mathrm{mm}$) and a $16.6\%$ reduction in the minimum droplet height across all surface configurations. This enhancement arises from delayed viscous dissipation and prolonged inertial spreading due to greater elastic memory. In contrast, increasing surface tension from $0.05~\mathrm{N/m}$ to $0.15~\mathrm{N/m}$ at a fixed relaxation time of $\lambda=0.05~\mathrm{s}$ suppresses the maximum spreading diameter by approximately $1.1\%$ (from $27.21~\mathrm{mm}$ to $26.90~\mathrm{mm}$) while increasing the minimum droplet height by $3.3\%$ (from $2.12~\mathrm{mm}$ to $2.20~\mathrm{mm}$), demonstrating the dominant role of capillary forces in limiting radial expansion and promoting faster vertical recovery.

On hybrid hydrophilic--hydrophobic surfaces ($\mathrm{WCA}=0^\circ$--$160^\circ$), the sharp wettability contrast induces pronounced asymmetric spreading and directional fluid migration toward the hydrophilic region. This leads to distinctive dustpan-like (top view) and shoe-like (side view) equilibrium morphologies, with viscoelastic stresses showing strong localization near the wettability discontinuity. The interplay between elastic stresses and the wettability gradient significantly amplifies asymmetry in both kinetic energy decay and first normal stress difference $N_1$, with higher relaxation times prolonging internal flow oscillations and sustaining asymmetric stress distributions.

The results demonstrate that fluid viscoelasticity and surface tension are strongly coupled with heterogeneous wettability. Longer relaxation times promote enhanced spreading and persistent elastic effects, whereas higher surface tension favors compact footprints and stronger recoil. These findings provide valuable quantitative insights for the design of advanced surfaces in inkjet printing, spray coating, and microfluidics, where precise control of droplet spreading, rebound, and final morphology is critical.

\section*{Acknowledgments}
The authors acknowledge the VEGA high-performance computing (HPC) system at Embry-Riddle Aeronautical University for providing the computational resources necessary to carry out this work.

\section*{Data Availability Statement}
The data that support the findings of this study are available from the corresponding author upon reasonable request.

\bibliographystyle{elsarticle-num} 
\bibliography{cas-refs}
\end{document}